\documentclass[journal]{IEEEtran}
\ifCLASSINFOpdf
\else
\fi

\usepackage{graphicx}
\usepackage{amsmath}
\def\R#1{(\ref{#1})}
\usepackage{tikz}
\usepackage{lipsum}
\usepackage{color}
\usepackage{cite}
\usepackage{multirow}
\usepackage{soul}
\usepackage{amsfonts}
\usepackage{amssymb}
\usepackage{algorithm}
\usepackage{algpseudocode}

\algnewcommand{\Initialize}[1]{%
  \State \textbf{Initialize:}
  \Statex \hspace*{\algorithmicindent}\parbox[t]{.8\linewidth}{\raggedright #1}
}
\algnewcommand{\REQUIRES}[1]{%
  \State \textbf{Require:}
  \Statex \hspace*{\algorithmicindent}\parbox[t]{.8\linewidth}{\raggedright #1}
}

\newcommand{\dquotes}[1]{``#1''}

\newcommand{\bs}{\boldsymbol}
\DeclareMathOperator*{\argmin}{arg\,min}
\newcommand{\ones}{\mathbf 1}

\makeatletter
\def\ps@IEEEtitlepagestyle{%
  \def\@oddfoot{\mycopyrightnotice}%
  \def\@evenfoot{}%
}

\def\mycopyrightnotice{
  {\footnotesize
  \begin{minipage}{\textwidth}
  \centering
  Copyright~\copyright~2019 IEEE. Personal use of this material is permitted. Permission from IEEE must be obtained for all other uses, in any current or future media, including reprinting/republishing this material for advertising or promotional purposes, creating new collective works, for resale or redistribution to servers or lists, or reuse of any copyrighted component of this work in other works. 
  \end{minipage}
  }
}

\begin{document}

\title{Improved low-count quantitative PET reconstruction with an iterative neural network}
\author{Hongki Lim,~\IEEEmembership{Student Member,~IEEE,}~Il Yong Chun,~\IEEEmembership{Member,~IEEE,}\\
        Yuni K. Dewaraja,~\IEEEmembership{Member,~IEEE,}~and~Jeffrey A. Fessler,~\IEEEmembership{Fellow,~IEEE}
\thanks{This work was supported by grant R01 EB022075, awarded by the National Institute of Biomedical Imaging and  Bioengineering,  National  Institute  of  Health,  U.S. Department of Health and Human Services.}
\thanks{Hongki Lim and Jeffrey A. Fessler are with the Department of Electrical Engineering and Computer Science, University of Michigan, Ann Arbor, MI 48109 USA (email: \{hongki, fessler\}@umich.edu).}
\thanks{Il Yong Chun is with the Department of Electrical Engineering, University of Hawai'i--M\=anoa, HI 96822 USA (iychun@hawaii.edu).}
\thanks{Yuni K. Dewaraja is with the Department of Radiology, University of Michigan, Ann Arbor, MI 48109 USA (yuni@med.umich.edu).}
        }
\maketitle

\begin{abstract}
Image reconstruction in low-count PET is particularly challenging because gammas from natural radioactivity in Lu-based crystals cause high random fractions that lower the measurement signal-to-noise-ratio (SNR). In model-based image reconstruction (MBIR), using more iterations of an unregularized method may increase the noise, so incorporating regularization into the image reconstruction is desirable to control the noise. New regularization methods based on learned convolutional operators are emerging in MBIR. We modify the architecture of an iterative neural network, \emph{BCD-Net}, for PET MBIR, and demonstrate the efficacy of the trained BCD-Net using XCAT phantom data that simulates the low true coincidence count-rates with high random fractions typical for Y-90 PET patient imaging after Y-90 microsphere radioembolization. Numerical results show that the proposed BCD-Net significantly improves CNR and RMSE of the reconstructed images compared to MBIR methods using non-trained regularizers, total variation (TV) and non-local means (NLM). Moreover, BCD-Net successfully generalizes to test data that differs from the training data. Improvements were also demonstrated for the clinically relevant phantom measurement data where we used training and testing datasets having very different activity distributions and count-levels.

\end{abstract}

\IEEEpeerreviewmaketitle
\vspace{-1.5em}
\section{Introduction}\label{sec:intro}
Image reconstruction in low-count PET is particularly challenging because dominant gammas from natural radioactivity in Lu-based crystals cause high random fractions,
lowering the measurement signal-to-noise-ratio (SNR) \cite{carlier:15:ypi}. To accurately reconstruct images in low-count PET, regularized model-based image reconstruction (MBIR) solves the following optimization problem consisting of \textit{1)} a data fidelity term $f(\bs{x})$ that models the physical PET imaging system, and \textit{2)} a regularization term $\mathsf{R}(\bs{x})$ that penalizes image roughness and controls noise\cite{ahn2015quantitative}:
\begin{align}
    \hat{\bs{x}} &= \arg \min_{\bs{x}\ge \bs{0}} f(\bs{x}) + \mathsf{R}(\bs{x}) \label{eq1}\\
    f(\bs{x}) &:= \ones^T (\bs{Ax}+\bar{\bs{r}}) - \bs{y}^T \log(\bs{Ax}+\bar{\bs{r}}). \nonumber
\end{align}

\noindent Here, $f(\bs{x})$ is the Poisson negative log-likelihood for measurement $\bs{y}$ and estimated measurement means $\bar{\bs{y}}(\bs{x}) = \bs{Ax} + \bar{\bs{r}}$, the matrix $\bs{A}$ denotes the system model, and $\bar{\bs{r}}$ denotes the mean background events such as scatter and random coincidences. Recently, applying learned regularizers to $\mathsf{R}(\bs{x})$ is emerging for MBIR \cite{wang:18:iri}. 

While there is much ongoing research on machine learning or deep-learning techniques applied to CT \cite{chen2017low, jin2017deep, ye2018deep, gupta2018cnn, Chun&etal:18Allerton} and MRI \cite{aggarwal:19:mmb, hammernik2018learning, sun2016deep, mardani2019deep, yang2018dagan} reconstruction problems, fewer studies have applied these techniques to PET. Most past PET studies used deep learning in image space without exploiting the physical imaging model in \R{eq1}. For example, \cite{xu2017200x} applied a deep neural network (NN) mapping between reconstructed PET images with normal dose and reduced dose and \cite{yang:18:ann} applied a multilayer perceptron mapping between reconstructed images using maximum a posteriori algorithm and a reference (true) image, and their framework uses the acquisition data only to form the initial image. Recently, \cite{haggstrom:19:dad} trained a NN to reconstruct a 2D image directly from PET sinogram and \cite{gong2019iterative, kim2018penalized} proposed a PET MBIR framework using a deep-learning based regularizer. Our proposed MBIR framework, \emph{BCD-Net}, also uses a regularizer that penalizes differences between the unknown image and ``denoised'' images given by a regression neural network in an iterative manner. In particular, whereas \cite{gong2019iterative, kim2018penalized} trained only a single image denoising NN, the proposed method is an iterative framework that includes multiple trained NNs. This iterative framework enables the NNs in the later stages to learn how to recover fine details. Our proposed BCD-Net also differs from \cite{gong2019iterative, kim2018penalized} in that our denoising NNs are defined by an optimization formulation with a mathematical motivation (whereas, for the trained regularizer, \cite{gong2019iterative, kim2018penalized} brought U-Net \cite{ronneberger2015u} and DnCNN that were \cite{zhang2017beyond} developed for medical image segmentation and general Gaussian denoising, respectively) and characterized by fewer parameters, thereby avoiding over-fitting and generalizing well to unseen data especially when training samples are limited.

Iterative NNs \cite{gregor2010learning,sun2016deep,aggarwal:19:mmb,chen2017trainable,hammernik2018learning, Chun&etal:18Allerton,chun2019momentum,ye2020momentum} are a broad family of methods that originate from an unrolling algorithm for solving an optimization problem and BCD-Net \cite{chun2018deep} is a specific example of an iterative NN. BCD-Net is constructed by unfolding a block coordinate descent (BCD) MBIR algorithm using ``learned'' convolutional analysis operators\cite{chun2019convolutional, Chun&etal:19SPL, crockett2019}, leading to significantly improved image recovery accuracy in extreme imaging applications, e.g., low-dose CT \cite{chun2019bcd}, dual-energy CT \cite{zhipeng2020}, highly undersampled MRI\cite{chun2018deep}, denoising low-SNR images\cite{chun2018deep}, etc. A preliminary version of this paper was presented at the 2018 Nuclear Science Symposium and Medical Imaging Conference \cite{lim2018application}. We significantly extended this work by applying our proposed method to measured PET data with newly developed techniques. We also added detailed analysis of our proposed method as well as comparisons to related works.

To show the efficacy of our proposed BCD-Net method in low-count PET imaging, we performed both digital phantom simulation and experimental measurement studies with activity distributions and count-rates that are relevant to clinical Y-90 PET imaging after liver radioembolization. Novel therapeutic applications have sparked growing interest in quantitative imaging of Y-90, an almost pure beta emitter that is widely used in internal radionuclide therapy. In addition to the FDA approved Y-90 microsphere radioembolization and Y-90 ibritumomab radioimmunotherapy, there are 50 active clinical trials for Y-90 labeled therapies (www.clinicaltrials.gov). However, the lack of gamma photons complicates imaging of Y-90; it involves SPECT via bremsstrahlung photons produced by the betas \cite{elschot2013quantitative} or PET via a very low abundance positron in the presence of bremsstrahlung that leads to low signal-to-noise \cite{pasciak2014radioembolization}.
This paper applies a BCD-Net that is trained for realistic low-count PET imaging environments and compares its performance with those of non-trained regularizers. Our proposed BCD-Net applies to PET imaging in general, particularly in other imaging situations that also have low counts. Using shorter scan times and lower tracer activity in diagnostic PET has cost benefits and reduces radiation exposure, but at the expense of reduced counts that makes traditional iterative reconstruction challenging.

Section~\ref{sec:method} develops the proposed BCD-Net architecture for PET MBIR. 
Section~\ref{sec:method} also explains the simulation studies in the setting of Y-90 radioembolization and provides details on how we perform the physical phantom measurement. 
Section~\ref{sec:result} presents how the different reconstruction methods perform with the simulation and measurement data. 
Section~\ref{sec:discussion} discusses what training and imaging factors most affect generalization performance of BCD-Net. Section~\ref{sec:conclusion} concludes with future works.

\begin{figure}[t]
    \begin{center}  
    \includegraphics[width=1\linewidth]{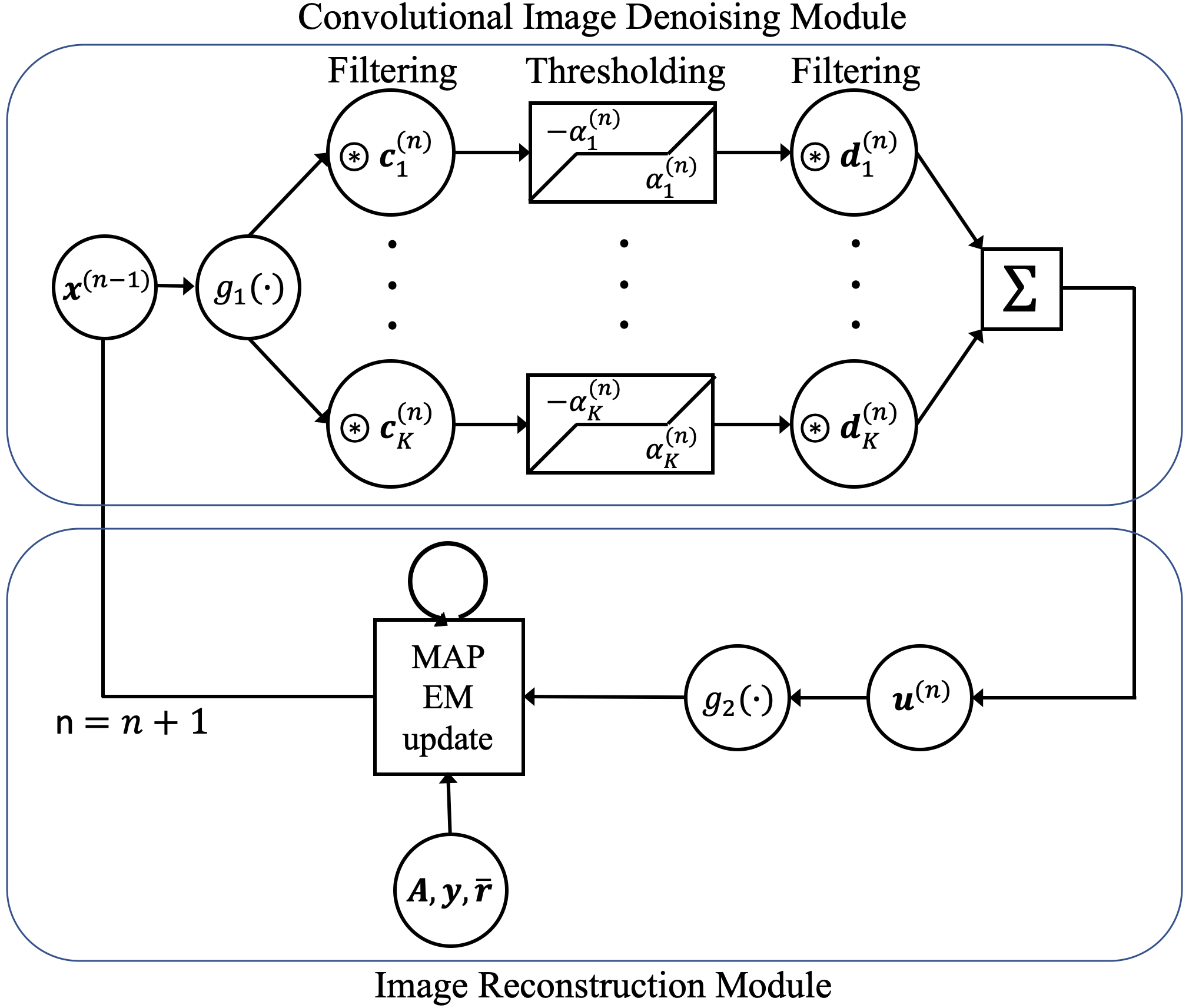}
    \caption{Architecture of the proposed BCD-Net for PET. The proposed BCD-Net has an iterative NN architecture: each BCD-Net iteration uses three inputs -- fixed measurement and mean background $\{ \bs{y}, \bar{\bs{r}} \}$, and the image $\bs{x}^{(n-1)}$ reconstructed at the previous BCD-Net iteration -- and provides the reconstructed image $\bs{x}^{(n)}$. A circular arrow above MAP EM update indicates inner iterations. $g_1(\cdot)$ and $g_2(\cdot)$ are the normalization and scaling functions described in Section~\ref{sec:method:bcdnet:meas}.
    }
    \label{fig1} 
    \end{center}
\end{figure}

\section{Methods}\label{sec:method}
This section presents the problem formulation of the BCD-Net and gives a detailed derivation that inspires the final form of BCD-Net. We also provide several techniques for BCD-Net that we specifically devised for PET data where each measurement has different count-level (and noise-level). Then we review the related works that we compare with BCD-Net such as MBIR methods using conventional non-trained regularizers. This section also describes the simulation setting and details on the measurement data and what evaluation metrics are used to assess the efficacy of each reconstruction algorithm. 

\subsection{BCD algorithm for MBIR using \dquotes{learned} convolutional regularization} \label{sec:method:bcdalg}

Conventional PET regularizers penalize differences between neighboring pixels \cite{nuyts2002concave}. That approach is equivalent to assuming that convolving the image with the [1,-1] finite difference filter along different directions produces sparse outputs. Using such ``hand-crafted'' filters is unlikely to be the best approach. A recent trend is to use training data to learn filters ${\bs{c}_k}$ that produce sparse outputs when convolved with images of interest \cite{chun2019convolutional, Chun&etal:19SPL, pfister2015learning, cai2014data}. Such learned filters can be used to define a regularizer that prefers images having sparse outputs, as follows\cite{Chun&Fessler:18Asilomar}:
\begin{equation}
\mathsf{R}(\bs{x}) = \min_{\{ \bs{z}_k \}}  \beta \left(\sum_{k=1}^K \frac{1}{2}\left\|\bs{c}_k * \bs{x} - \bs{z}_k\right\|_2^2 + \alpha_k \left\|\bs{z}_k\right\|_1\right), \label{eq:problem} 
\end{equation}
where $\beta$ is regularization parameter, $\{\bs{c}_k \in \mathbb{R}^{R}: k=1,\ldots,K\}$ is a set of convolutional filters, $\{\bs{z}_k \in \mathbb{R}^{n_p}: k=1,\ldots,K\}$ is a set of sparse codes, $\{\alpha_k \in \mathbb{R}: k=1,\ldots,K\}$ is a set of thresholding parameters controlling the sparsity of $\{\bs{z}_k\}$, $n_p$ is the number of image voxels, and $R$ and $K$ denote the size and number of learned filters, respectively. BCD-Net is inspired by this type of ``learned'' regularizer. Ultimately, we hope that the learned regularizer can better separate true signal from noisy components compared to hand-crafted filters\cite{chun2019bcd}.

A natural BCD algorithm solves (\ref{eq1}) with regularizer (\ref{eq:problem}) by alternatively updating $\{\bs{z}_k\}$ and $\bs{x}$ :
\begingroup
\allowdisplaybreaks
\begin{align}
    \{\bs{z}_k^{(n+1)}\} &= \argmin_{\{\bs{z}_k\}} \frac{1}{2} \big\|\bs{c}_k * \bs{x}^{(n)}- \bs{z}_k\big\|_2^2 + \alpha_k \left\|\bs{z}_k\right\|_1 \nonumber  \\&= \mathcal{T}(\bs{c}_k * \bs{x}^{(n)}, \alpha_k ) \label{eq:z-update} \\
    \bs{x}^{(n+1)} &= \argmin_{\bs{x}\ge \bs{0}} f(\bs{x}) + \frac{\beta}{2} \left( \sum_{k=1}^K  \left\|\bs{c}_k * \bs{x} - \bs{z}_k^{(n+1)}\right\|_2^2   \right), \label{eq:x-update}
\end{align}
\endgroup
where $\mathcal{T}\left(\cdot,\cdot\right)$ is the element-wise soft thresholding operator: $\mathcal{T} (\bs{t},q )_j := \mathrm{sign}(t_j) \max (|t_j| - q, 0 )$.

Assuming that learned filters $\{ \bs{c}_k \}$ satisfy the tight-frame condition, $\sum_{k=1}^K \|\bs{c}_k * \bs{x}\|_2^2 = \|\bs{x}\|_2^2$~$\forall \bs{x}$ \cite{chun2019convolutional}, we rewrite the updates in (\ref{eq:z-update})-(\ref{eq:x-update}) as follows: 
\begin{align}
\bs{u}^{(n+1)} &= \sum_{k=1}^K \tilde{\bs{c}}_k * \left( \mathcal{T}\left(\bs{c}_k * \bs{x}^{(n)}, \alpha_k  \right)\right) \label{eq:denoising} \\
    \bs{x}^{(n+1)} &= \argmin_{\bs{x}\ge \bs{0}} f(\bs{x}) + \frac{\beta}{2}   \left\| \bs{x} - \bs{u}^{(n+1)}\right\|_2^2 \label{eq:reg-x},
\end{align}
where $\tilde{\bs{c}}_k$ denotes a rotated version of $\bs{c}_k$. The operations of convolution, soft thresholding and then filtering again with summation typically have the effect of denoising the image $\bs{x}^{(n)}$.

For efficient image reconstruction (\ref{eq:reg-x}) in PET, we use the standard EM-surrogate of Poisson log-likelihood function \cite{depierro:95:ame}:
\begingroup
\setlength{\thinmuskip}{1.5mu}
\setlength{\medmuskip}{2mu plus 1mu minus 2mu}
\setlength{\thickmuskip}{2.5mu plus 2.5mu}
\begin{align*}
    &~ f(\boldsymbol{x}) +  \frac{\beta}{2}  \left\|\boldsymbol{x}-\bs{u}^{(n+1)}\right\|_2^2 \\ &= \sum_{i=1}^{n_d} [\boldsymbol{Ax}]_i + \bar{r}_i - y_i\log([\boldsymbol{Ax}]_i + \bar{r}_i) +  \frac{\beta}{2}  \sum_{j=1}^{n_p} (x_j - u_j^{(n+1)})^2\\
    &\le \sum_{j=1}^{n_p} \big\{-e_j(\boldsymbol{x}^{(n')})(x_j^{(n')})\log(x_j) + a_j x_j +  \frac{\beta}{2} (x_j - u_j^{(n+1)})^2 \big\}\\
    &= \sum_{j=1}^{n_p} Q_j(x_j) 
\end{align*}
\endgroup
where $n'$ denotes $n'$th inner-iteration in (\ref{eq:reg-x}), $e_j(\boldsymbol{x}^{(n')}) = \sum_{i=1}^{n_d} a_{ij} \frac{y_i}{\bar{y}_i(\boldsymbol{x}^{(n')})}$, $a_{ij}$ denotes an element of the system model at $i$th row and $j$th column, and $n_d$ is the number of rays. Equating $\frac{\partial Q_j(x_j)}{\partial x_j}$ to zero is equivalent to finding the root of the following quadratic formula: 
\begingroup
\setlength{\thinmuskip}{1.5mu}
\setlength{\medmuskip}{2mu plus 1mu minus 2mu}
\setlength{\thickmuskip}{2.5mu plus 2.5mu}
\begin{align*}
    \beta x_j^2 + \left(a_j - \beta u_j^{(n+1)}\right)x_j - e_j(\boldsymbol{x}^{(n')})x_j^{(n')} = 0,
\end{align*}
\endgroup 
and finding the root \cite{press:88} leads to the minimizer:
\begin{align*}
    x_j^{(n'+1)} &= \begin{cases} \frac{\sqrt{\lambda^2+\beta\nu}-\lambda}{\beta}, & \lambda <0 \\
    \frac{\nu}{\sqrt{\lambda^2+\beta\nu}+\lambda},& \lambda \ge0 ,
    \end{cases}
\end{align*}
where $\lambda = \frac{1}{2} (a_j- \beta  u_j^{(n+1)}), ~\nu = e_j(\boldsymbol{x}^{(n')})x_j^{(n')}, ~ a_j = \sum_{i=1}^{n_d} a_{ij}$. 

\tiny{}
\begin{algorithm}[t]
  \caption{BCD-Net for PET MBIR}\label{algo: algo1}
  \begin{algorithmic}
    \REQUIRES{$\{ \bs{c}_k^{(n)}, \bs{d}_k^{(n)}, \alpha_k^{(n)}: n=1,\ldots,T \},~\bs{y},~\bar{\bs{r}},~\bs{A},~c$}
    \Initialize{$\bs{x}^{(0)}$ using EM algorithm}
    \text{Calculate} $a_j = \sum_{i=1}^{n_d} a_{ij}$
    \For{$n = 0,\ldots,T-1$}
    \State $\bs{u}^{(n+1)} = \sum_{k=1}^K \bs{d}^{(n+1)}_k * \left( \mathcal{T}\left(\bs{c}_k^{(n+1)} * g_1\left(\bs{x}^{(n)}\right), \alpha_k^{(n+1)}  \right)\right)$
    \State $\beta^{(n+1)} = \frac{\left\|a_j - \sum_{i=1}^{n_d} a_{ij} \frac{y_i}{\bar{y}_i(\boldsymbol{x}^{(n)})} \right\|_2}{\left\| \bs{x}^{(n)} - g_2(\bs{u}^{(n+1)})\right\|_2} \cdot c$
    \For{$n' = 0,\ldots,T'-1$}
    \State $\lambda = \frac{1}{2} \left(a_j- \beta^{(n+1)} g_2\left( u_j^{(n+1)} \right) \right)$
    \State $\nu = x_j^{(n')} \left( \sum_{i=1}^{n_d} a_{ij} \frac{y_i}{\bar{y}_i(\boldsymbol{x}^{(n')})} \right) $
    \State $x_j^{(n'+1)} = \begin{cases} \frac{\sqrt{\lambda^2+\beta^{(n+1)}\nu}-\lambda}{\beta^{(n+1)}}, & \lambda <0 \\
    \frac{\nu}{\sqrt{\lambda^2+\beta^{(n+1)}\nu}+\lambda},& \lambda \ge0 
    \end{cases}$
    \EndFor
    \State $\bs{x}^{(n+1)} = \bs{x}^{(T')}$
    \EndFor
  \end{algorithmic}
\end{algorithm}
\normalsize{}
\subsection{BCD-Net for PET MBIR and training its denoising module}
\label{sec:method:bcdnet}

To further improve denoising capability by providing more trainable parameters, we extend the convolutional image denoiser (CID) in \R{eq:denoising} \cite{chun2018deep}, 
by replacing $\{\tilde{\bs{c}}_k\}$ with separate decoding filters $\{\bs{d}_k\}$.
We \textit{define} BCD-Net to use the following updates for each iteration:
\begin{align}
    \bs{u}^{(n+1)} &= \sum_{k=1}^K \bs{d}^{(n+1)}_k * \left( \mathcal{T}\left(\bs{c}_k^{(n+1)} * \bs{x}^{(n)}, \alpha_k^{(n+1)}  \right)\right) \label{eq:u-update} \\
    \bs{x}^{(n+1)} &= \argmin_{\bs{x}\ge\bs{0}} f(\bs{x}) + \frac{\beta}{2} \left\| \bs{x} - \bs{u}^{(n+1)}\right\|_2^2, \label{eq:reg-x-new}
\end{align}
where separate encoding and decoding filters $\{\bs{c}_k\}$ and $\{\bs{d}_k\}$ are learned for each iteration. Fig.~{\ref{fig1}} shows the corresponding BCD-Net architecture.
We refer to the $\bs{u}$ and $\bs{x}$ updates in (\ref{eq:u-update})-(\ref{eq:reg-x-new}) as two \emph{modules}: 1) image denoising module and 2) image reconstruction module. The final output image is from the reconstruction module. 

The image denoising module consists of encoding and decoding filters $\{\bs{c}_k^{(n+1)}\}, \{\bs{d}_k^{(n+1)}\}$ and thresholding values $\{\alpha_k^{(n+1)}\}$. We train these parameters to ``best map'' from noisy images into high-quality reference images (e.g., true images if available) in the sense of mean squared error:
\begin{align}
    \argmin_{ \{\bs{c}_k\}, \{\bs{d}_k\}, \{\alpha_k\} } \sum_{l=1}^{L} \left\| \bs{x}_{\text{true},l} - \sum_{k=1}^K \bs{d}_k * \left( \mathcal{T}\left(\bs{c}_k * \bs{x}_l^{(n)},  \alpha_k \right)  \right)\right\|_2^2, \label{eq:training}
\end{align}
where $L$ is the total number of training samples, $\{\bs{x}_{\text{true},l} \in \mathbb{R}^{n_p}: l=1,\ldots,L\}$ is a set of true images and $\{\bs{x}_l^{(n)}\in \mathbb{R}^{n_p}: l=1,\ldots,L\}$ is a set of images estimated by image reconstruction module in the $n$th iteration. We train the set of filters and thresholding values iteration-by-iteration and do not include the system matrix or sinograms for training as shown in (\ref{eq:training}). Moreover, we do not enforce the tight-frame condition when training the filters.

One can further extend the CID in \R{eq:u-update} to a general regression NN, e.g., a deep U-Net \cite{ronneberger2015u}. We investigated if the iterative BCD-Net combined with U-Net denoisers (by replacing the denoising module in \R{eq:u-update} with a U-Net) performs better than the proposed BCD-Net using CID \R{eq:u-update}. Section~\ref{sec:method:train:unet} gives the details of the U-Net implementation.

\subsection{Adaptive BCD-Net generalizing to various count-levels} 
\label{sec:method:bcdnet:meas}

\subsubsection{Normalization and scaling scheme}
Different PET images can have very different intensity values due to variations in scan time and activity, and it is important for trained methods to be able to generalize to a wide range of count levels. Towards this end, we implemented normalization and scaling techniques in BCD-Net. \cite{kim2018penalized} extended \cite{gong2019iterative} by implementing ``local linear fitting'' to ensure that the denoising NN output has similar intensity as the input patch from the current estimated image. Our approach is different in that we normalize and scale the image with a global approach, not a patch-based approach. In particular, we modify the architecture in (\ref{eq:u-update})-(\ref{eq:reg-x-new}) as:
\begin{align}
    \bs{u}^{(n+1)} &= \sum_{k=1}^K \bs{d}^{(n+1)}_k * \left( \mathcal{T}_{\alpha_k^{(n+1)}}\left(\bs{c}_k^{(n+1)} * g_1(\bs{x}^{(n)})\right)\right) \label{eq:u-update-meas} \\
    \bs{x}^{(n+1)} &= \argmin_{\bs{x}\ge \bs{0}} f(\bs{x}) + \frac{\beta}{2} \left\| \bs{x} - g_2(\bs{u}^{(n+1)})\right\|_2^2, \label{eq:x-update-meas}
\end{align}
where the normalization function $g_1(\cdot)$ is defined by $g_1(\bs{v}) := \frac{1}{\sum_j v_j} \bs{v}$ to ensure that $\ones^T g_1(\bs{v}) = 1$, and the scaling function $g_2(\cdot)$ is defined by $g_2(\bs{v}) :=  \{ \argmin_{s} f(s \cdot \bs{v}) \} \bs{v}$. We solve the optimization problem over $s$ using Newton's method: 
\begin{align}
    s^{(n+1)} &= s^{(n)} - \frac{\nabla_s f(s^{(n)}\cdot\bs{v})}{\nabla_s^2 f(s^{(n)}\cdot\bs{v})} \nonumber \\
    &= s^{(n)} - \frac{\sum_{i=1}^{n_d} [\bs{Av}]_i - y_i \frac{[\bs{Av}]_i}{s^{(n)}[\bs{Av}]_i + \bar{r}_i}}{\sum_{i=1}^{n_d} y_i \left(\frac{[\bs{Av}]_i}{(s^{(n)}[\bs{Av}]_i + \bar{r}_i)}\right)^2 }. \label{eq:scaling}
\end{align}
To be consistent with the modified CID in (\ref{eq:u-update-meas}), we also apply this image-based normalization technique when training the convolutional filters and thresholding values:
\begin{align*}
\footnotesize{
    \argmin_{ \{\bs{c}_k\}, \{\bs{d}_k\}, \{\alpha_k\} } \sum_{l=1}^{L} \left\| g_1(\bs{x}_{\text{true},l}) - \sum_{k=1}^K \bs{d}_k * \left( \mathcal{T}_{\alpha_k}\left(\bs{c}_k * g_1(\bs{x}_l^{(n)})\right)  \right)\right\|_2^2.}
\end{align*}

\subsubsection{Adaptive regularization parameter scheme}
\label{sec:method:meas:reg}

The best regularization parameter value can also vary greatly between scans, depending on the count level. Therefore, instead of choosing one specific value for the regularization parameter, we set the $\beta$ value for each iteration based on evaluation on current gradients of data-fidelity term and regularization term: 
\begin{align}
    \beta^{(n+1)} &= \frac{\left\|\nabla_{\bs{x}} f(\bs{x}^{(n)})\right\|_2}{\left\|\nabla_{\bs{x}} \mathsf{R}(\bs{x}^{(n)})\right\|_2} \cdot c \nonumber \\
    &= \frac{\left\|a_j - e_j(\boldsymbol{x}^{(n)}) \right\|_2}{\left\| \bs{x}^{(n)} - g_2(\bs{u}^{(n+1)})\right\|_2} \cdot c,\quad n = 0,\ldots,T-1, \label{eq:beta_sel}
\end{align}
where $c$ is a constant specifying how we balance between the data-fidelity term and regularization term and $n$ denotes $n$th outer-iteration. Algorithm \R{algo: algo1} gives detailed pseudocode of the proposed method. $T$ denotes the total number of outer-iterations and $T'$ denotes the number of inner iterations used for (\ref{eq:reg-x-new}). We use $\bs{x}^{(n)}$ as the initial image when solving (\ref{eq:x-update-meas}). 


\begin{table}[t]
    \begin{center}
    \caption{Details on XCAT simulation data: variations between training and testing data. }
    \begin{tabular}{|l|l|l|}
\cline{1-3}
  & Training data & Testing data \\ \cline{1-3}
Concentration ratio (hot:warm) & 9:1  & 4:1    \\\cline{1-3}
Total net trues & 200 K & 500 K \\ \cline{1-3}
Random fraction (\%) & 90.9 & 87.5 \\ \cline{1-3}
\end{tabular}
\label{table1}
    \end{center}
\end{table}

\begin{table}[t]
    \begin{center}
    \caption{Details on phantom measurement data: activity concentration ratio between hot and warm regions and randoms fractions for two phantom studies. }
    \begin{tabular}{|l|l|l|}
\cline{1-3}
  & Sphere & Liver-torso  \\ \cline{1-3}
Total activity (GBq) & 0.65  &  1.9       \\\cline{1-3}
Concentration ratio (hot:warm) & 8.9:1  & 5.4:1       \\\cline{1-3}
Total prompts & 3.2 - 6.3 M & 2.3 M      \\ \cline{1-3}
Total randoms & 2.9 - 5.7 M & 2.1 M   \\ \cline{1-3}
Total net trues & 308 - 599 K & 220 K \\ \cline{1-3}
Random fraction(\%) & 90.3 - 90.5 & 90.7 \\ \cline{1-3}
\end{tabular}
\label{table2}
    \end{center}
\end{table}

\begin{table}[t]
    \begin{center}
    \caption{Details on typical patient measurement data: total trues and randoms fractions. }\vspace{-0.4em}
    \begin{tabular}{|l|l|}
\cline{1-2}
  & Patient A \\ \cline{1-2} 
Total activity (GBq) & 2.55      \\ \cline{1-2}
Total prompts & 2.7 M      \\ \cline{1-2}
Total randoms & 2.3 M  \\ \cline{1-2}
Total net trues & 380 K  \\ \cline{1-2}
Random fraction(\%) & 85.8 \\ \cline{1-2}
\end{tabular}
\label{table3}
    \end{center}
\end{table}
\subsection{Conventional MBIR methods: Non-trained regularizers}
We compared the proposed BCD-Net with two MBIR methods that use standard non-trained regularizers.
\subsubsection{Total-variation (TV)}
TV regularization penalizes the sum of absolute value of differences between adjacent voxels:
\begin{align*}
\mathsf{R}(\bs{x})=\beta \left\|\bs{Cx}\right\|_1,
\end{align*}
where $\bs{C}$ is finite differencing matrix. Recent work \cite{zhang2018optimization} applied Primal-Dual Hybrid Gradient (PDHG) \cite{chambolle2016introduction} for PET MBIR using TV regularization and demonstrated that PDHG-TV is superior than clinical reconstruction (e.g., OS-EM) for low-count datasets in terms of several image quality evaluation metrics such as contrast recovery and variability.

\subsubsection{Non-local means (NLM)}
NLM regularization penalizes the differences between nearby patches in image:
\begin{align*}
    \mathsf{R}(\bs{x})= \beta \sum_{i,j \in S_i}p\left(\left\|\bs{N}_i \bs{x} - \bs{N}_j \bs{x}\right\|_2^2\right),
\end{align*}
where $p(t)$ is a potential function of a scalar variable $t$, $S_i$ is the search neighborhood around the $i$th voxel, and $\bs{N}_i$ is a patch extraction operator at the $i$th voxel. We used the Fair potential function for $p(t)$:
\begin{align*}
    p(t) = \sigma_f^2 \left( \sqrt{\frac{t}{\sigma_f^2 N_f}} + \log\left(1+\sqrt{\frac{t}{\sigma_f^2 N_f}}\right) \right),
\end{align*}
where $\sigma_f$ is a design parameter and $N_f$ is the number of voxels in the patch $\bs{N}_i \bs{x}$.
Unlike conventional local filters that assume similarity between only adjacent voxels, NLM filters can average image intensities over distant voxels. As in \cite{chun2014alternating}, we used ADMM to accelerate algorithmic convergence with an adaptive penalty parameter selection method \cite{boyd:10:doa}. 

\subsection{Experimental setup: Digital phantom simulation and experimental measurement}

\subsubsection{Y-90 PET/CT XCAT simulations}
We used the XCAT \cite{segars20104d} phantom (Fig.~\ref{fig2}) to simulate Y-90 PET following radioembolization. We set the image size to 128$\times$128$\times$100 with a voxel size 4.0$\times$4.0$\times$4.0 (mm$^3$) and chose 100 slices ranging from lung to liver. To simulate extremely low count scans with high random fractions, typical for Y-90 PET, we set total true coincidences and random fractions based on numbers from patient PET imaging performed after radioembolization \cite{lim:18:apr}. To test the generalization capability of the trained BCD-Net, we changed all imaging factors between training and testing dataset. Here, imaging factors include activity distribution (shape and size of tumor and liver background, concentration ratio between hot and warm region) and count-level (total true coincidences and random fraction). Fig.~\ref{fig2} and Table~\ref{table1} provide details on how we changed the testing dataset from the training dataset. We trained BCD-Net using five pairs ($L=5$) of 3D true images and estimated images at each iteration (1 true image, 5 realizations). We generated multiple (5) realizations to train the denoising NN to deal with the Poisson noise. We also generated 5 realizations (1 true image, 5 realizations) as a testing dataset to evaluate the noise across realizations.

\subsubsection{Y90 PET/CT physical phantom measurements and patient scan}

For training BCD-Net, we used PET measurements of a sphere phantom (Fig.~\ref{fig4}) where six `hot' spheres (2,4,8,16,30 and 113 mL, 0.5 MBq/ml) are placed in a `warm' background (0.057 MBq/ml) with total activity of 0.65 GBq. The phantom was scanned for 40 (3 acquisitions) - 80 (1 acquisition) $(L=4)$ minutes on a Siemens Biograph mCT PET/CT. For testing BCD-Net and other reconstruction algorithms, we used an anthropomorphic liver/lung torso phantom (Fig.~\ref{fig4}) with total activity and distribution that is clinically realistic for imaging following radioembolization with Y-90 microspheres: 5\% lung shunt, 1.17 MBq/mL in liver, 3 hepatic lesions (4 and 16 mL spheres, 29 mL ovoid) of 6.6 MBq/ml. The phantom with total activity of 1.9 GBq was scanned 5 times (each 30 minutes) on a Siemens Biograph mCT PET/CT. Fig.~\ref{fig4} and  Table~\ref{table2} provide details on the count-level (random fraction) and activity distribution differences between training (sphere phantom) and testing (liver phantom) dataset. We also tested BCD-Net with an actual Y-90 patient scan and  Table~\ref{table3} provides count-level information.

We acquired all measurement data with time of flight TOF information. The measurement data size is 200$\times$168$\times$621$\times$13. The last dimension of measurement indicates the number of time bin. The reconstructed image size is 200$\times$200$\times$112 with a voxel size 4.07$\times$4.07$\times$2.03 (mm$^3$). To reconstruct the image with measurement data, we used a SIEMENS TOF system model ($\bs{A}$ in (\ref{eq1})) along with manufacturer given attenuation/normalization correction, PSF modelling, and randoms/scatters estimation.
\vspace{-1em}

\subsection{Evaluation metrics}
\begin{figure*}[t]
\addtolength{\tabcolsep}{-6.5pt}
\centering
\begin{tabular}{ccccc}
\hspace{-1.3em} \small{Attenuation map (coronal)}  & \hspace{-1.3em} \small{Attenuation map (axial)} & \hspace{-2em} \small{True activity (training)}  & \hspace{-2em} \small{True activity (testing)} & \hspace{-2em} \small{Zoomed in}    \\
\includegraphics[scale=0.19, trim=4em 0em 0em 2em, clip]{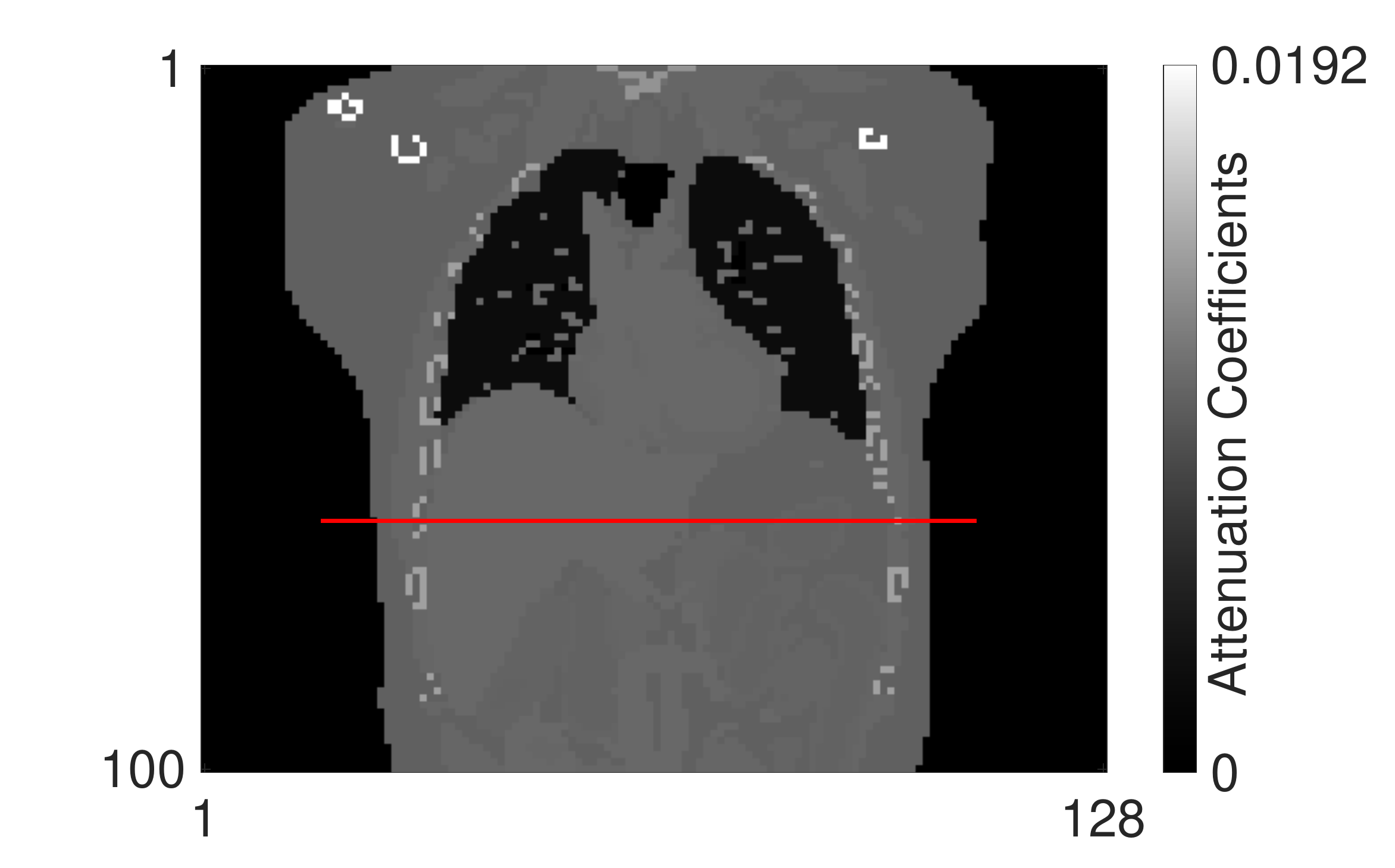} &
\includegraphics[scale=0.19, trim=4em 0em 0em 2em, clip]{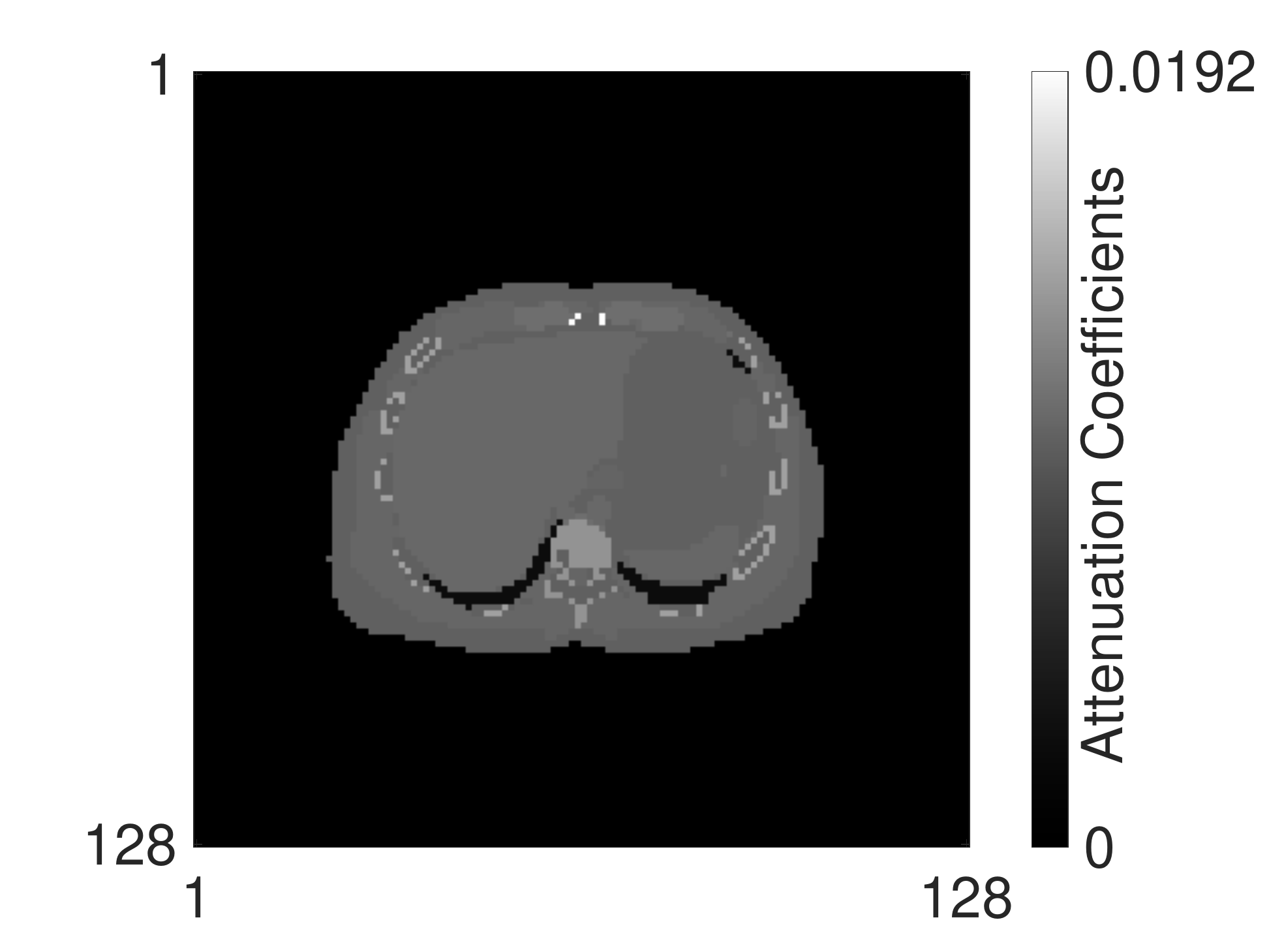} &
\includegraphics[scale=0.19, trim=4em 0em 0em 2em, clip]{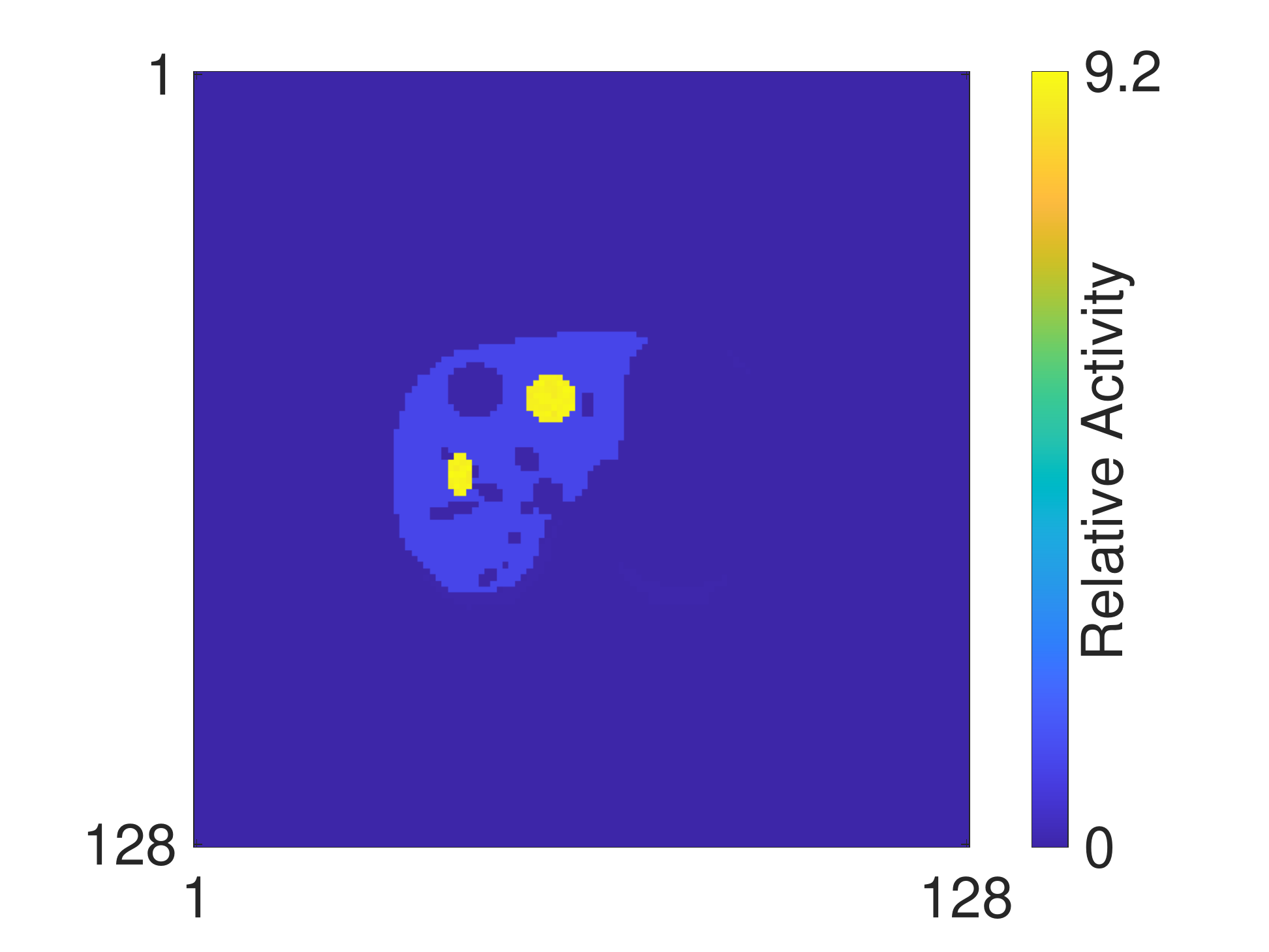} &
\includegraphics[scale=0.19, trim=4em 0em 0em 2em, clip]{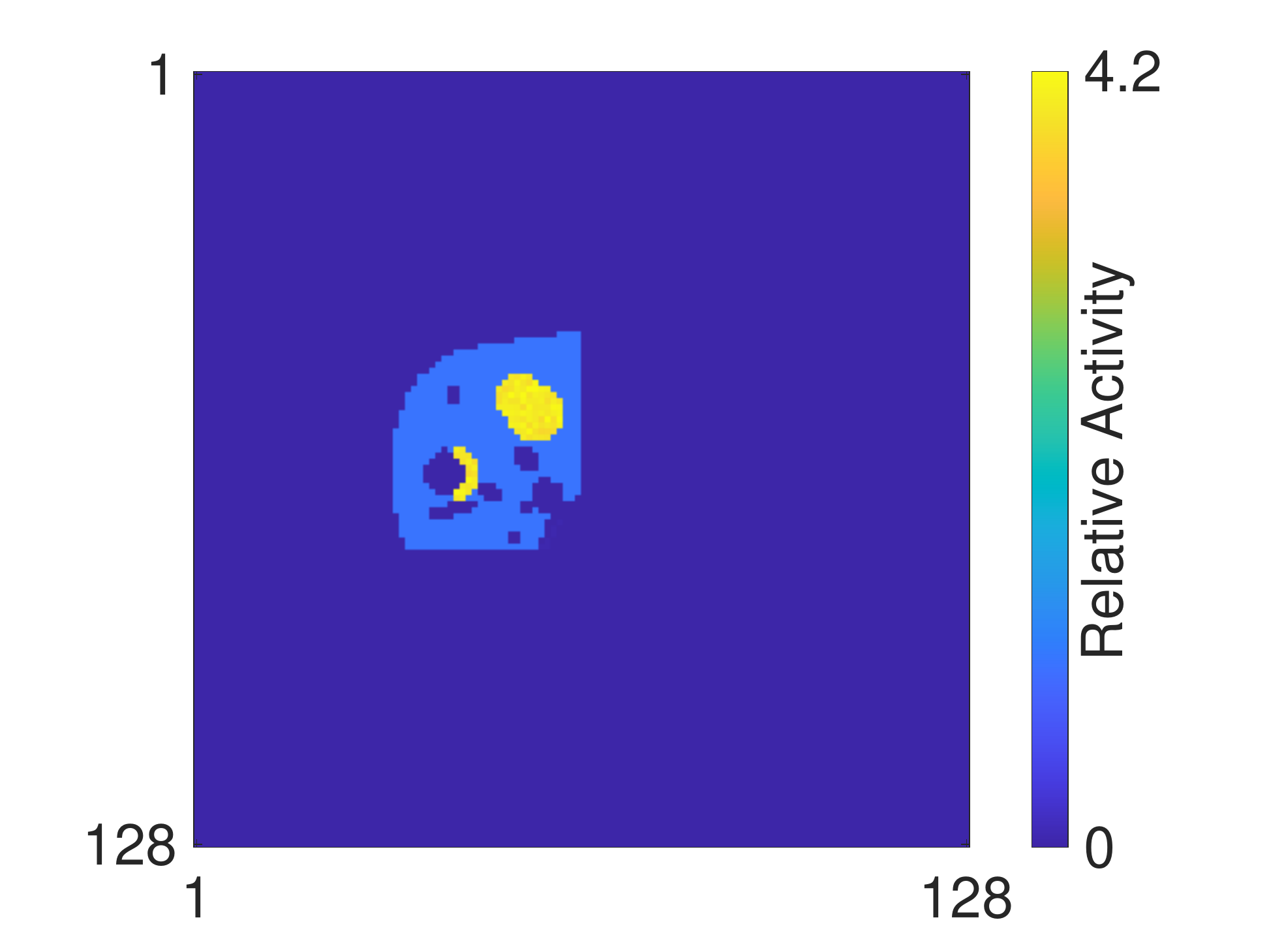} &
\includegraphics[scale=0.19, trim=4em 0em 0em 2em, clip]{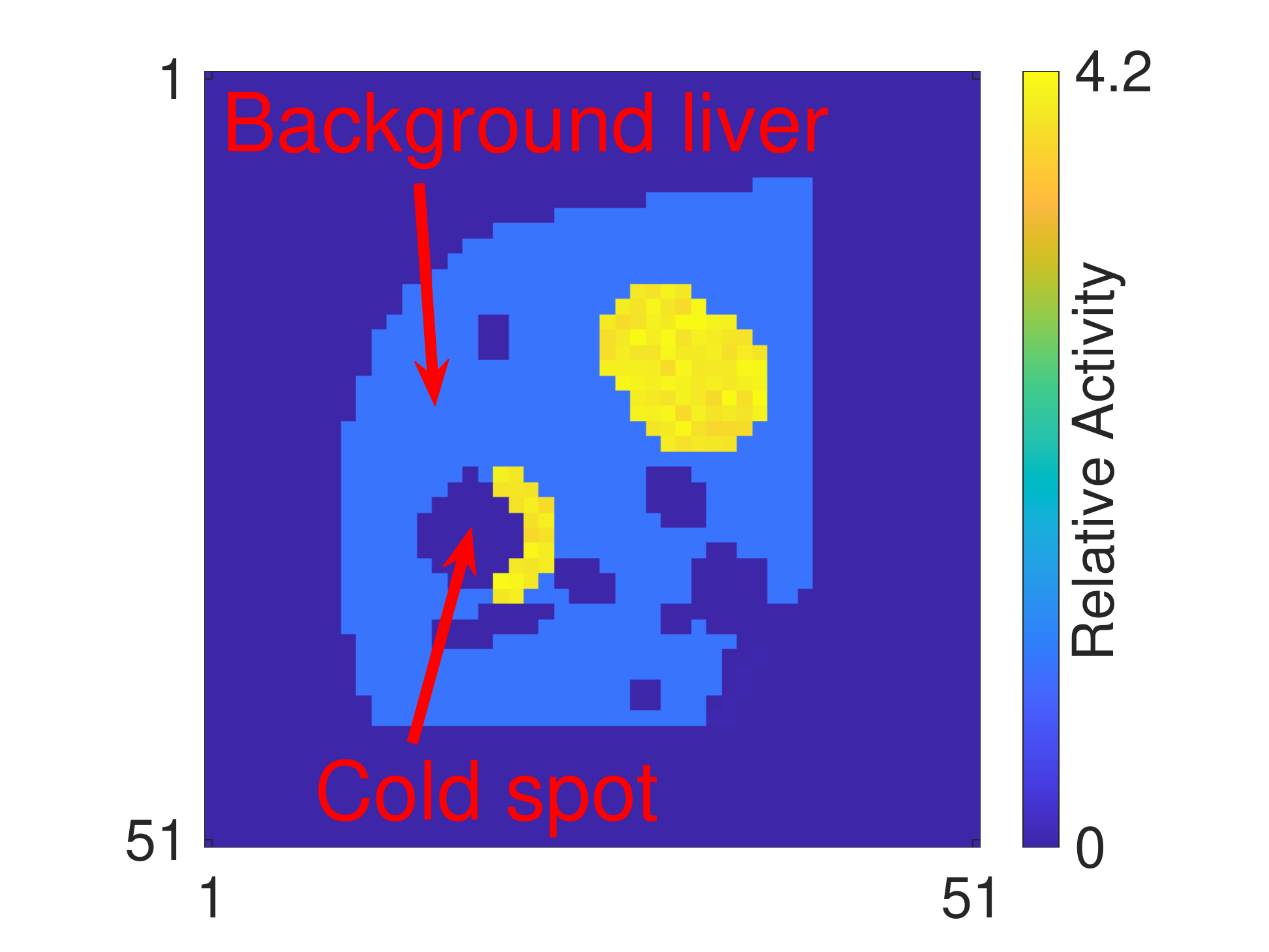}  
\end{tabular}
\begin{tabular}{cccccc}
\hspace{-1em}EM & \hspace{-1em}TV  & \hspace{-1em} NLM & \hspace{-1em} BCD-Net-CID& \hspace{-1em} BCD-Net-UNet& \hspace{-1em} BCD-Net-UNet  \\
 &   & & (params: 4K) & (params: 4K) & (params: 1.4M)  \\
\includegraphics[scale=0.18, trim=5em 0em 4em 2em, clip]{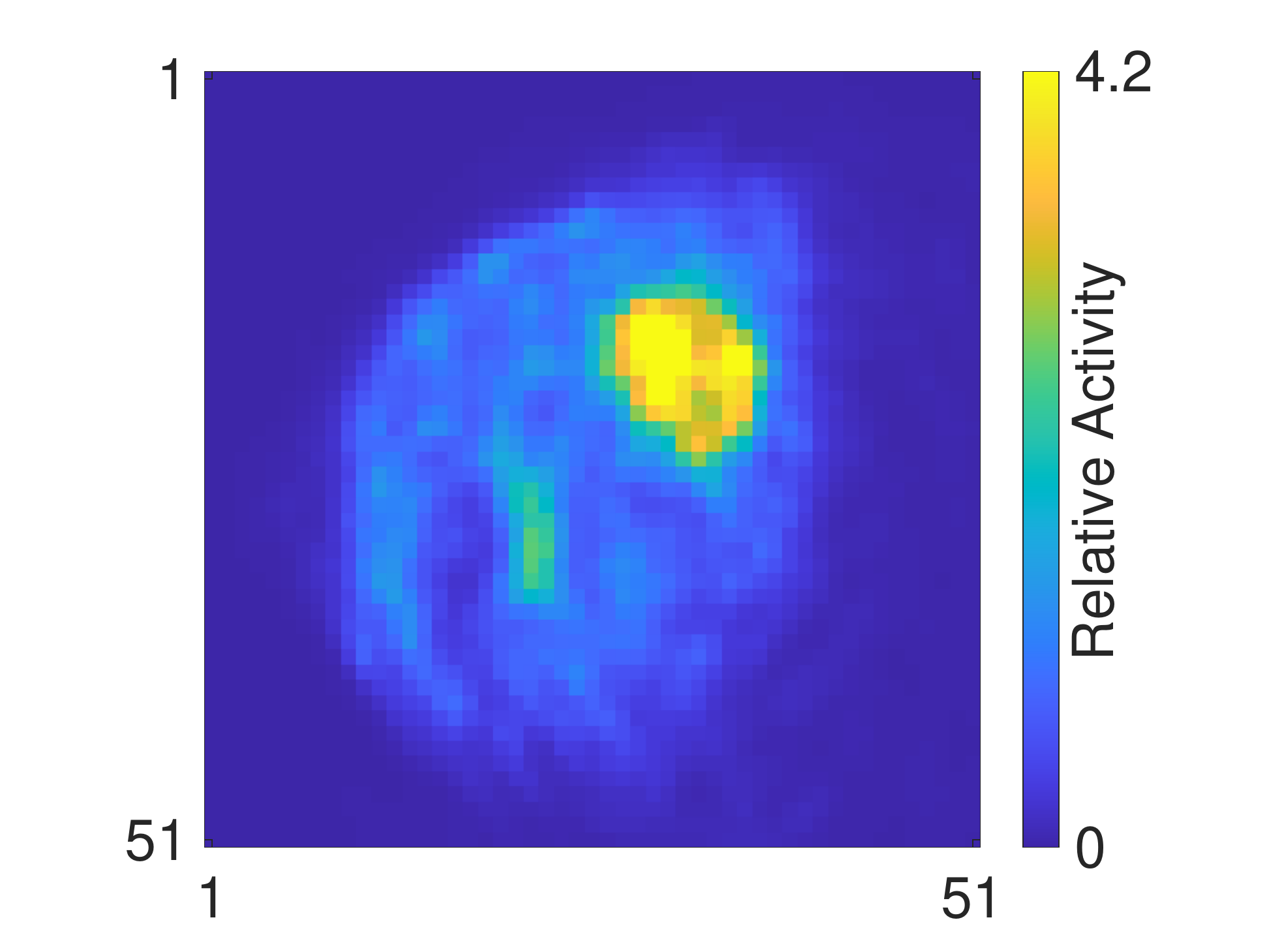} &
\includegraphics[scale=0.18, trim=5em 0em 4em 2em, clip]{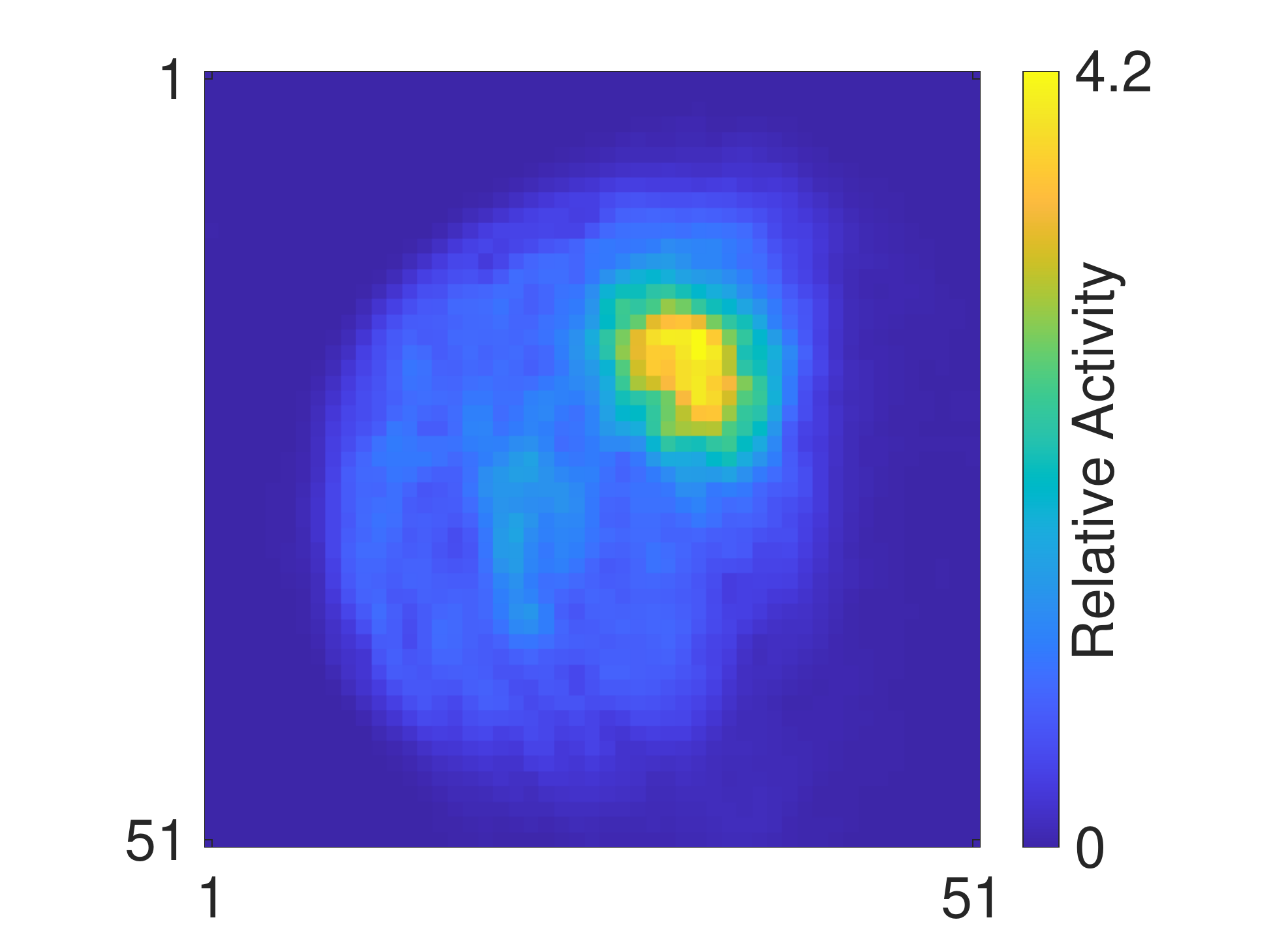} &
\includegraphics[scale=0.18, trim=5em 0em 4em 2em, clip]{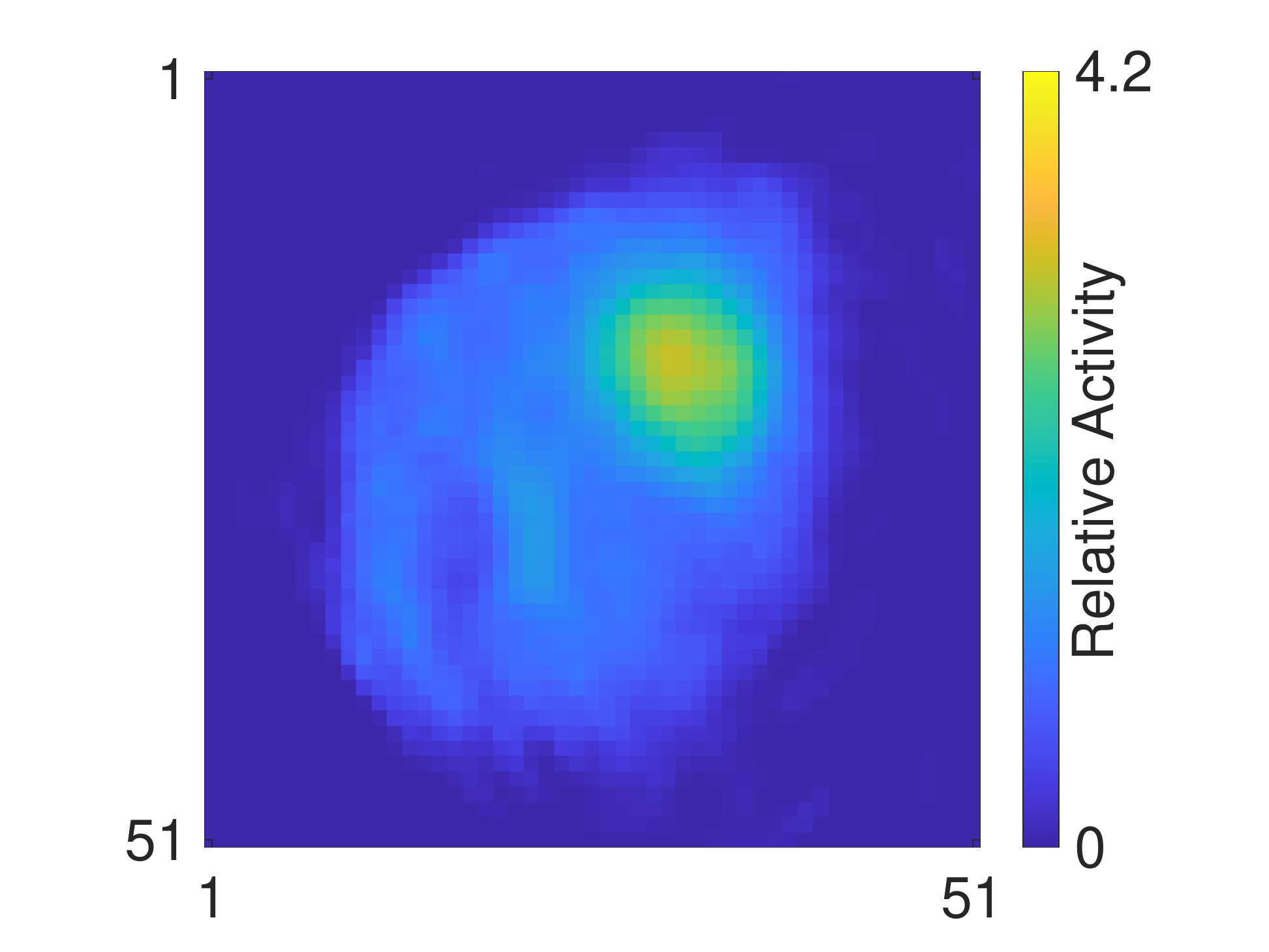} &
\includegraphics[scale=0.18, trim=5em 0em 4em 2em, clip]{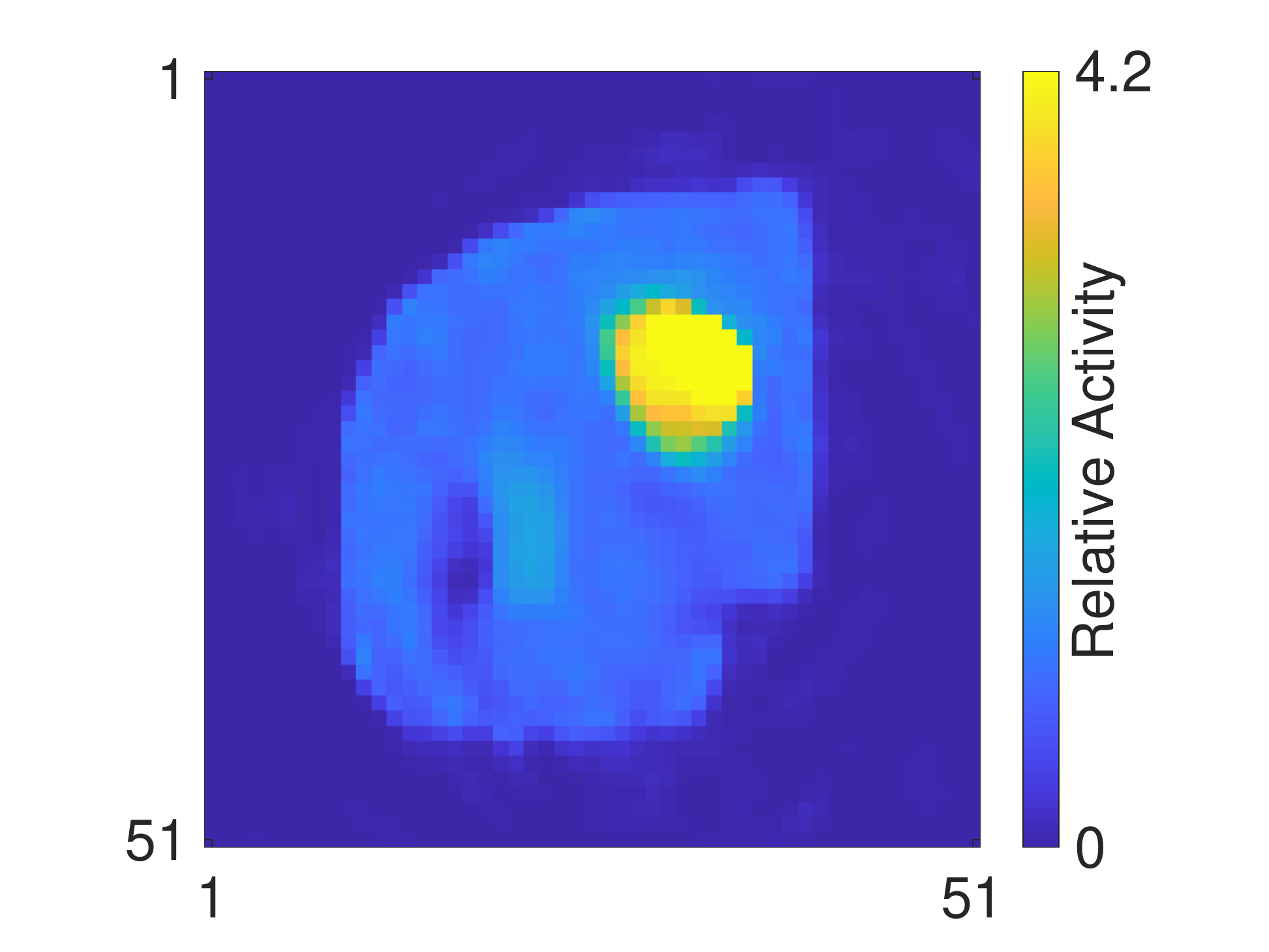} & 
\includegraphics[scale=0.18, trim=5em 0em 4em 2em, clip]{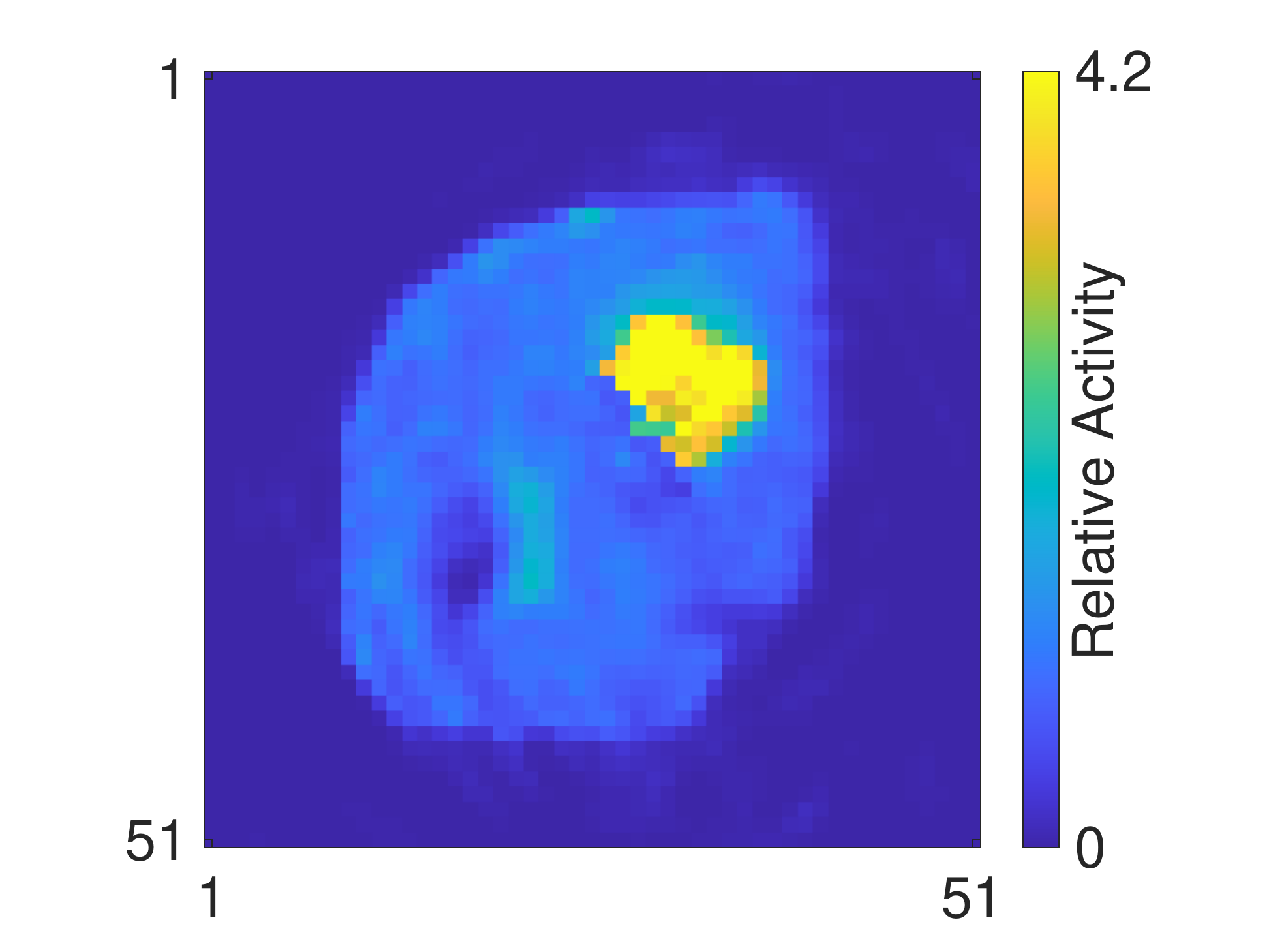}& 
\includegraphics[scale=0.18, trim=5em 0em 4em 2em, clip]{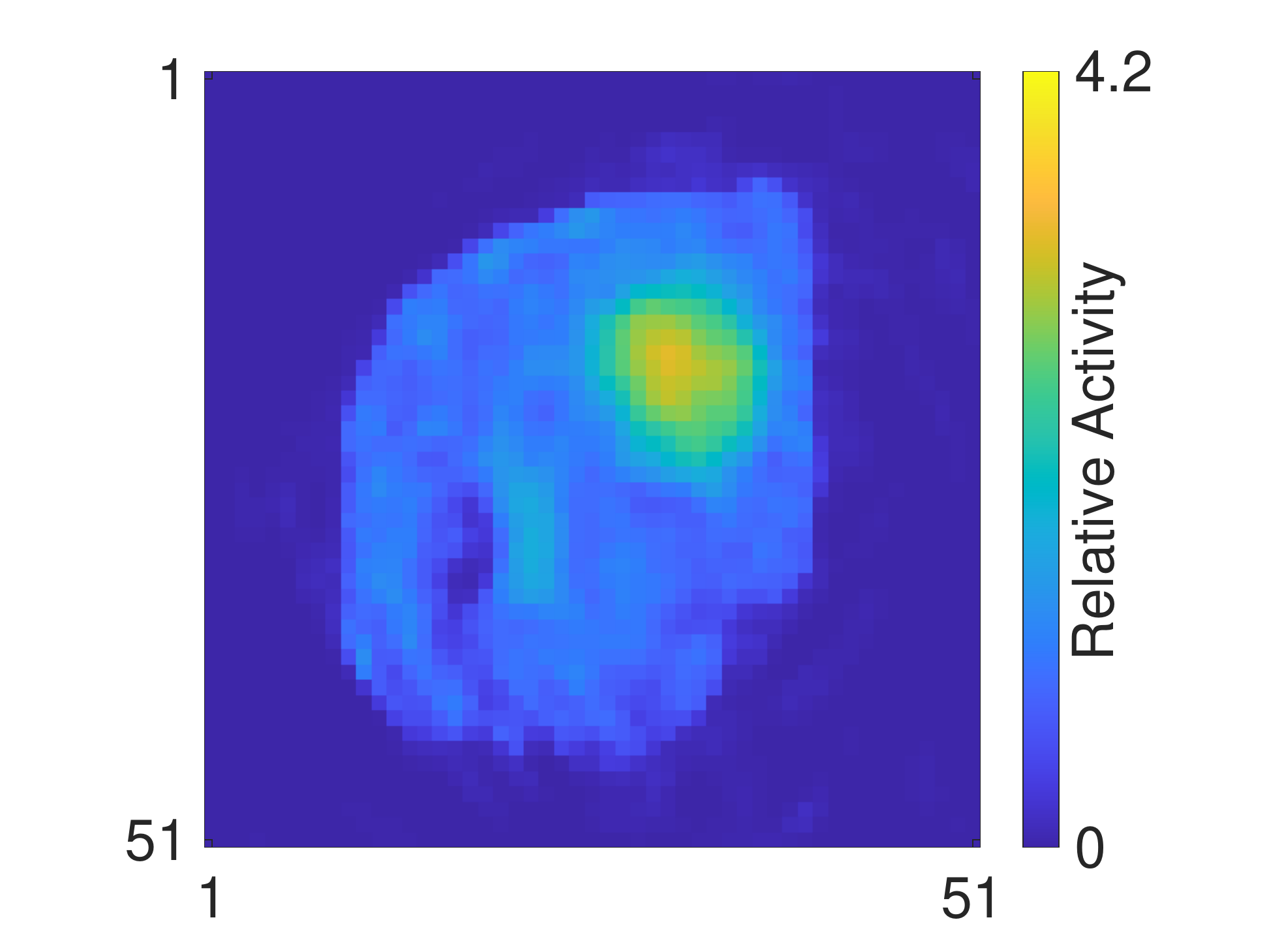}  \\
\end{tabular}
\caption{XCAT phantom simulation: (First row) coronal and axial view of attenuation map and true relative activity distribution corresponding to axial attenuation map. (Second row) reconstructed images of one slice from different reconstruction methods. BCD-Net-CID/UNet is the BCD-Net with CID/UNet and params indicates the number of trainable parameters.}
\label{fig2} 
\end{figure*}

For the XCAT phantom simulation, we evaluated each reconstruction with contrast recovery (CR) (volume-of-interest (VOI): cold spot indicated in Fig.~\ref{fig2}), noise across realizations, root mean squared error (RMSE), and contrast to noise ratio (CNR). For the physical phantom measurement, we used CR (VOI: hot spheres) and CNR averaged over multiple hot spheres. We define each VOI's mask based on attenuation map interpolated to PET voxel size. For the patient measurement, we used CNR and the field of view (FOV) activity bias since the total activity in FOV is known (equal to the injected activity because the microspheres are trapped) wheareas the activity distribution is unknown: 
\begin{align*} 
    &\text{CR (VOI: cold spot)} = \bigg(1- \frac{C_{\text{VOI}}}{C_{\text{BKG}}} \bigg) \times 100~(\%)  \\
    &\text{CR (VOI: hot sphere)} = \frac{\frac{C_{\text{VOI}}}{C_{\text{BKG}}}-1}{R_{\text{True}}-1} \times 100~(\%)  \\
    &\text{Noise} = \tiny{} \frac{\sqrt{\frac{1}{J_\text{Liver}}\sum_{j \in \text{Liver}} \bigg(\frac{1}{M-1} \sum_{m=1}^{M} (\hat{\boldsymbol{x}}_m[j] - \frac{1}{M}\sum_{m'=1}^M \hat{\boldsymbol{x}}_{m'}[j])^2 \bigg)}}{\frac{1}{J_\text{Liver}}\sum_{j \in \text{Liver}}  \boldsymbol{x}_{\mathrm{true}}[j]} \times 100~\%\\\normalsize{}
    &\text{RMSE} = \sqrt{ \frac{\sum_j (\bs{x}_{\text{true}}[j] - \hat{\bs{x}}[j])^2}{J_{\text{FOV}}}} \times 100~(\%)\\
    &\text{CNR} = \frac{C_{\text{Lesion}}-C_{\text{BKG}}}{\mathrm{STD}_{\text{BKG}}}\\
    &\text{FOV bias} =  \frac{\sum_j \hat{\bs{x}}[j] - \bs{x}_{\text{true}}[j]}{\sum_j \bs{x}_{\text{true}}[j]} \times 100~(\%),
\end{align*} 
where $C_{\text{VOI}}$ is mean counts in the VOI, $R_{\text{True}}$ is true ratio between hot and warm region, $\bs{x}[j]$ denotes the $j$th voxel of an image $\bs{x}$, $M$ is the number of realizations ($M=5$ in both XCAT phantom simulation and physical phantom measurement) and $J_{\text{Liver}}$ is the number of voxels in the volume of liver, $\mathrm{STD}_{\text{BKG}}$ is standard deviation between voxel values in uniform background liver (indicated in Fig.~\ref{fig2}), and $J_{\text{FOV}}$ is the total number of voxels in the FOV. As the background region when calculating the patient CNR, we used a part of liver region that has relatively uniform activity distribution.

\begin{figure*}[t]
\addtolength{\tabcolsep}{-6.5pt}
\centering
\begin{tabular}{ccc}
\includegraphics[scale=0.25, trim=3em 0em 5em 2em, clip]{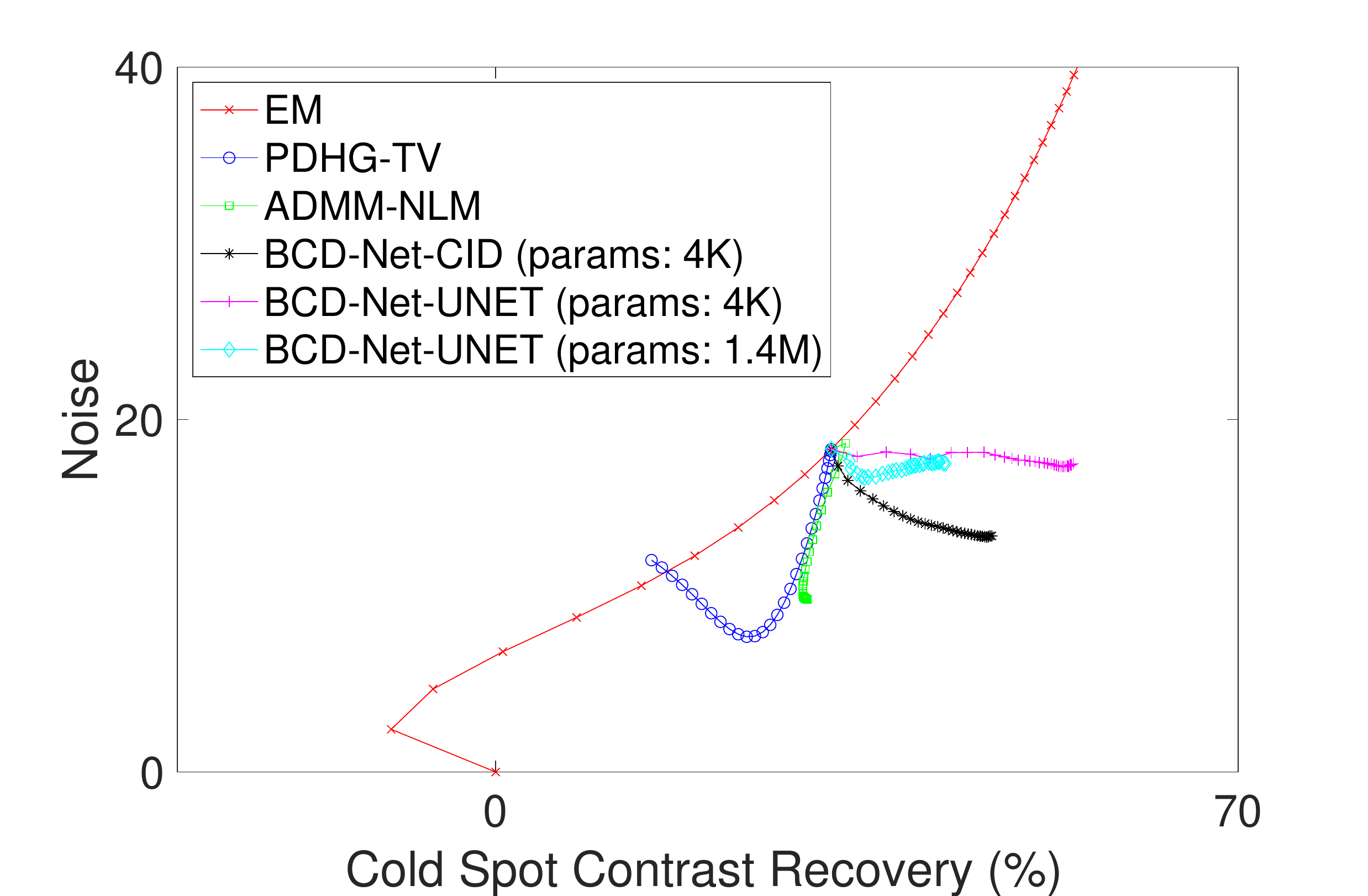} &
\includegraphics[scale=0.25, trim=3em 0em 5em 2em, clip]{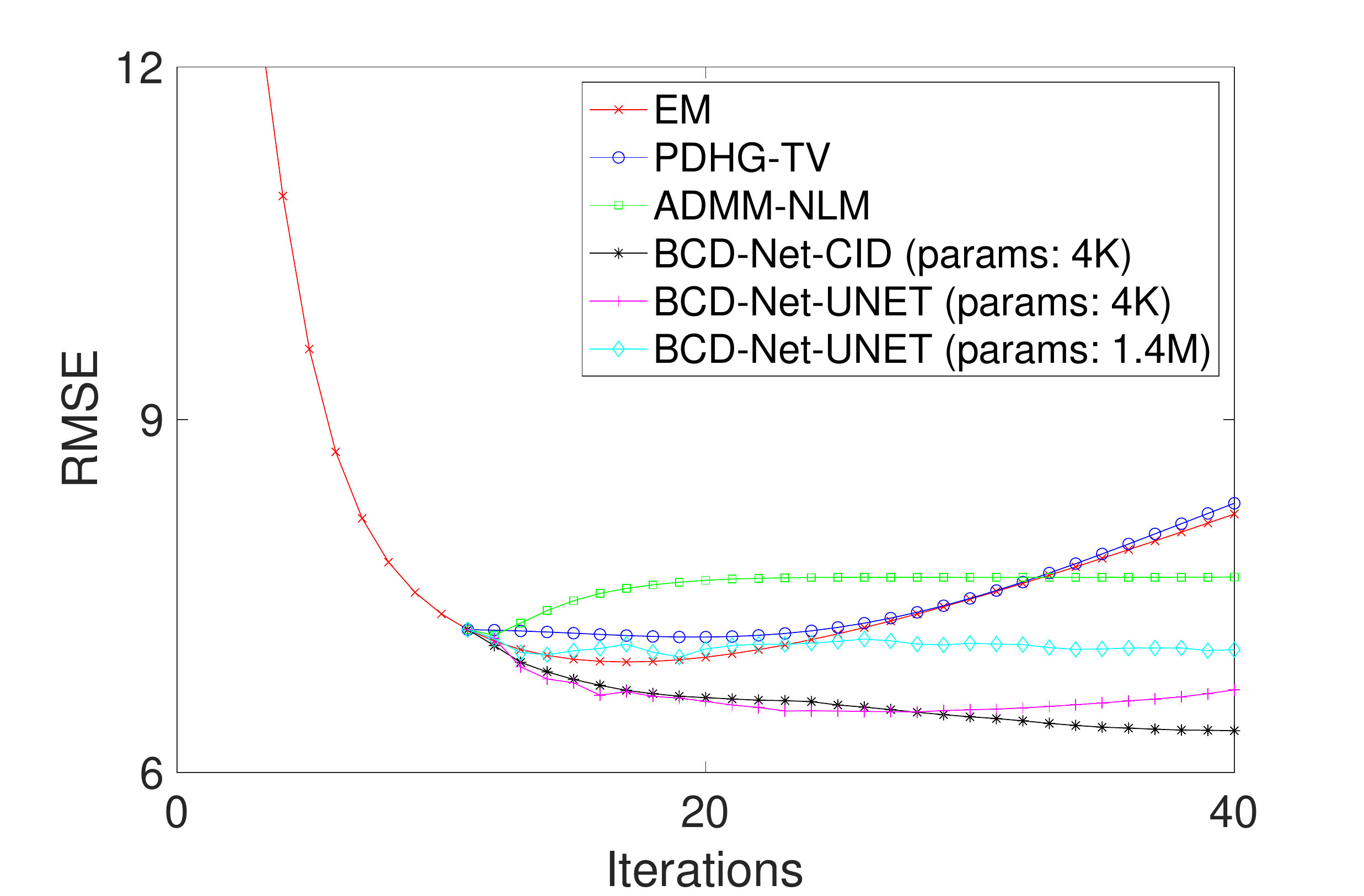} &
\includegraphics[scale=0.25, trim=3em 0em 5em 2em, clip]{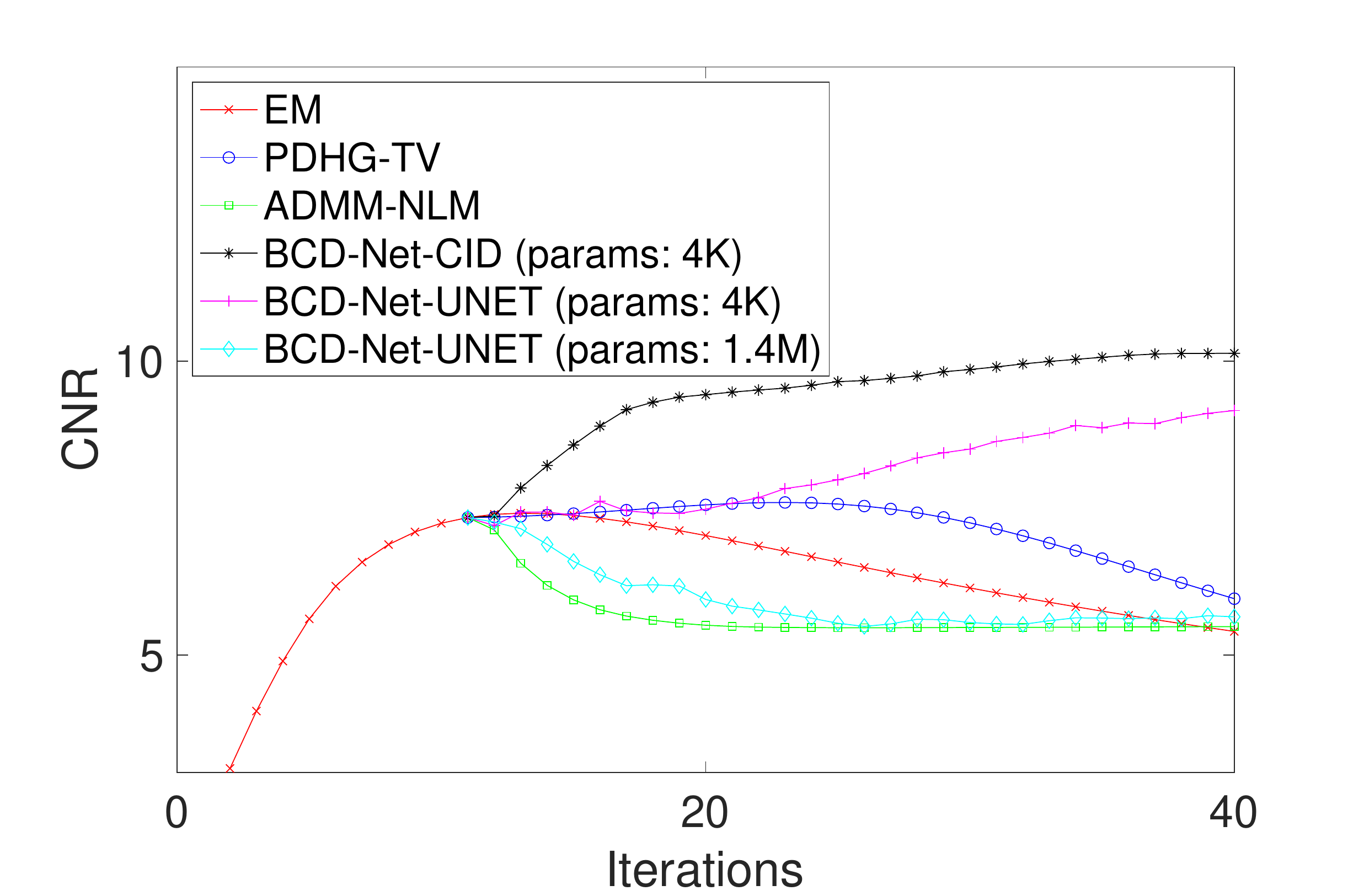} \vspace{-0.0em}\\
 (a)&(b)&(c)
\end{tabular}
\caption{(a) Plot of noise in background liver vs contrast recovery in cold spot (b) RMSE vs iteration (c) Contrast to noise ratio vs iteration. We initialized regularized methods with the 10th iterate of EM reconstruction.}
\label{fig3} 
\end{figure*}

\begin{figure*}[t]
\small\addtolength{\tabcolsep}{-6.5pt}
\centering
\begin{tabular}{cccc}
Sphere phantom attenuation map (coronal) &  Attenuation map (axial) & True activity image & $\bs{x}^{(0)}$ of regularized methods (EM) \\
\includegraphics[scale=0.20,  trim=18em 0em 0em 2em, clip]{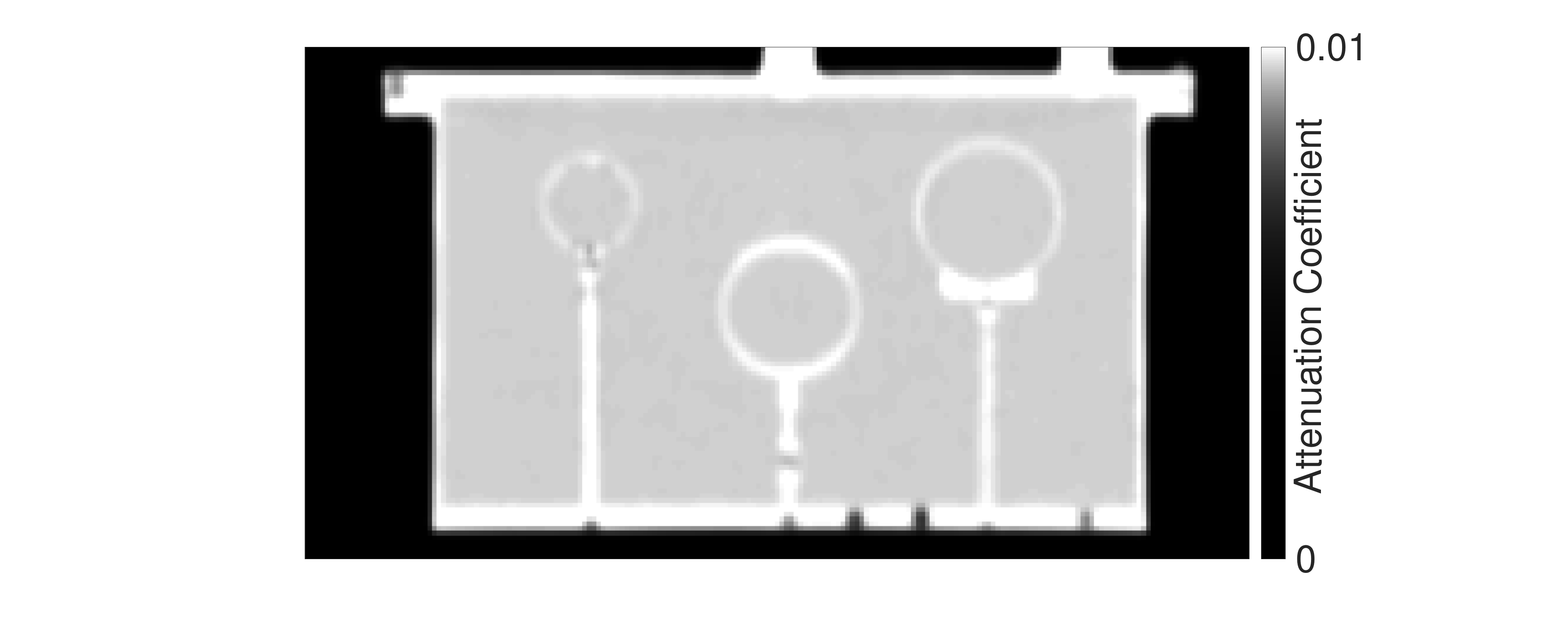} &
\includegraphics[scale=0.20, trim=4em 0em 0em 2em,  clip]{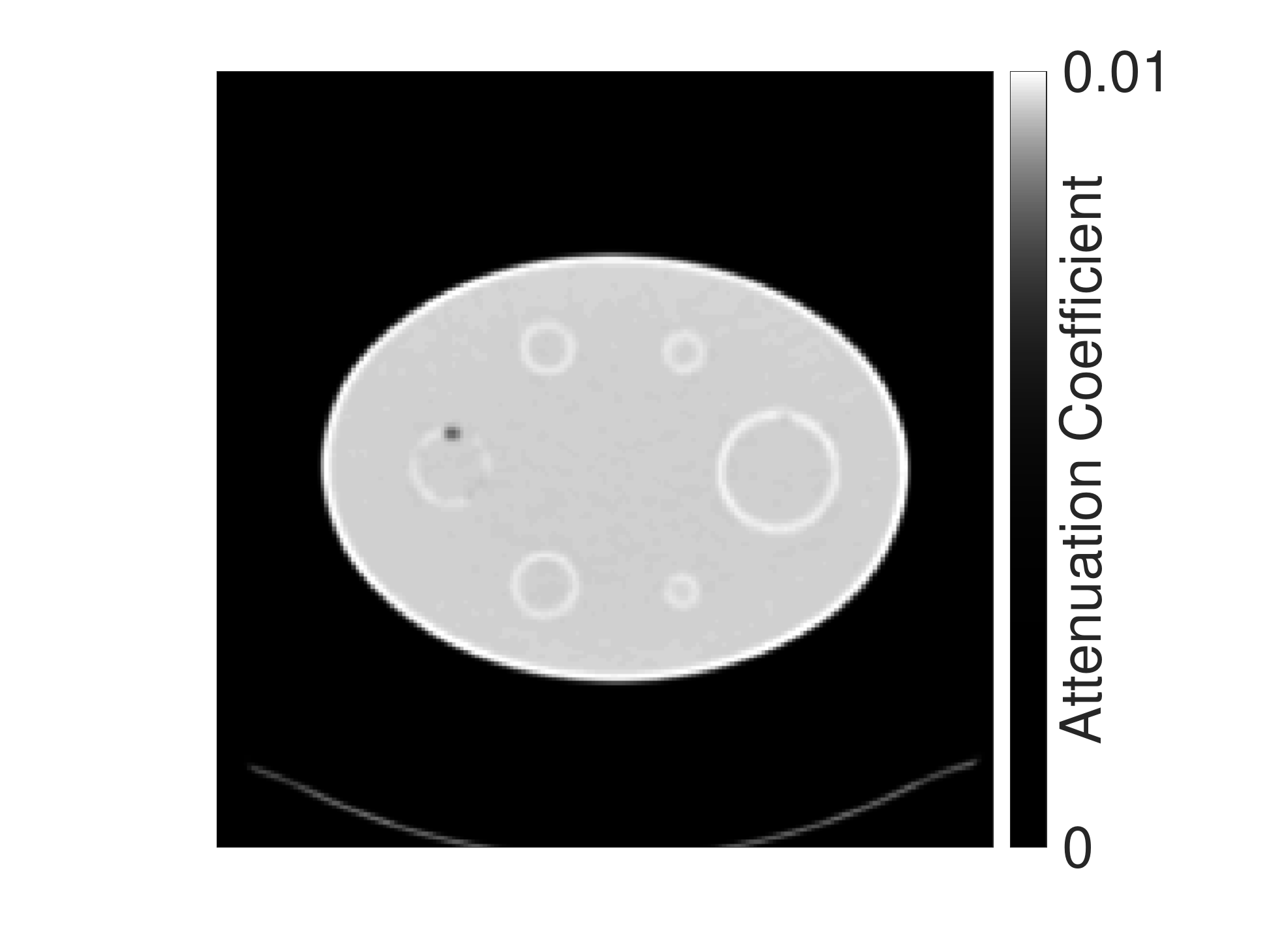} & 
\includegraphics[scale=0.20,  trim=4em 0em 0em 2em, clip]{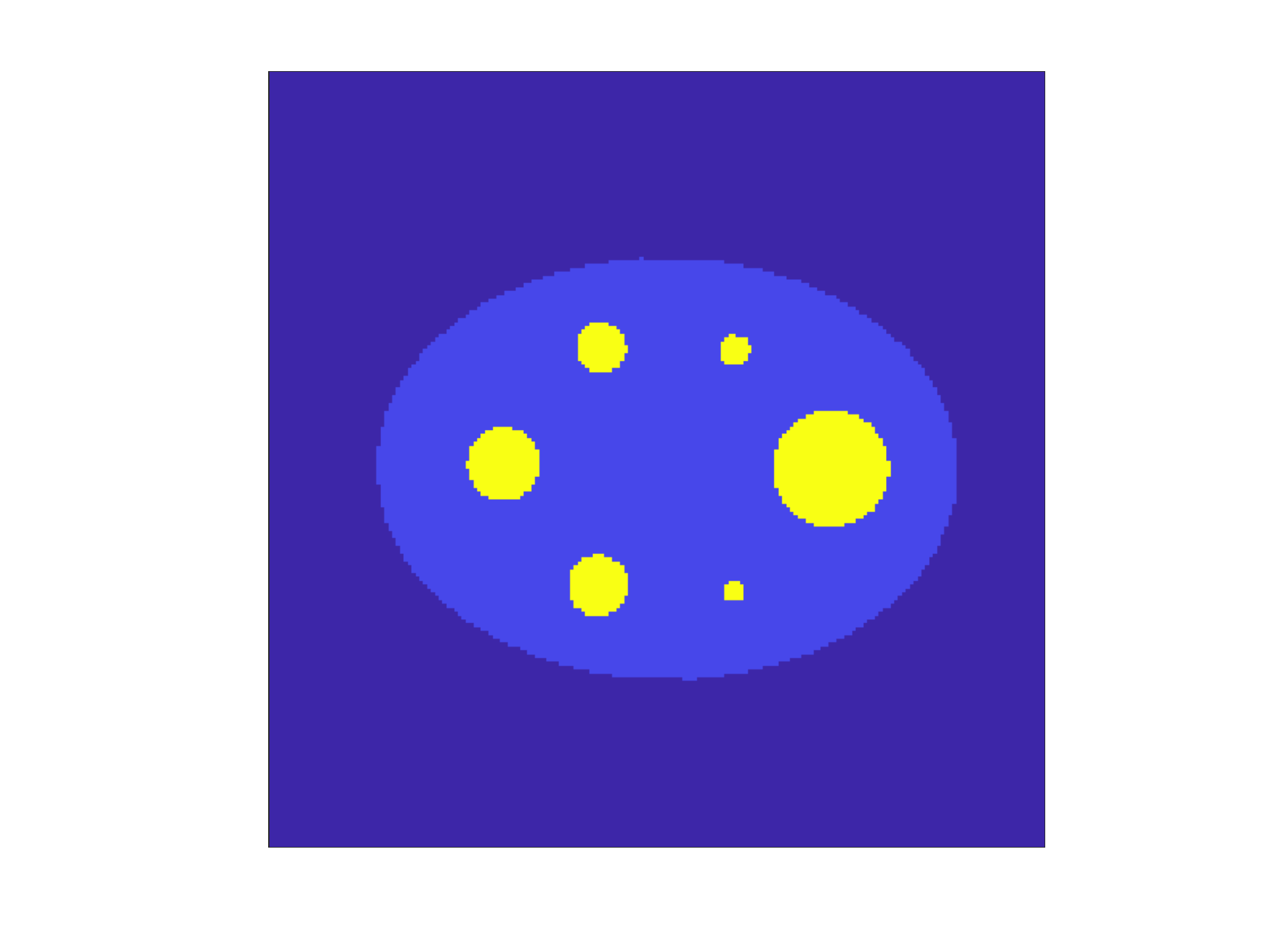} &
\includegraphics[scale=0.20, trim=4em 0em 0em 2em,  clip]{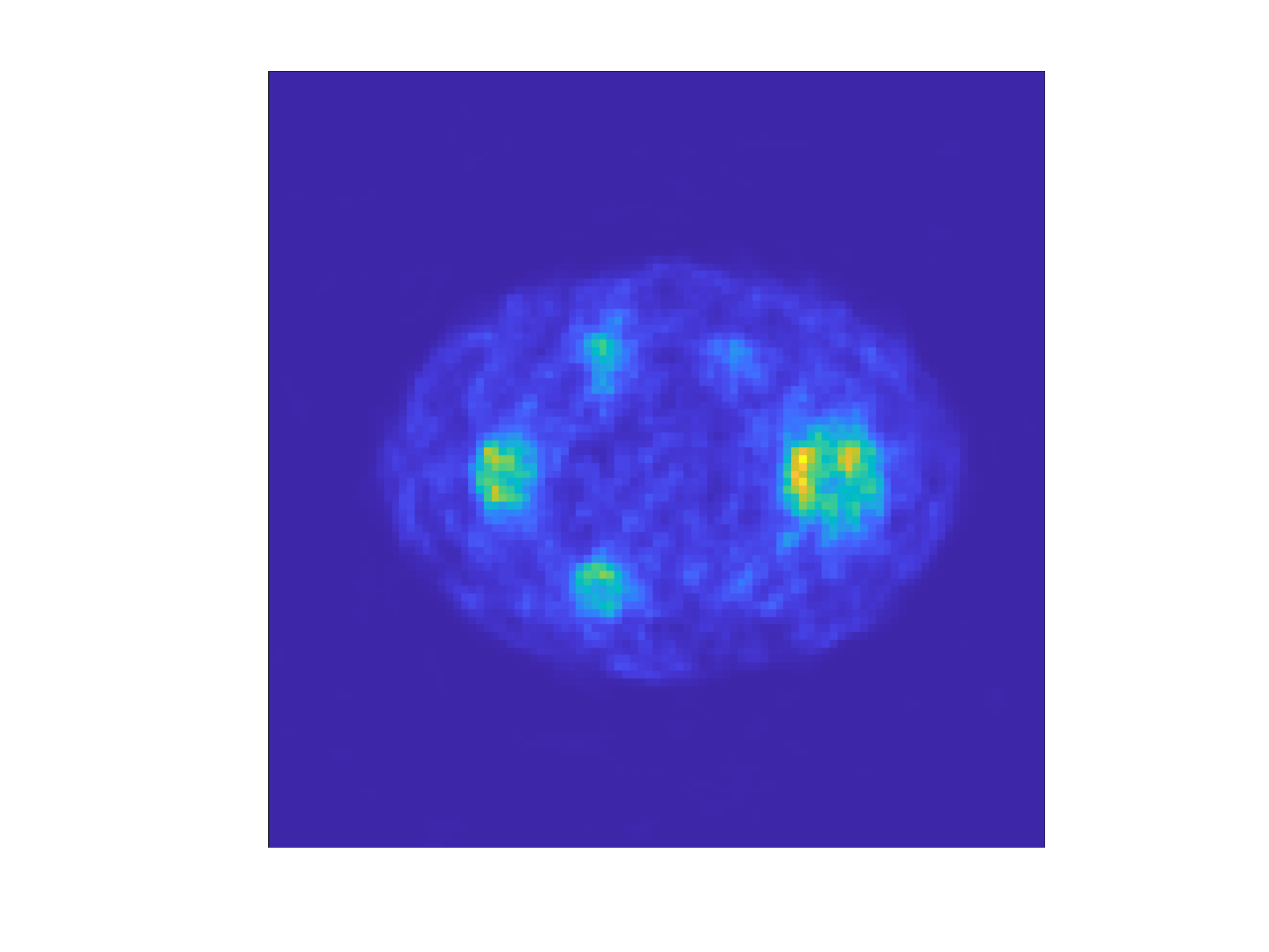} \vspace{-0.5em}\\
Liver phantom attenuation map (coronal)&   Attenuation map (axial) & True activity image & $\bs{x}^{(0)}$ of regularized methods (EM) \\
\includegraphics[scale=0.20,  trim=4em 0em 0em 2em, clip]{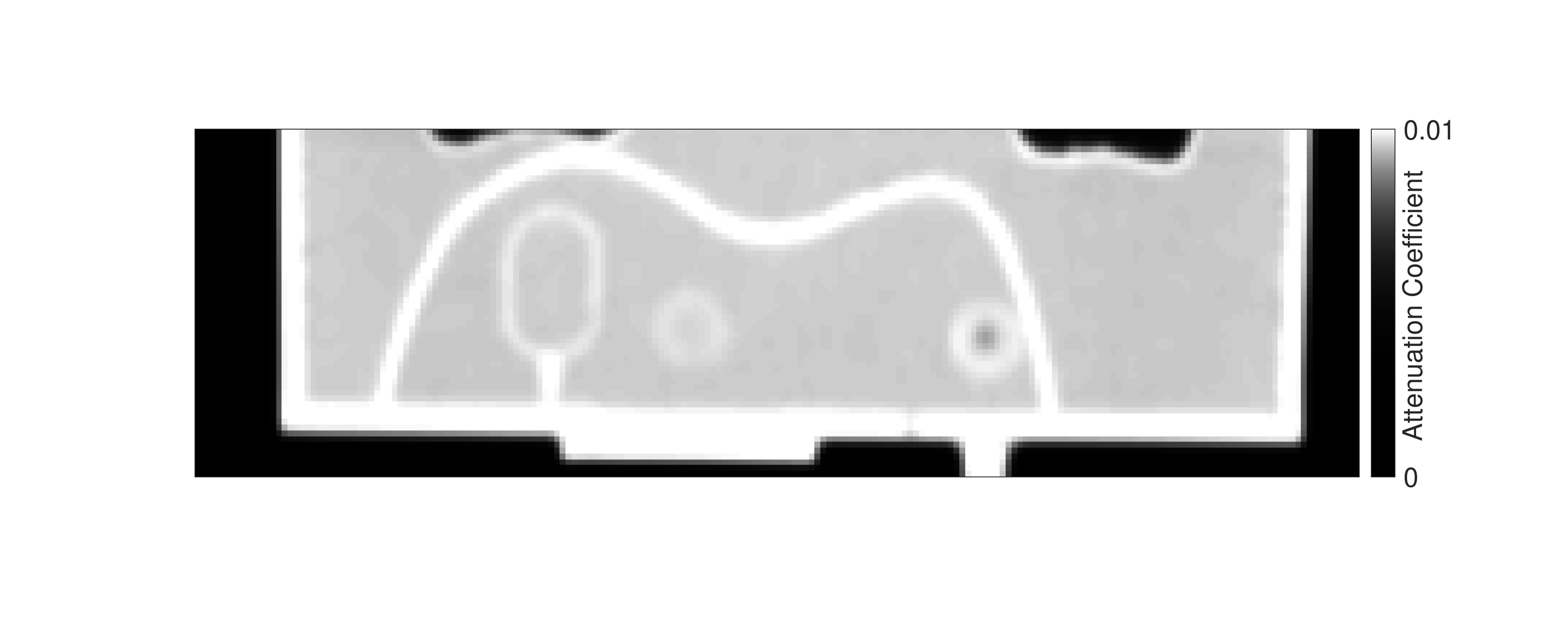} &
\includegraphics[scale=0.20,  trim=4em 0em 0em 2em, clip]{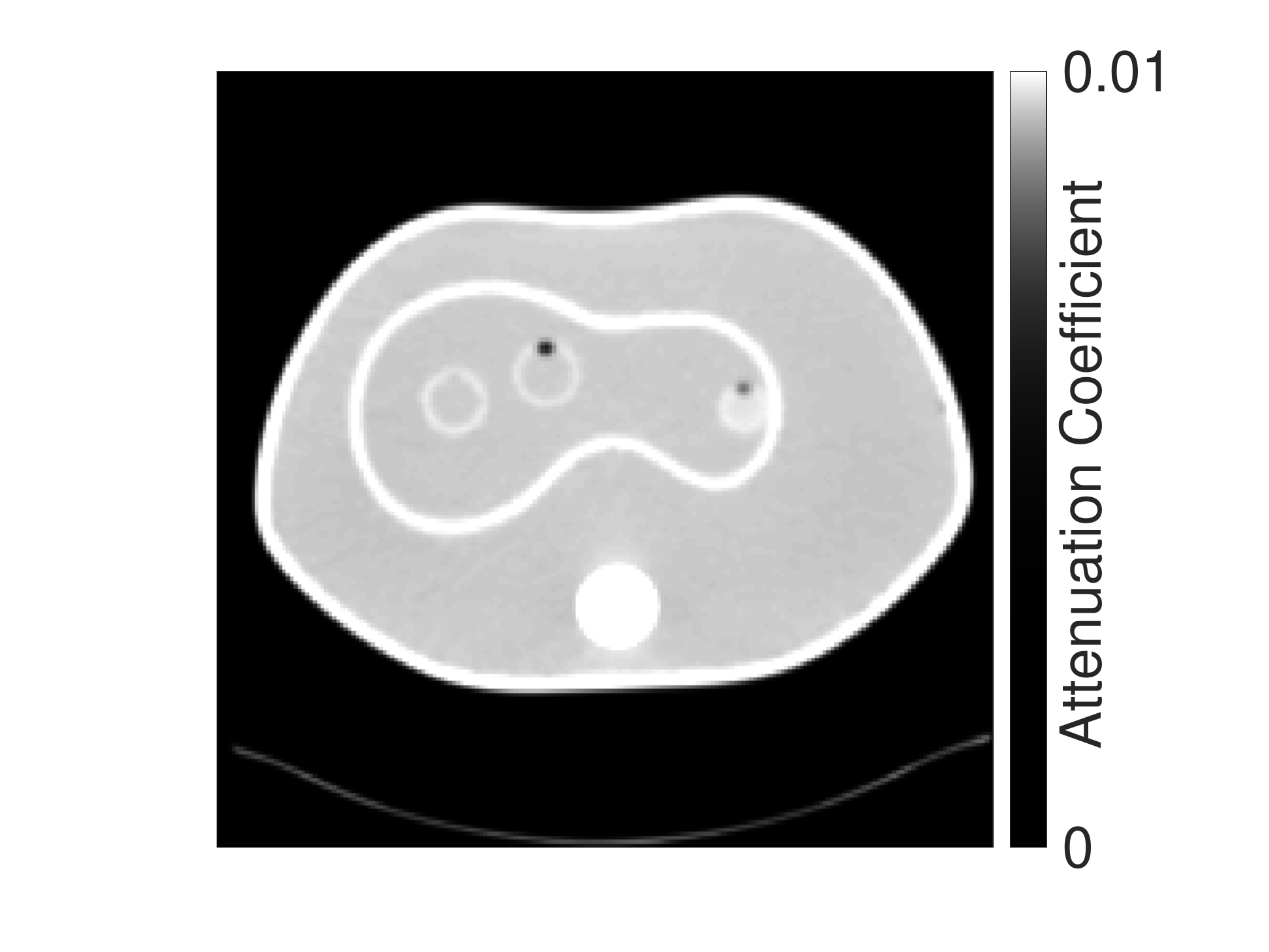} &
\includegraphics[scale=0.20,  trim=4em 0em 0em 2em, clip]{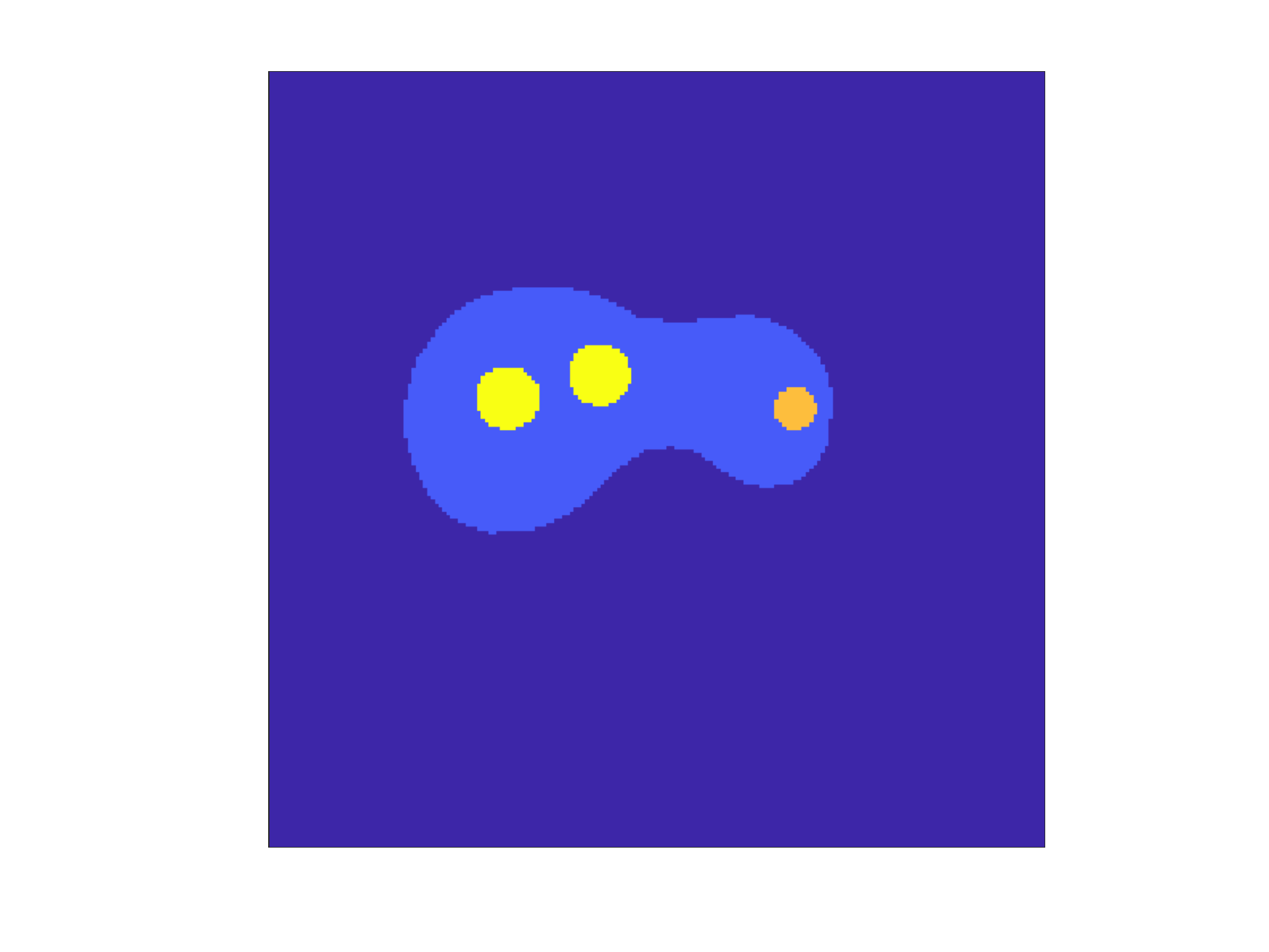} &
\includegraphics[scale=0.20, trim=4em 0em 0em 2em,  clip]{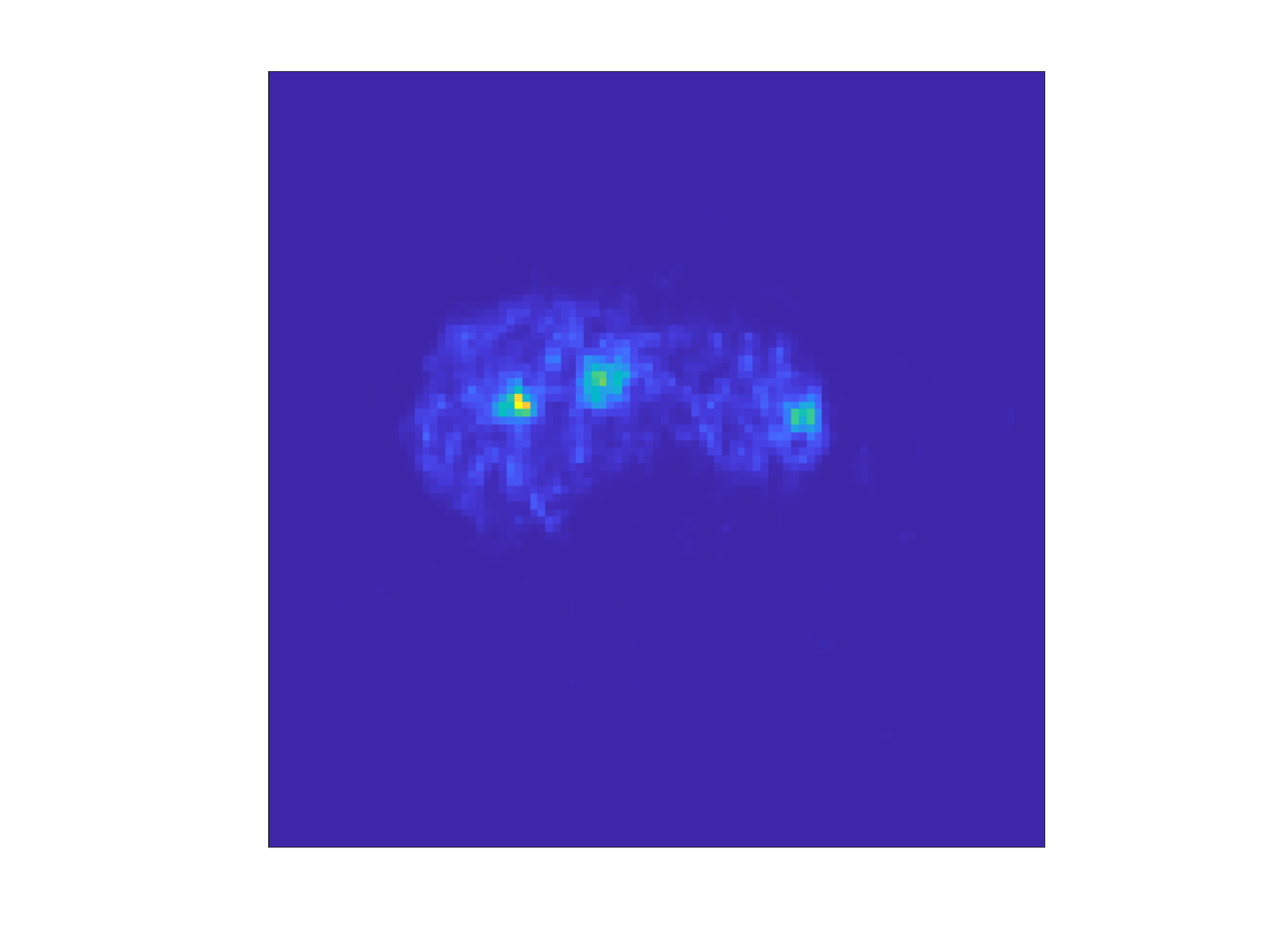} \vspace{-0.5em}\\
\end{tabular}
\begin{tabular}{ccccc}
\hspace{-1em}TV  & \hspace{-1em} NLM & \hspace{-1em} BCD-Net-CID& \hspace{-1em} BCD-Net-UNet& \hspace{-1em} BCD-Net-UNet  \\
     & & (params: 4K) & (params: 4K) & (params: 1.4M)  \\
\includegraphics[scale=0.20,  trim=3em 0em 3em 5em, clip]{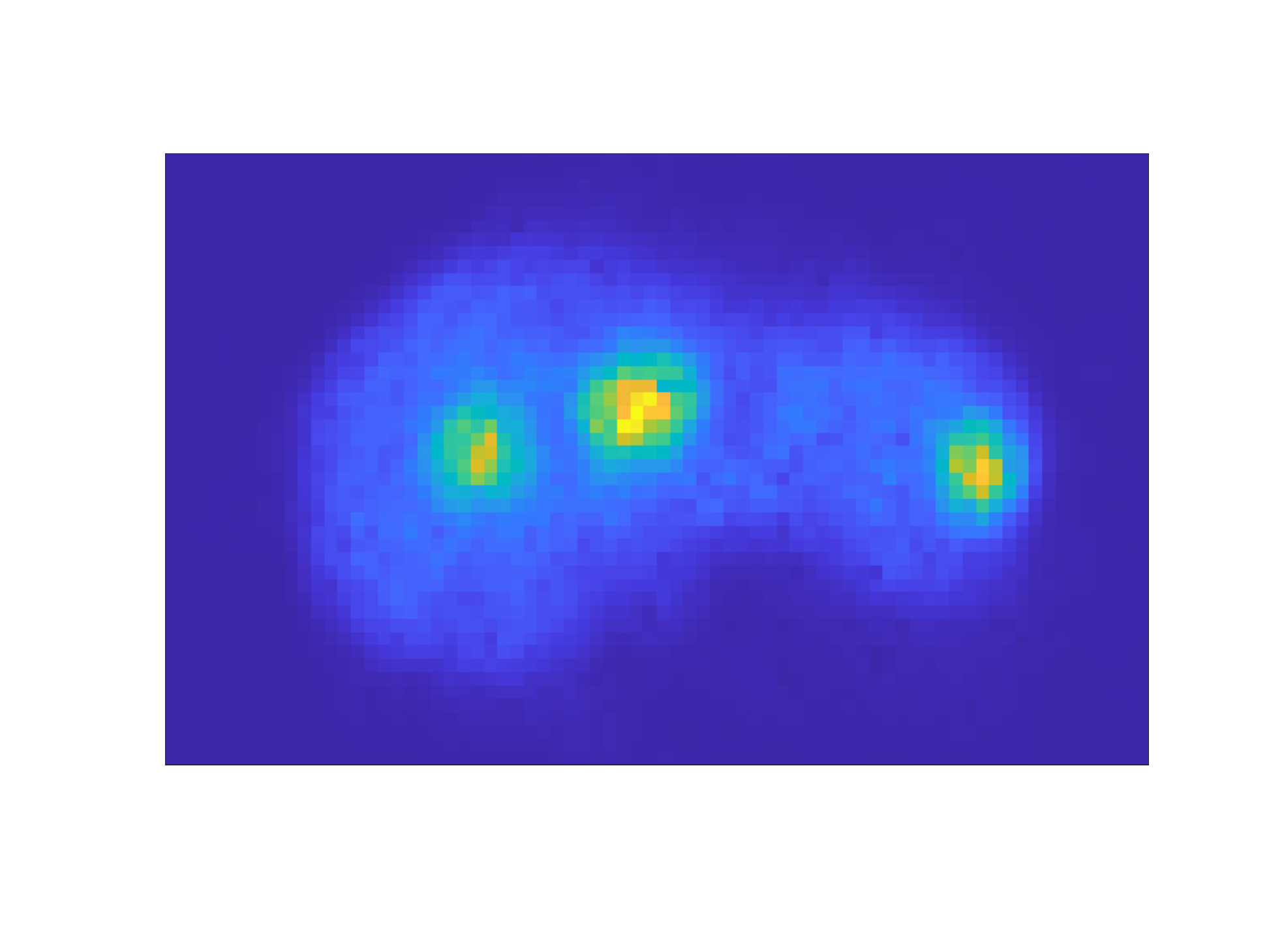} &
\includegraphics[scale=0.20,  trim=3em 0em 3em 5em, clip]{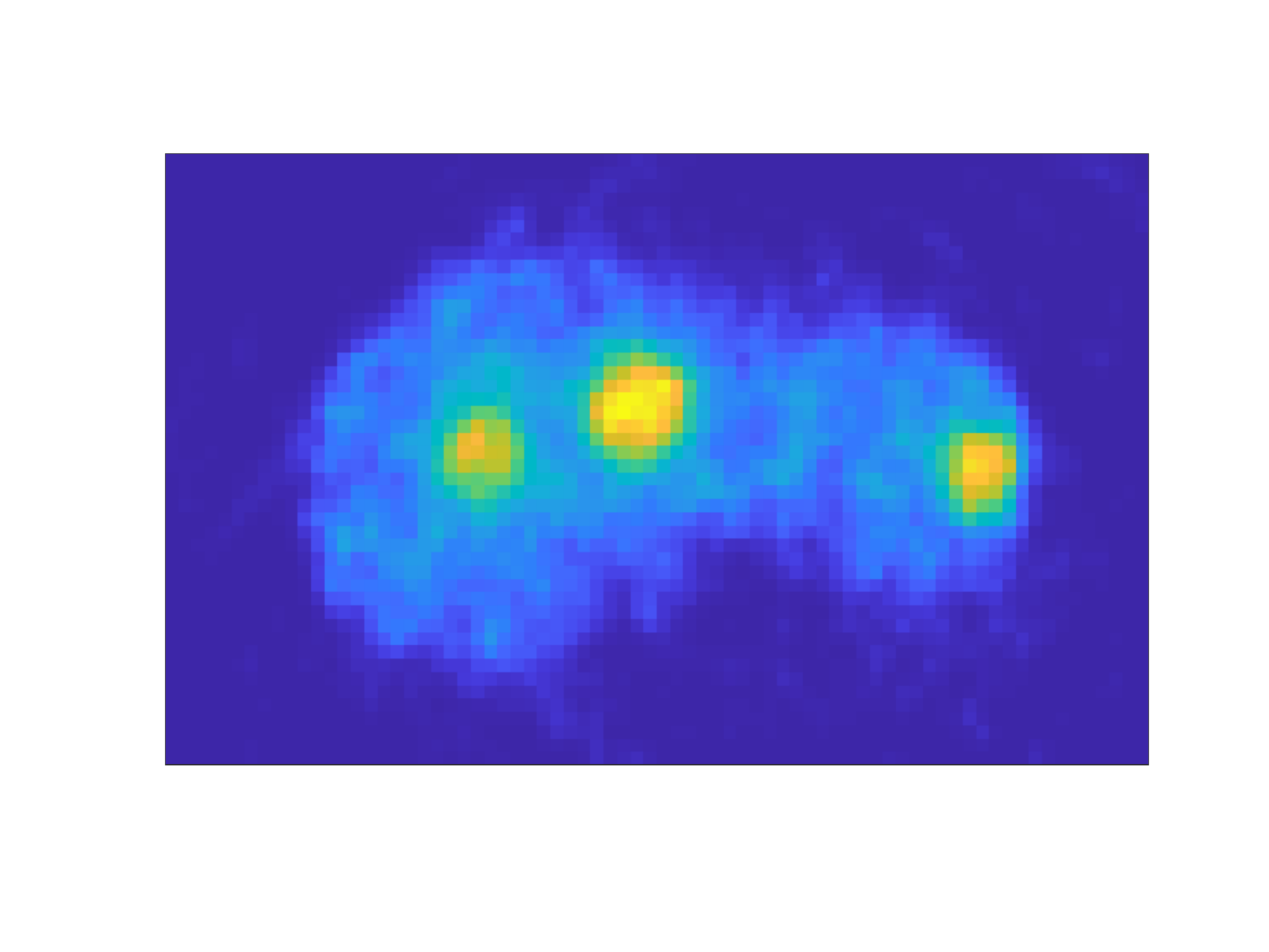} &
\includegraphics[scale=0.20,  trim=3em 0em 3em 5em, clip]{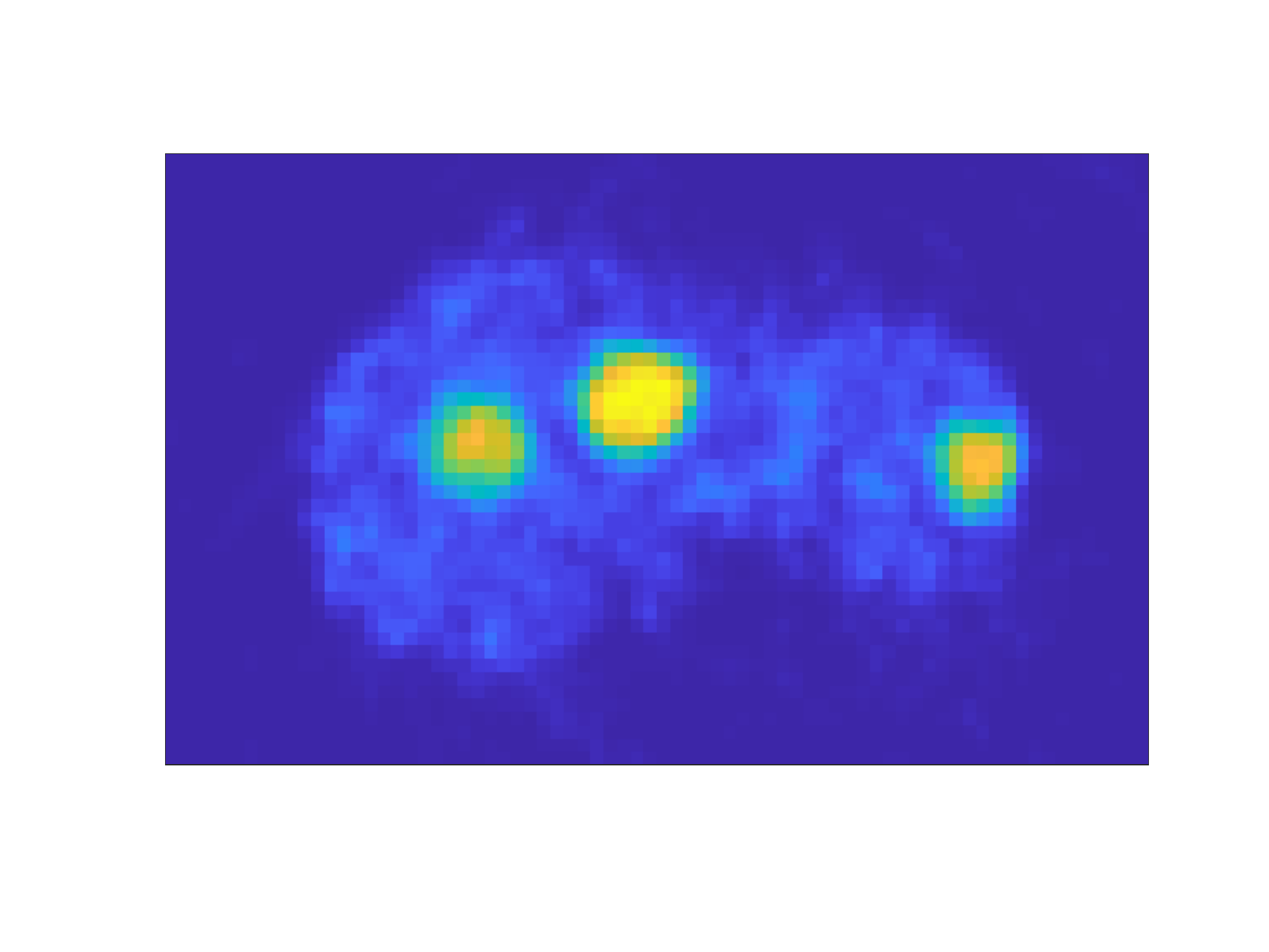} &
\includegraphics[scale=0.20,  trim=3em 0em 3em 5em, clip]{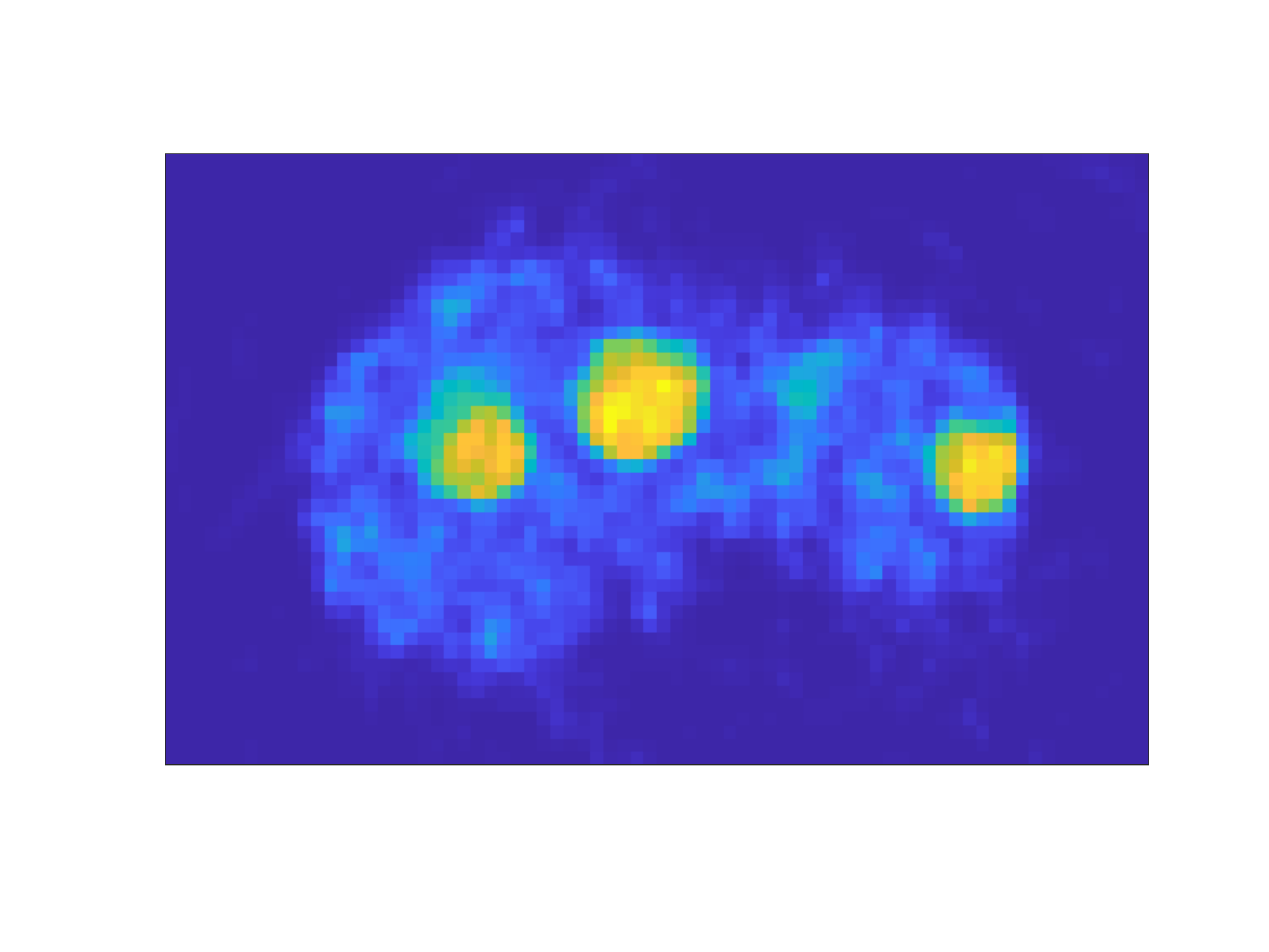} &
\includegraphics[scale=0.20,  trim=3em 0em 3em 5em, clip]{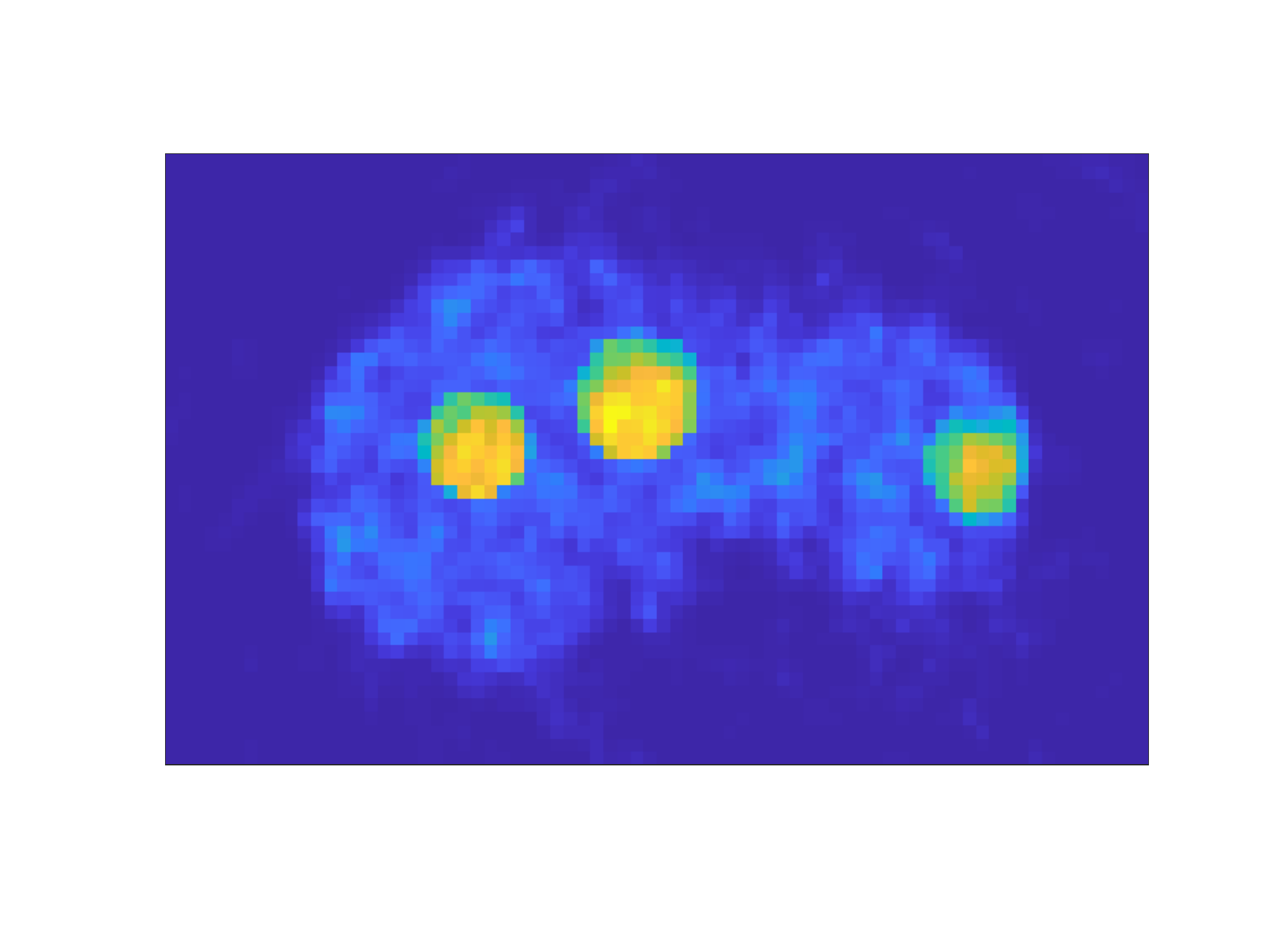} \vspace{-0.5em}\\
\end{tabular}

\caption{Y90 PET/CT physical phantom measurement: (First row: training data, Second row: testing data) Attenuation map, true activity, and $\bs{x}^{(0)}$ of regularized methods of sphere and liver phantom used for training and testing BCD-Net. (Third row) Reconstructed images of one slice from different reconstruction methods.  }
\label{fig4} 
\end{figure*}

\begin{figure*}[t]
\addtolength{\tabcolsep}{-6.5pt}
\centering
\begin{tabular}{cc}
\includegraphics[scale=0.30, trim=3em 0em 5em 0em, clip]{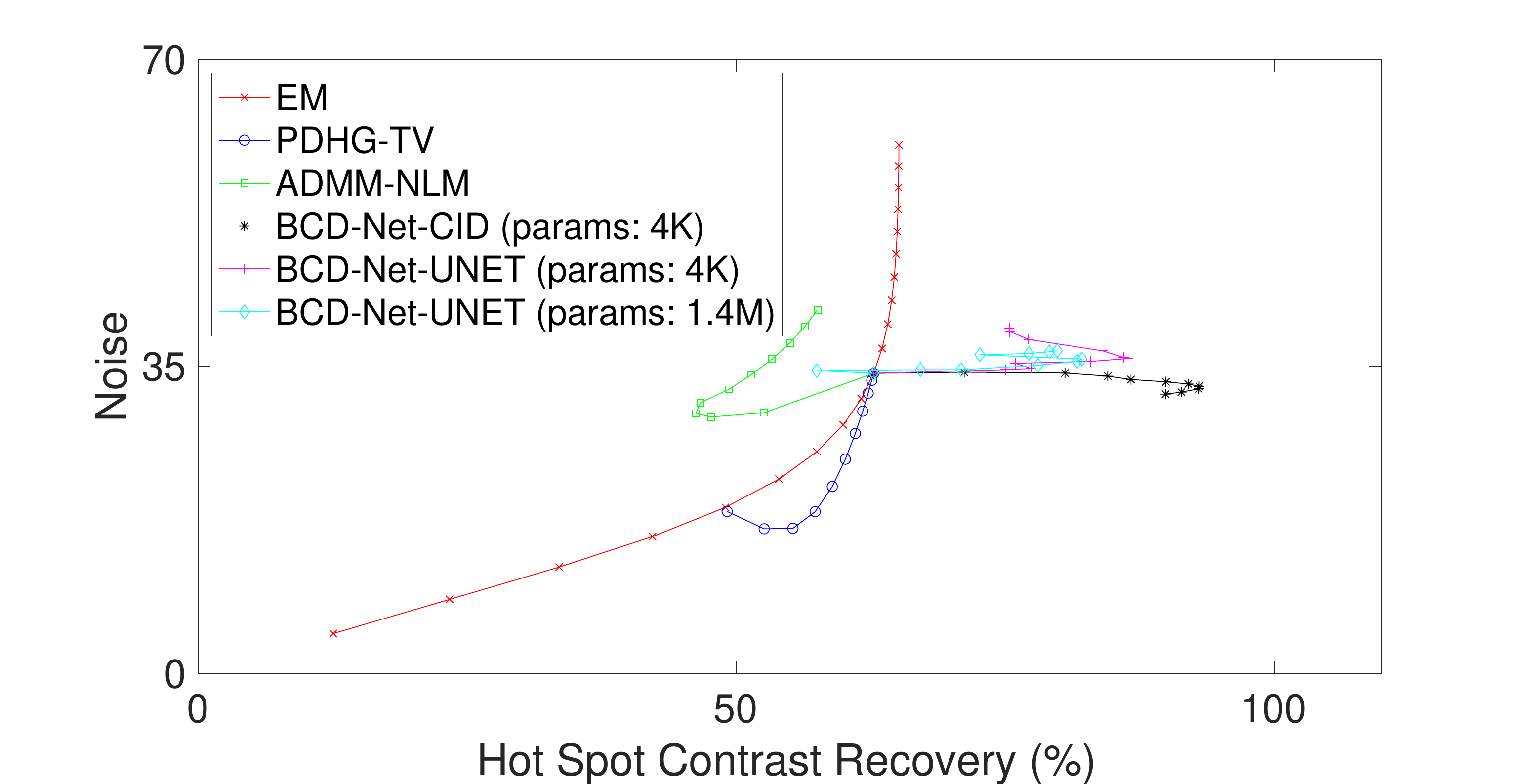} &
\includegraphics[scale=0.30, trim=3em 0em 5em 0em, clip]{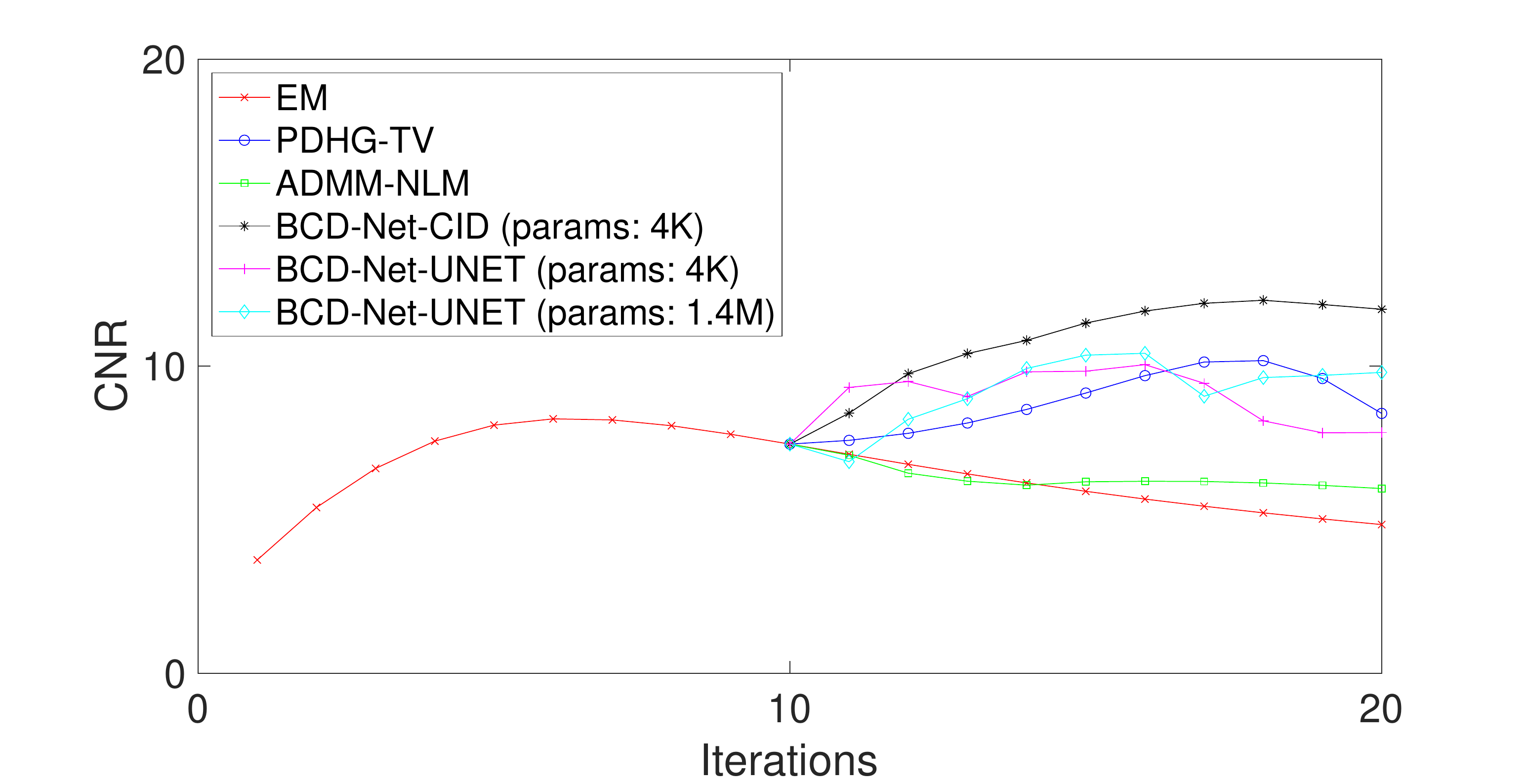}\\
 (a)&(b)
\end{tabular}
\caption{Liver phantom measurement: (a) Plot of noise in background liver vs contrast recovery in hot spheres (b) Contrast to noise ratio vs iteration. We initialized regularized methods with the 10th iterate of EM reconstruction.}
\label{fig5} 
\end{figure*}

\begin{figure*}[t]
\small\addtolength{\tabcolsep}{-6.5pt}
\centering
\begin{tabular}{cccc}
Patient attenuation map (coronal) &  Attenuation map (axial) &   OSEM w/ filter (coronal) &   OSEM w/ filter (axial) \\
\hspace{-3em}\includegraphics[scale=0.20,  trim=20 0em 10em 2em, clip]{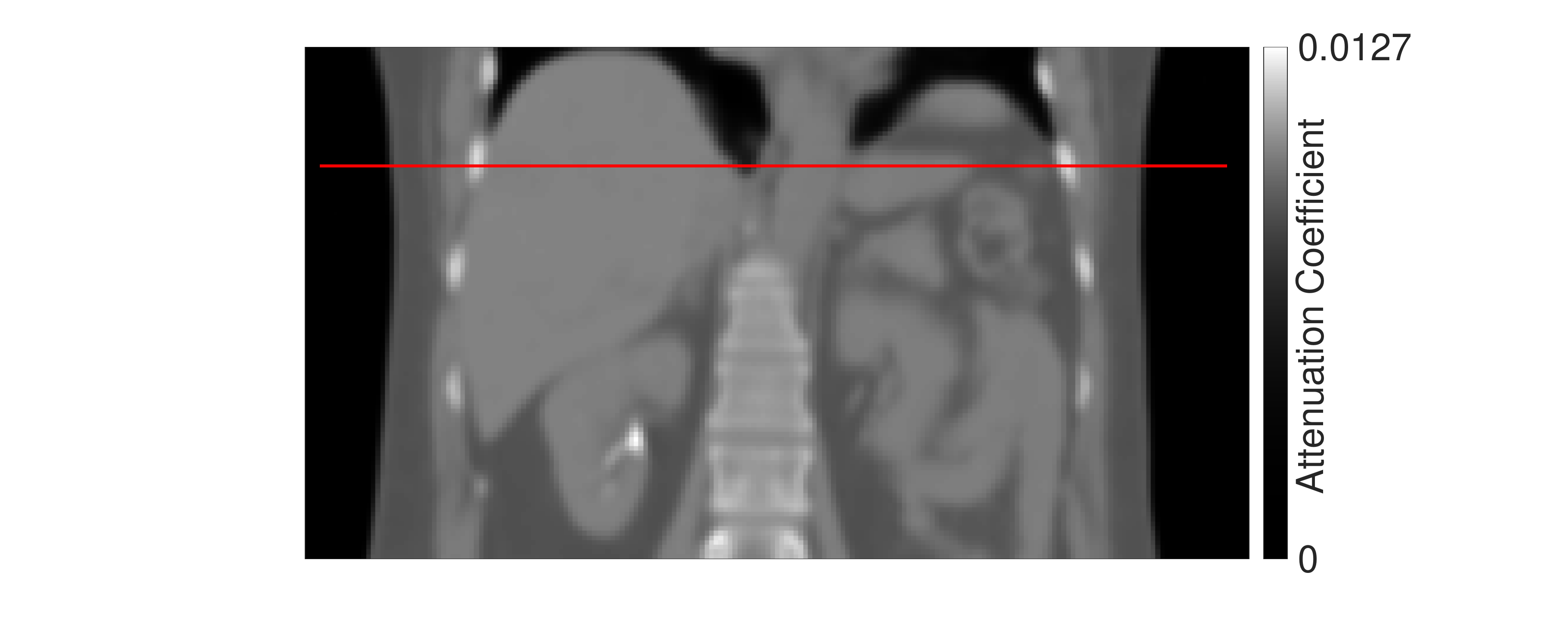} &
\includegraphics[scale=0.20, trim=10em 0em 1em 2em,  clip]{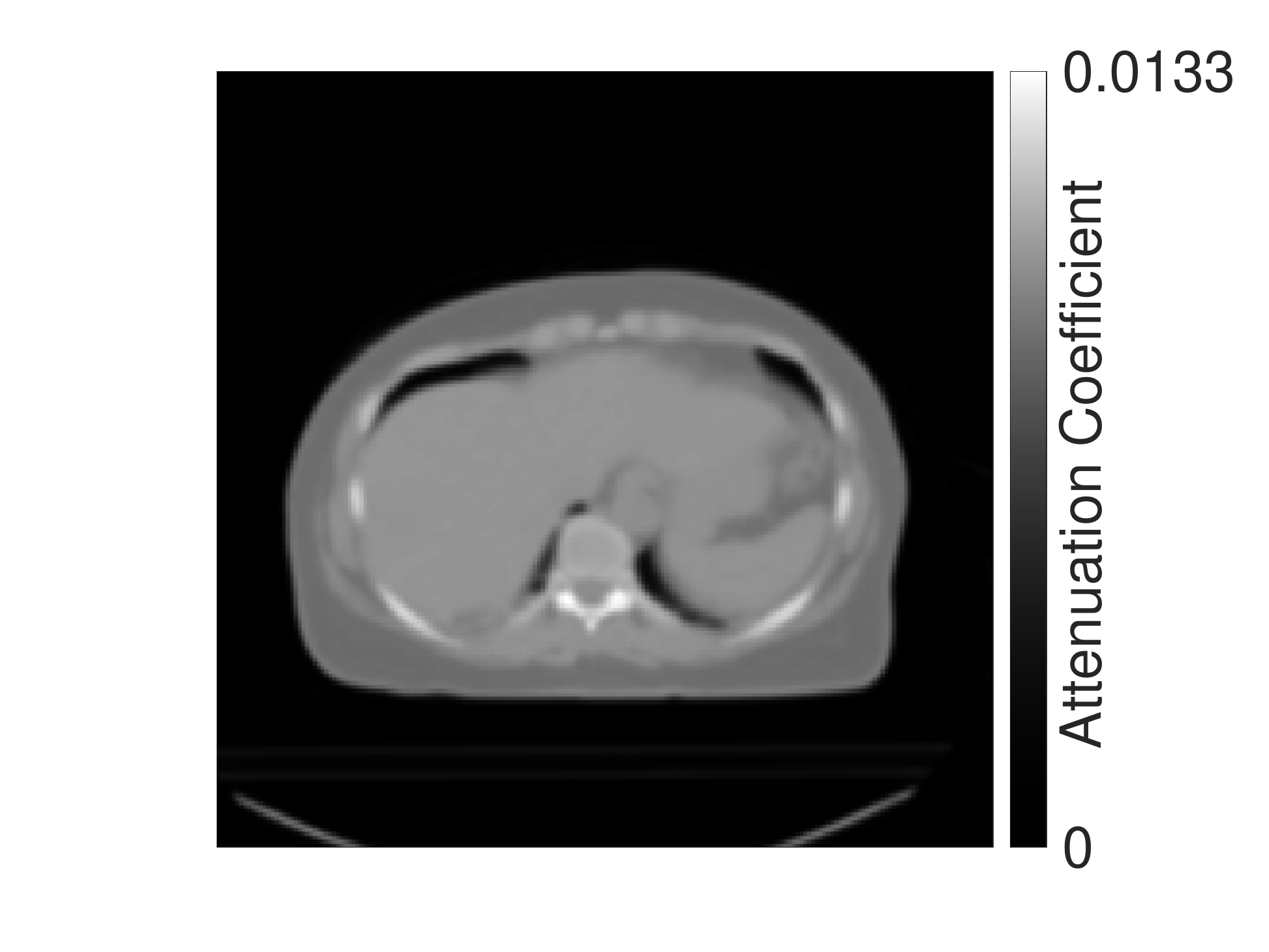} & 
\includegraphics[scale=0.20,  trim=20em 0em 20em 2em, clip]{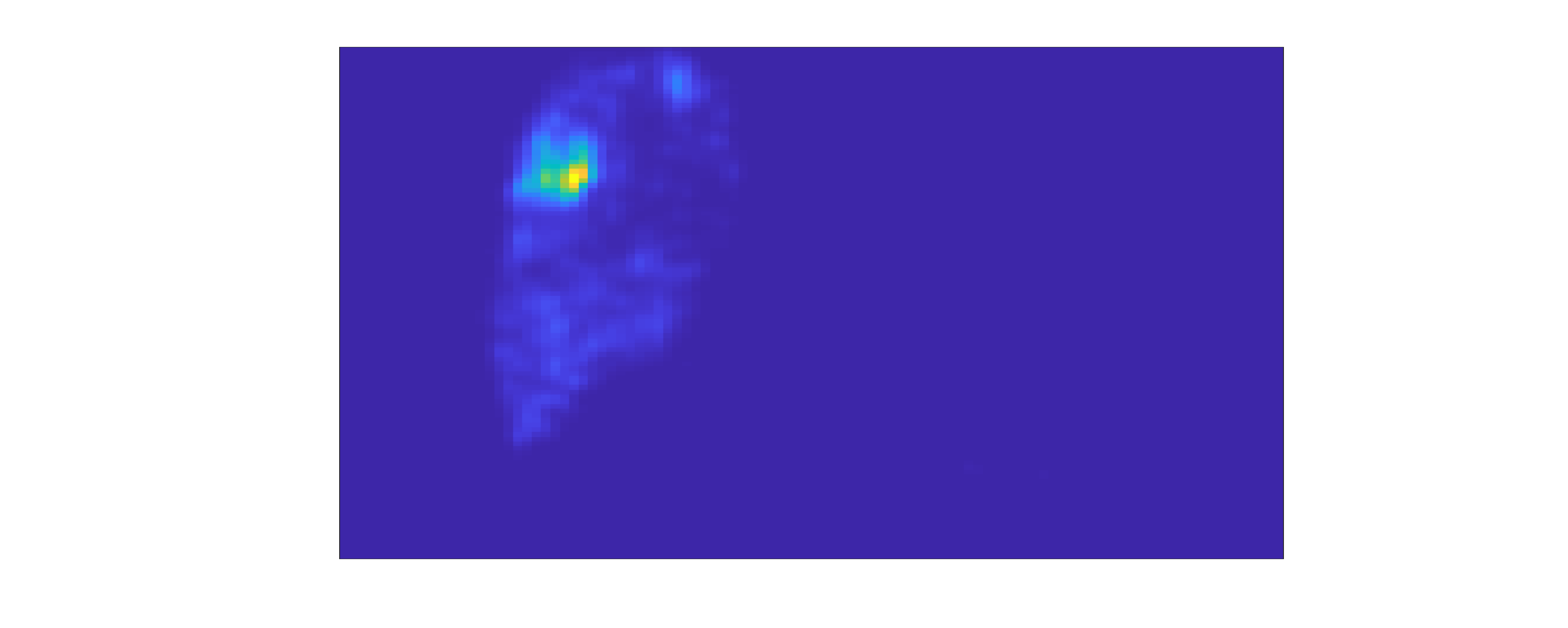} &
\includegraphics[scale=0.20, trim=10em 0em 10em 2em,  clip]{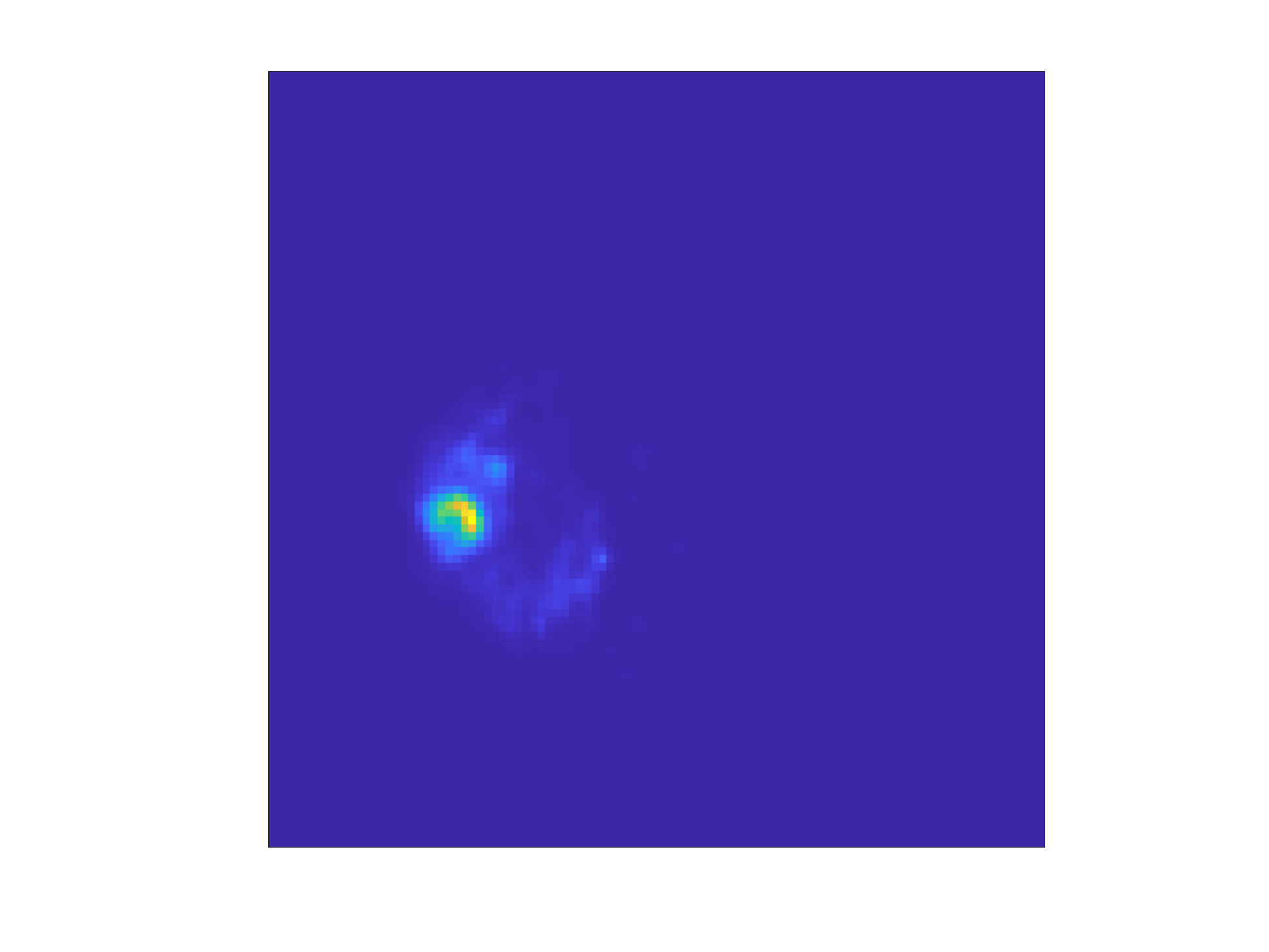} \vspace{-0.5em}\\
TV (coronal) & TV (axial) &  NLM (coronal) &  NLM (axial) \\
\hspace{-2.5em}\includegraphics[scale=0.20,  trim=20em 0em 20em 2em, clip]{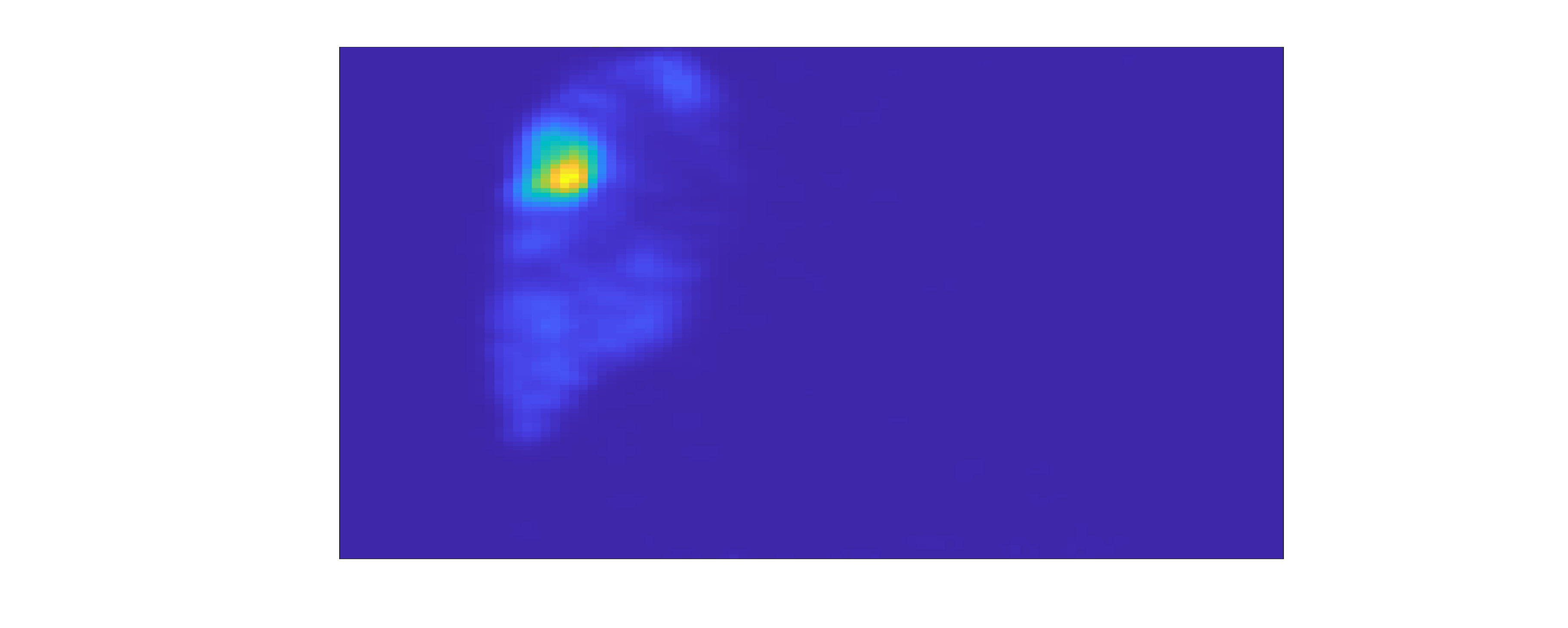} &
\hspace{-2.5em}\includegraphics[scale=0.20, trim=10em 0em 10em 2em,  clip]{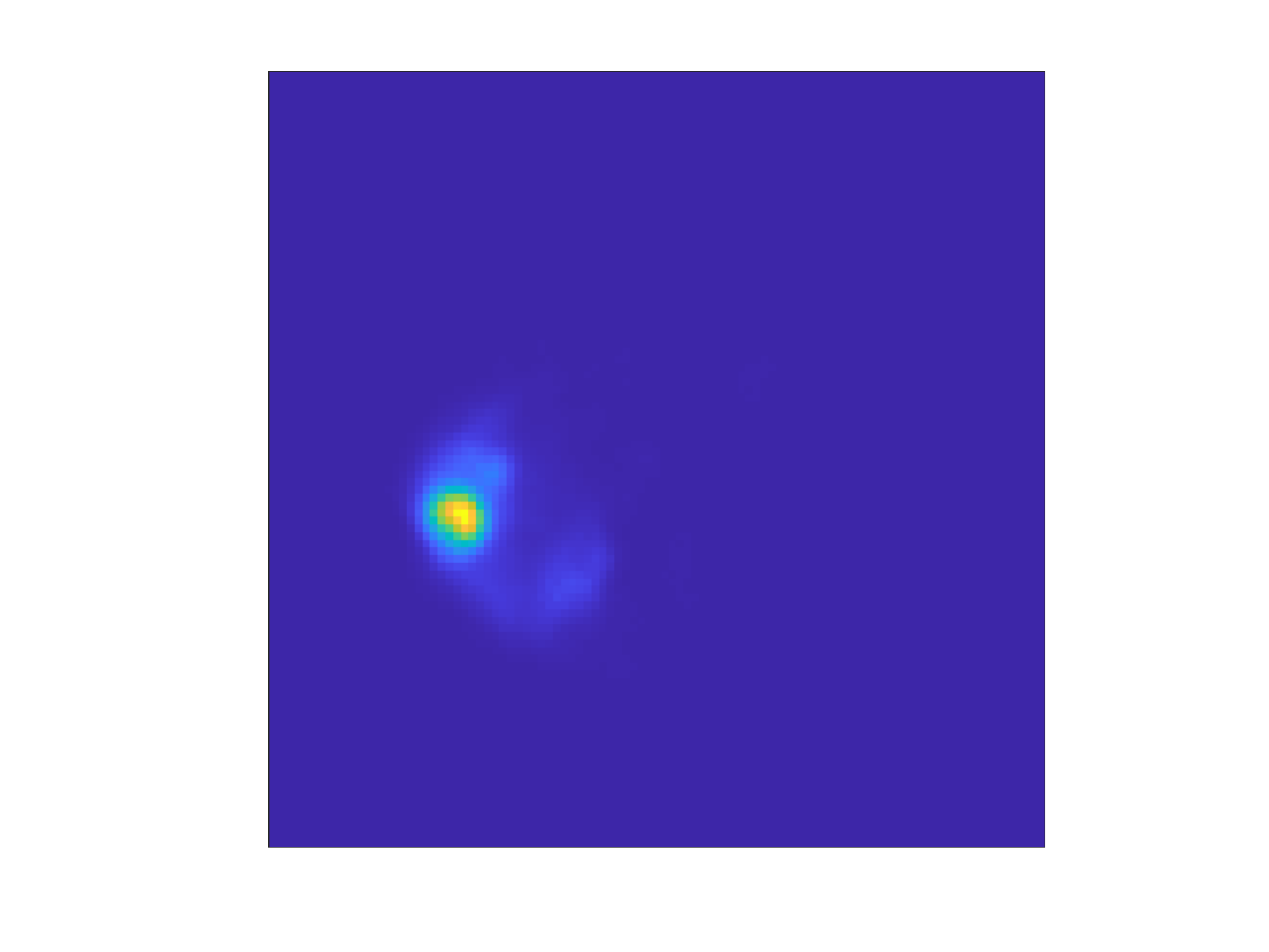}& 
\includegraphics[scale=0.20, trim=20em 0em 20em 2em, clip]{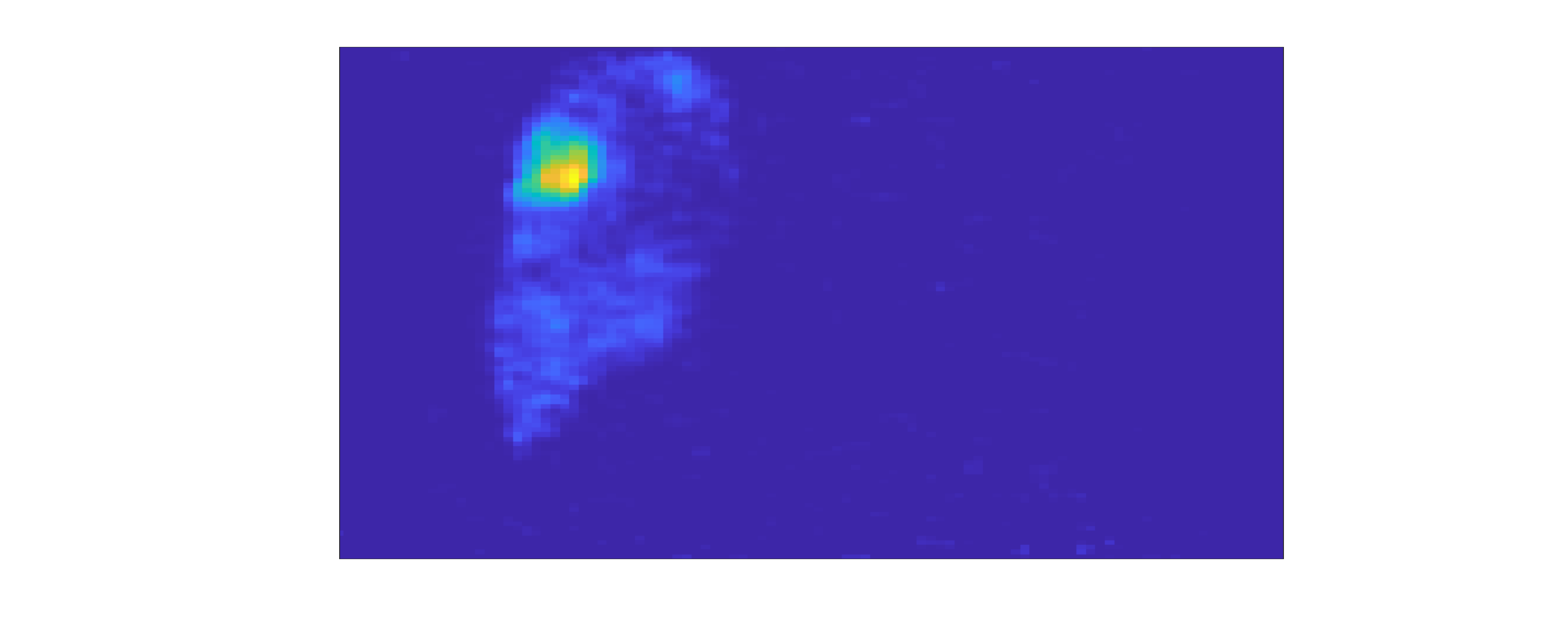} &
\includegraphics[scale=0.20, trim=10em 0em 10em 2em,  clip]{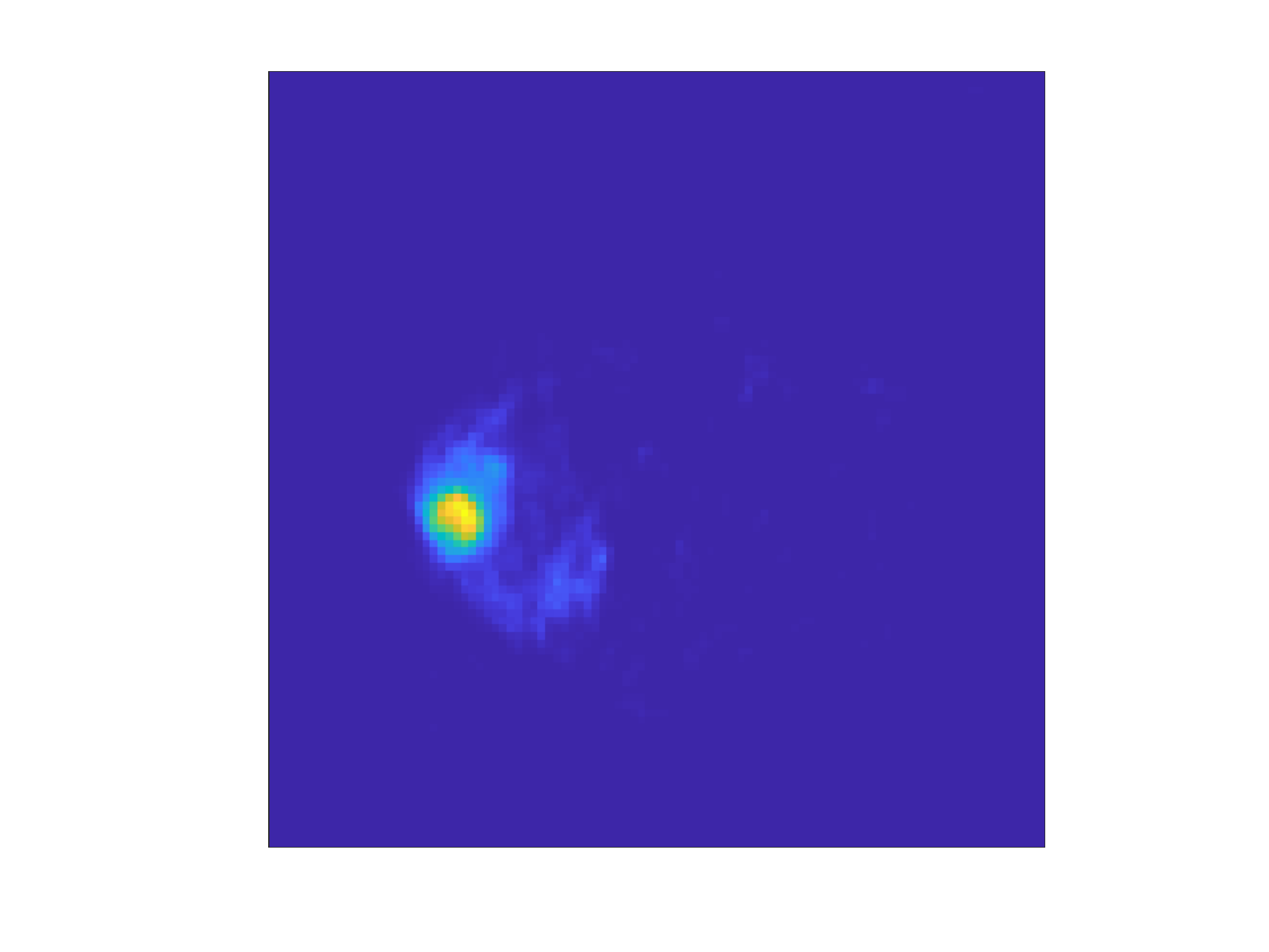} \vspace{-0.5em}\\
BCD-Net-CID (coronal) &  BCD-Net-CID (axial) & BCD-Net-UNet (coronal) & BCD-Net-UNet (axial)  \\
\hspace{-2.5em}\includegraphics[scale=0.20,  trim=20em 0em 20em 2em, clip]{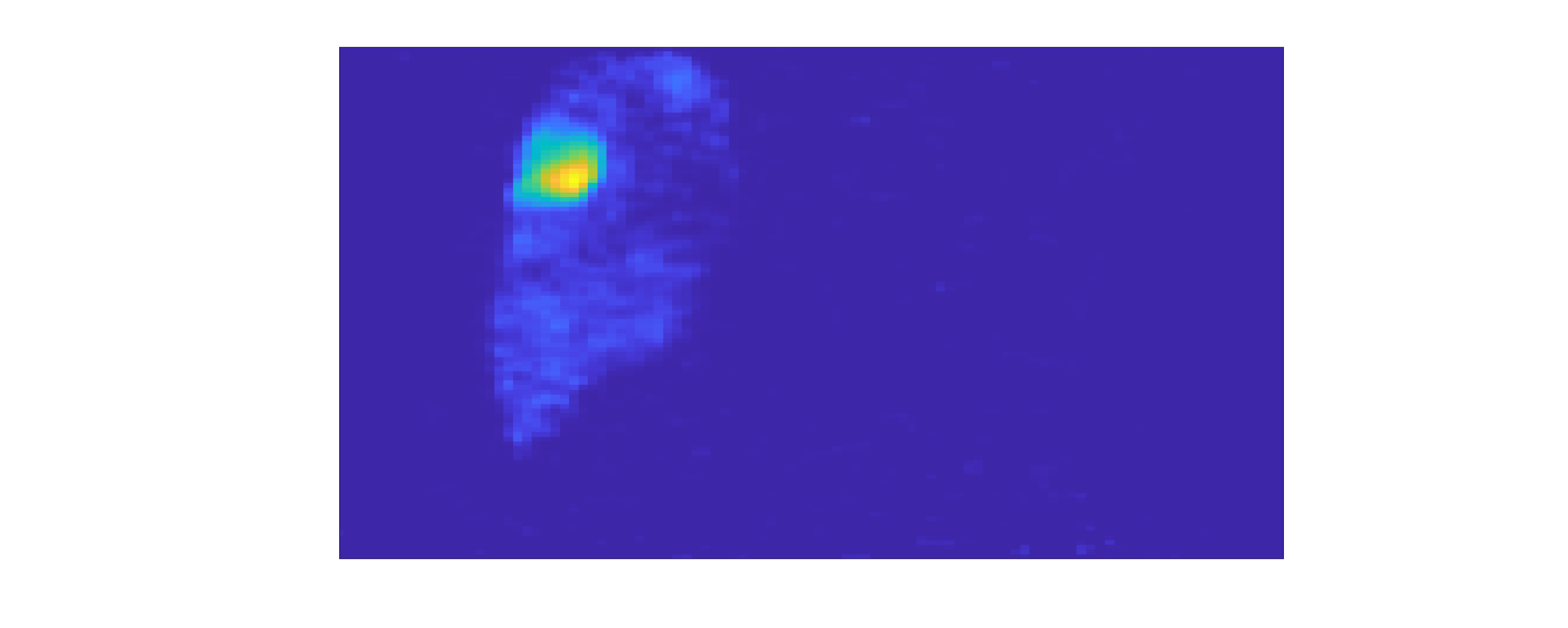} &
\hspace{-2.5em}\includegraphics[scale=0.20, trim=10em 0em 10em 2em,  clip]{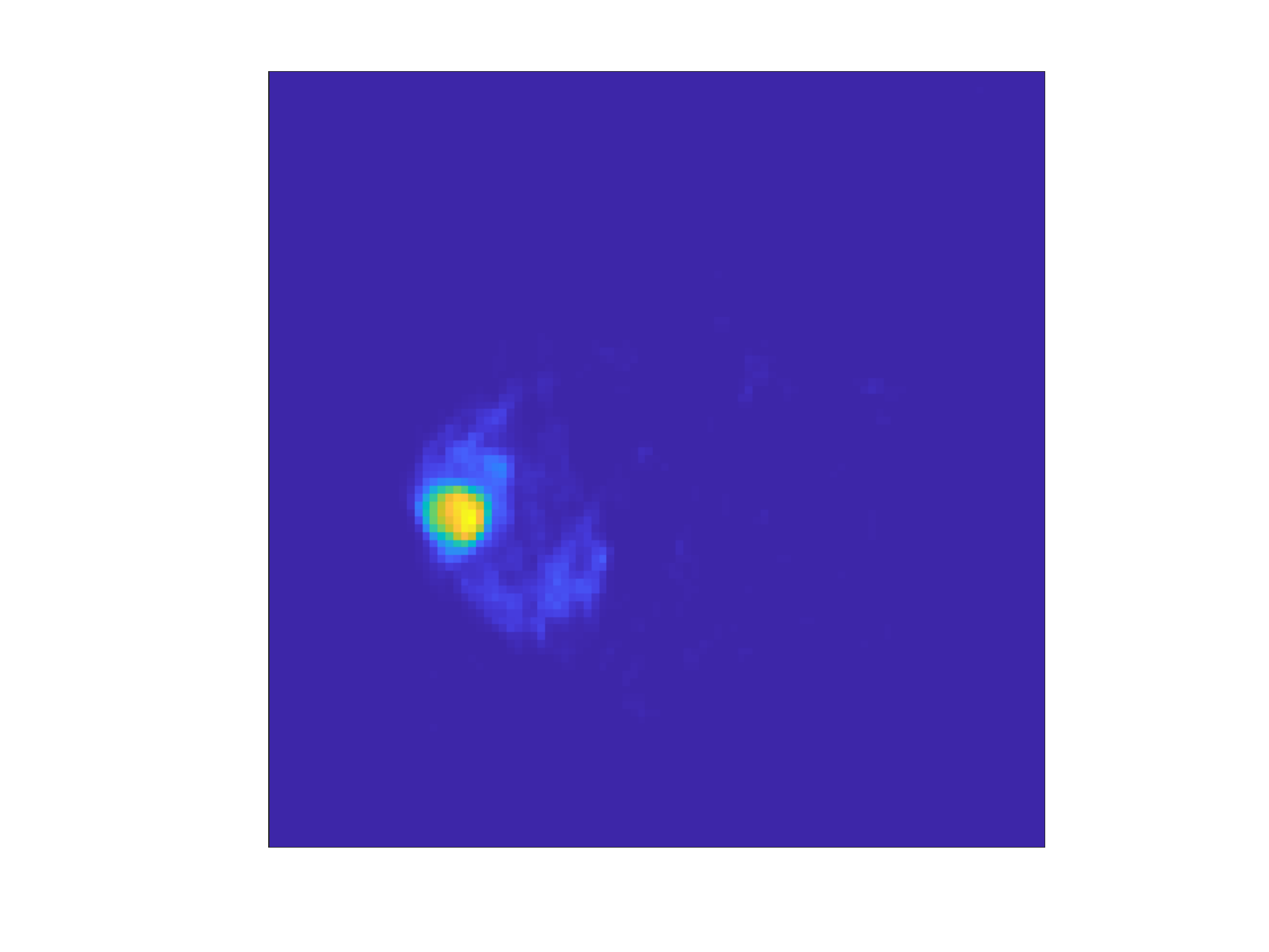}& 
\includegraphics[scale=0.20,  trim=20em 0em 20em 2em, clip]{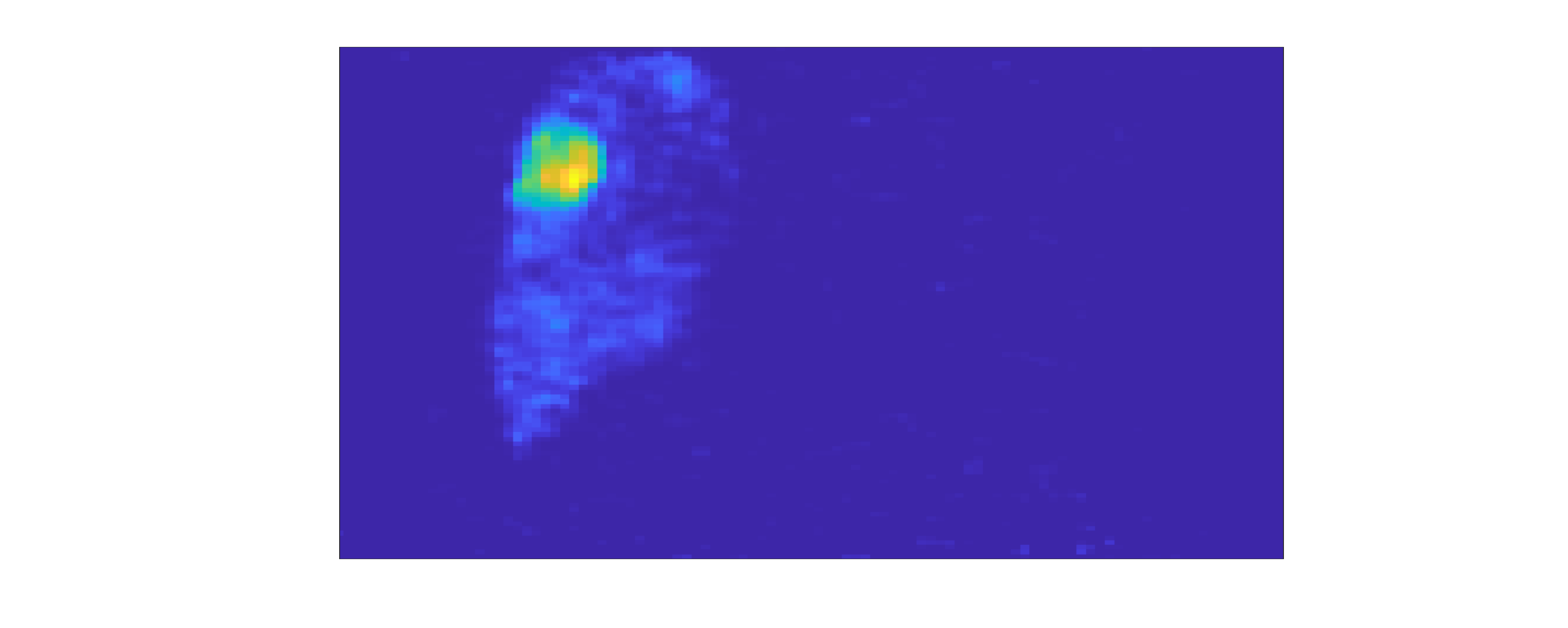} &
\includegraphics[scale=0.20, trim=10em 0em 10em 2em,  clip]{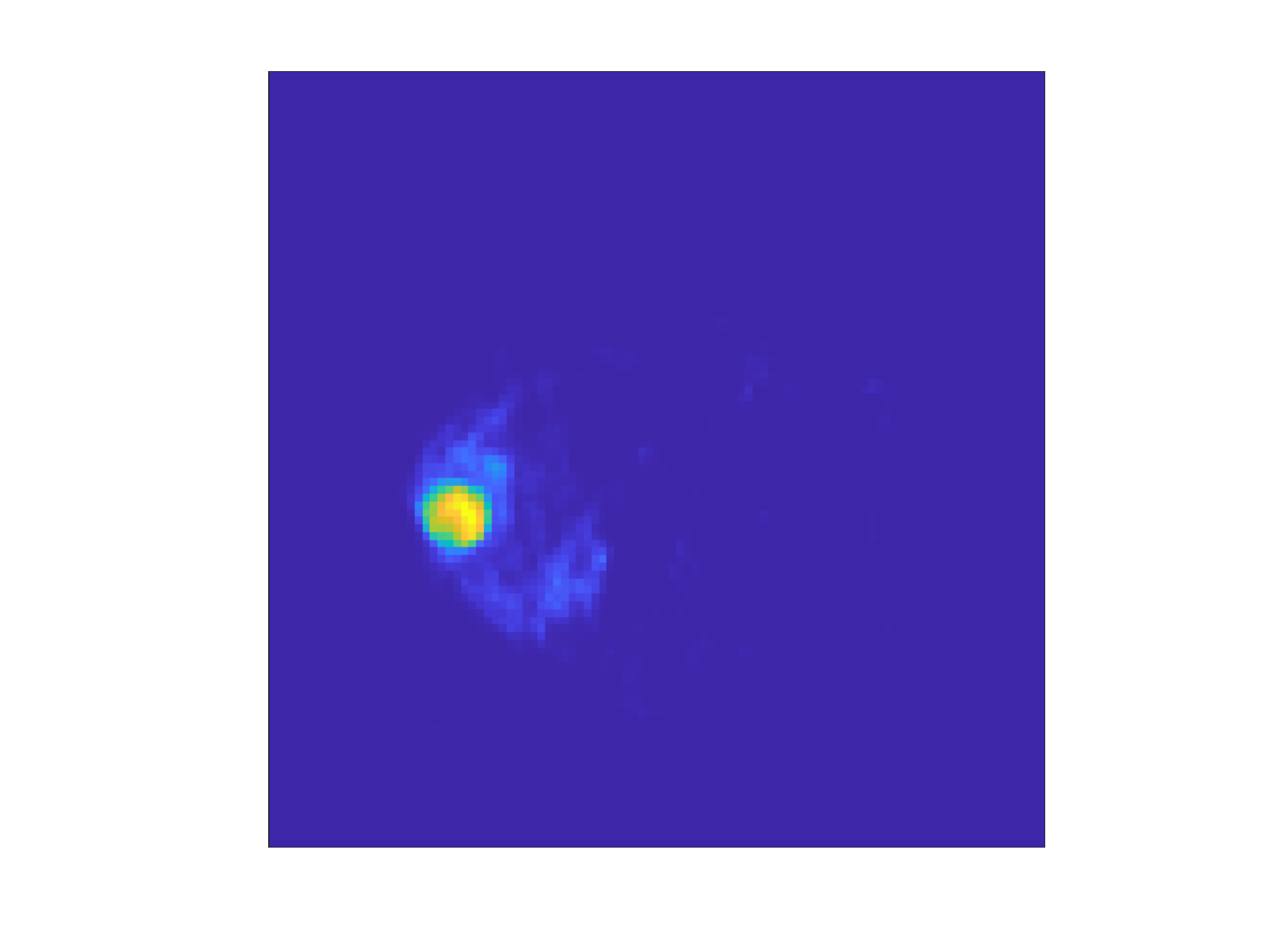} \vspace{-0.5em}\\
\end{tabular}
\caption{Y90 PET/CT patient measurement: Attenuation map and reconstructed images of one slice (coronal and axial view) using OSEM, TV, NLM, and BCD-Net. We visualized the reconstructed image of BCD-Net-UNet with 4 K parameters}
\label{fig6} 
\end{figure*}

\begin{figure*}[t]
\addtolength{\tabcolsep}{-6.5pt}
\centering
\begin{tabular}{cc}
 \\
\includegraphics[scale=0.30, trim=1em 0em 2em 0em, clip]{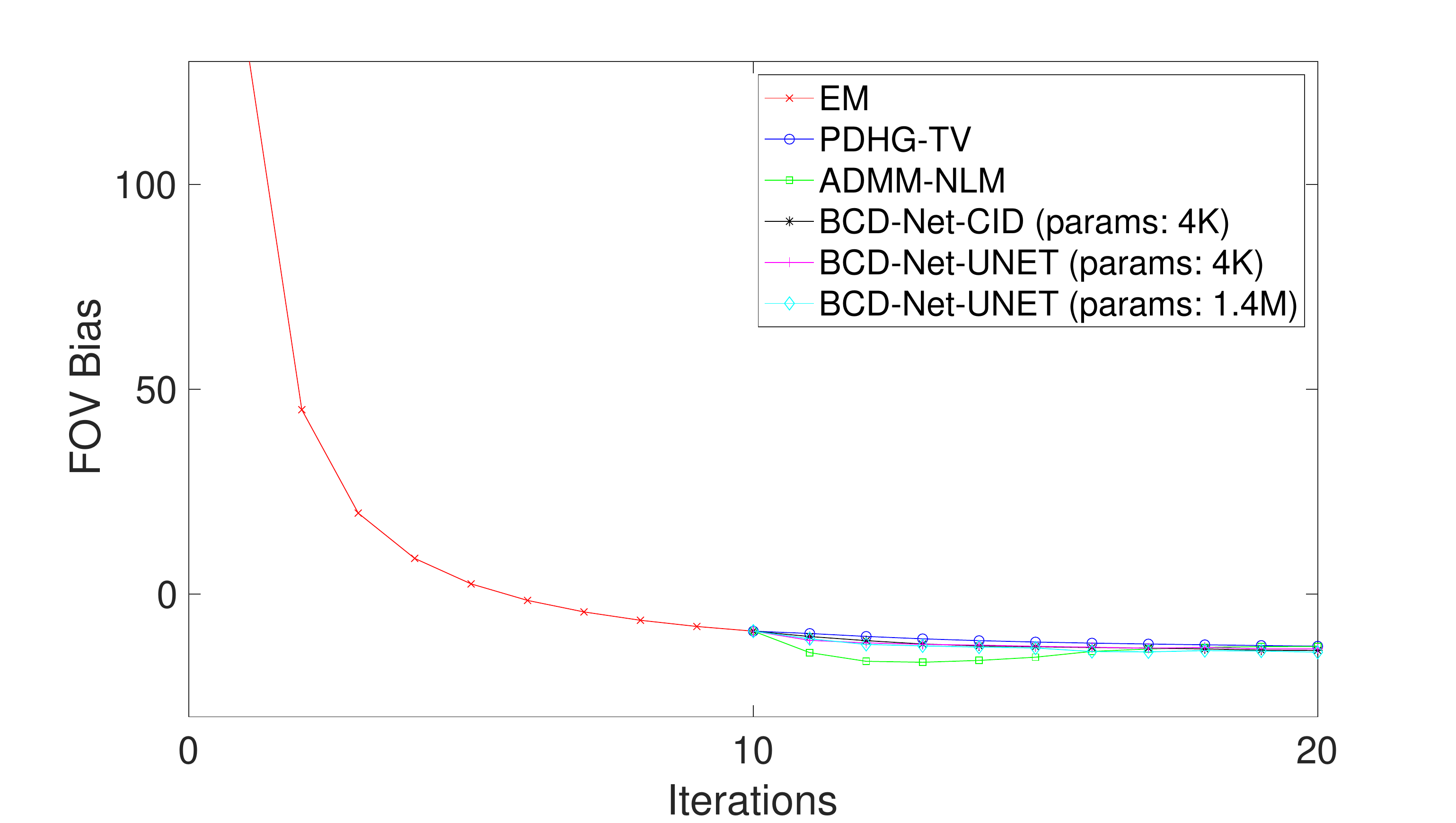} 
& \includegraphics[scale=0.30, trim=1em 0em 2em 0em, clip]{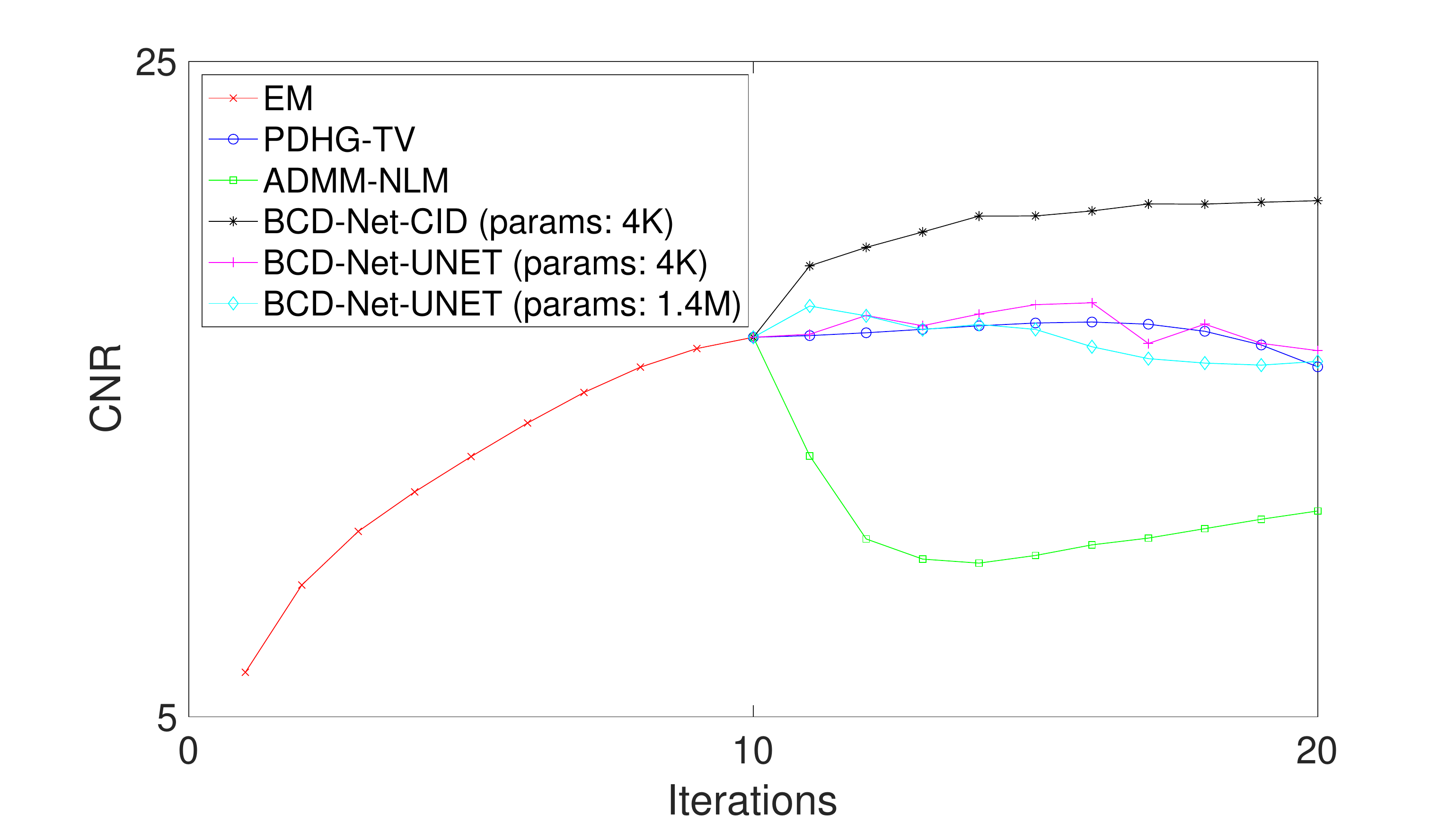} 
\end{tabular}
\caption{Patient scan: (a) Field of view bias vs iteration. BCD-Net shows similar results compared to other methods. (b) Contrast to noise ratio vs iteration. }
\label{fig7} 
\end{figure*}

\subsection{Training details}
We trained the denoising network in each iteration with a stochastic gradient descent method using the PyTorch \cite{paszke2017automatic} deep-learning library.



\subsubsection{BCD-Net with CID}
We trained a set of CID for the denoising module in BCD-Net where each iteration has 78 sets of thresholding values and convolutional encoding/decoding filters $(K=78)$. We set the size of each filter as $3\times3\times3~(R=3^3)$, and set the initial thresholding values by sorting the initial estimate of image and getting a 10\% largest value of sorted initial image. We used the Adam optimization method \cite{kingma2014adam} to train the NN. We applied the learning rate decay scheme. Due to the large size of 3D input, we set the batch size as 1.

\subsubsection{BCD-Net with U-Net}\label{sec:method:train:unet}
We implemented a 3-D version of U-Net  by modifying a shared code\footnote{https://github.com/facebookresearch/fastMRI} (implemented for denoising 2-D MRI images) for fastMRI challenge\cite{zbontar2018fastMRI}. We used a batch normalization layer instead of the instance normalization layer used in the baseline code. The `encoder' part of U-Net consists of multiple sets of 1) max pooling layer, 2) 3$\times$3$\times$3 convolutional layer, 3) batch normalization (BN) layer, 4) ReLU layer and the `decoder' part of U-Net consists of multiple sets of 1) upsampling with trilinear interpolation \cite{gong2019iterative}, 2) 3$\times$3$\times$3 convolutional layer, 3) BN layer, 4) ReLU layer. For training the U-Net, we used the same training dataset that we used for training the CID. We also used the Adam optimization method and identical settings (number of epochs, learning rate decay, batch size) as those of the CID. We trained and tested two different U-Nets sizes. At each BCD-Net iteration, the U-Net has either about 4~K (similar size to the CID) or 1.4~M trainable parameters. We set the number of convolutional filter channels of the first encoder layer as 12 with 4 times of contraction/expansion for the U-Net with 1.4~M parameters and 5 with 1 time of contraction/expansion for the U-Net with 4~K parameters.

\section{Results}\label{sec:result}
\subsection{Reconstruction setup}

We compared the proposed BCD-Net method to the standard EM (1 subset), TV-based MBIR with PDHG algorithm (PDHG-TV), and NLM-based MBIR with ADMM algorithm (ADMM-NLM). For regularized MBIR methods including BCD-Net, we used 10 EM algorithm iterations to get the initial image $\bs{x}^{(0)}$. For each regularization method, we finely tuned the regularization parameter $\beta$ (within range $[2^{-15}, 2^{15}]$) by considering the recovery accuracy and noise. For NLM, we additionally tuned the window and search sizes. For the XCAT simulation data, we used 40 iterations for EM and 30 iterations $(T=30)$ for PDHG-TV, ADMM-NLM, and BCD-Net. We used 1 inner-iteration $(T'=1)$ for the reconstruction module (\ref{eq:x-update-meas}) for each outer-iteration of BCD-Net.  For the measured data, we used 20 iterations for EM and 10 iterations $(T=10)$ for PDHG-TV, ADMM-NLM, and BCD-Net. We used 1 inner-iteration $(T'=1)$ for the reconstruction module (\ref{eq:x-update-meas}). We set $c = 0.01$ in (\ref{eq:beta_sel}) in the XCAT simulation study and $c = 0.005$ in both the phantom measurement and patient studies.

\subsection{Results: Reconstruction (testing) on simulation data} \label{sec:result:sim} 
Fig.~\ref{fig2}-\ref{fig3} shows that the proposed iterative NN, BCD-Net, significantly improves overall reconstruction performance over the other non-trained regularized MBIR methods. Fig.~\ref{fig3} reports averaged evaluation metrics over realizations. Fig.~\ref{fig3} shows that BCD-Net with a trained CID achieves the best results in most evaluation metrics. In particular, BCD-Net with a CID improves CNR and RMSE compared to PDHG-TV and ADMM-NLM. BCD-Net also improved contrast recovery in the cold region while not increasing noise compared to the initial EM reconstruction,  whereas PHDG-TV and ADMM-NLM improved noise while degrading the CR. For Fig.~\ref{fig2}, we selected the iteration number for EM to obtain the highest CNR and the last iteration number for other methods. Fig.~\ref{fig2} shows that BCD-Net's reconstructed image with a CID is closest to the true image whereas PHDG-TV and ADMM-NLM exceedingly blur the cold region. BCD-Net with the U-Net denoiser shows good recovery for the cold region, however, it blurs the hot region. Moreover, the larger sized U-Net (params: 1.4$~$M) denoiser worsens the performance of BCD-Net possibly due to over-fitting the training dataset.

\subsection{Results: Reconstruction (testing) on measurement data} \label{sec:result:meas}

\subsubsection{Phantom study}
Similar to the simulation results,
Fig. 4-5 shows that, BCD-Net improved overall reconstruction performance over the other reconstruction methods. Fig.~\ref{fig4} shows that reconstructed images using PHDG-TV and ADMM-NLM show uniform texture in background liver compared to EM, however, those exceedingly blur around hot spheres. The blurred hot region is more evident in the quantification results in Fig.~\ref{fig5}. BCD-Net gives more visibility for hot spheres with noisier texture in uniform liver region. Fig.~\ref{fig5} shows that BCD-Net with a CID improves CNR compared to PDHG-TV and ADMM-NLM. BCD-Net with CID also improved contrast recovery in hot spheres while slightly increasing noise compared to the initial EM reconstruction. In Fig.~\ref{fig5} (a), BCD-Net with U-Net denoiser shows a fluctuation with iterations, however, the plot trend is similar to that of BCD-Net with CID. 

\subsubsection{Patient study}
Because of the unknown true activity distribution, we quantitatively evaluated each reconstruction method with FOV activity bias. In this quantitative evaluation, BCD-Net showed similar results compared to other methods. See Fig.~\ref{fig6}-\ref{fig7}. Fig.~\ref{fig6} shows that the quality of image using different methods in patient study is similar to that of phantom measurement study shown in Fig.~\ref{fig4}. Fig.~\ref{fig7} (b) shows that the CNR trend in the patient study is similar to that of the XCAT simulation and the liver phantom measurement.

\section{Discussion} \label{sec:discussion}

In this study we showed the efficacy of trained BCD-Net on both qualitative and quantitative Y-90 PET/CT imaging and compared between conventional non-trained regularizers. The proposed approach uses learned denoising NNs to lift estimated signals and thresholding operations to remove unwanted signals. In particular, the iterative framework of BCD-Net enables one to train the filters and thresholding values to deal with the different image roughness at its each iteration. We experimentally demonstrate its generalization capabilities with simulation and measurement data. In the XCAT PET/CT simulation with activity distributions and count-rates mimicking Y-90 PET imaging, total counts in the cold spot were overestimated with standard reconstruction and other MBIR methods using non-trained regularization, yet approached the true value with the proposed approach. Improvements were also demonstrated for the measurement data where we used training and testing datasets having very different activity distribution and count-levels. The architecture and size of denoising NN significantly affect the performance of BCD-Net. In both simulation and measurement experiments, the CID outperformed the U-Net architectures. Using a U-Net with more trainable
parameters degraded the performance, especially in the simulation study, due to the small size of dataset. Size of the denoising NN should be set with consideration of training dataset size.





\begin{figure}[t!]
\small\addtolength{\tabcolsep}{-6.5pt}
\centering

\begin{tabular}{cc}
(a) Impact of number of filter & (b) Impact of size of filter\\
(Fixed $R = 3^3$)  & (Fixed $K = 2^5$)\\
\includegraphics[scale=0.24, trim=1em 2em 2em 0em, clip]{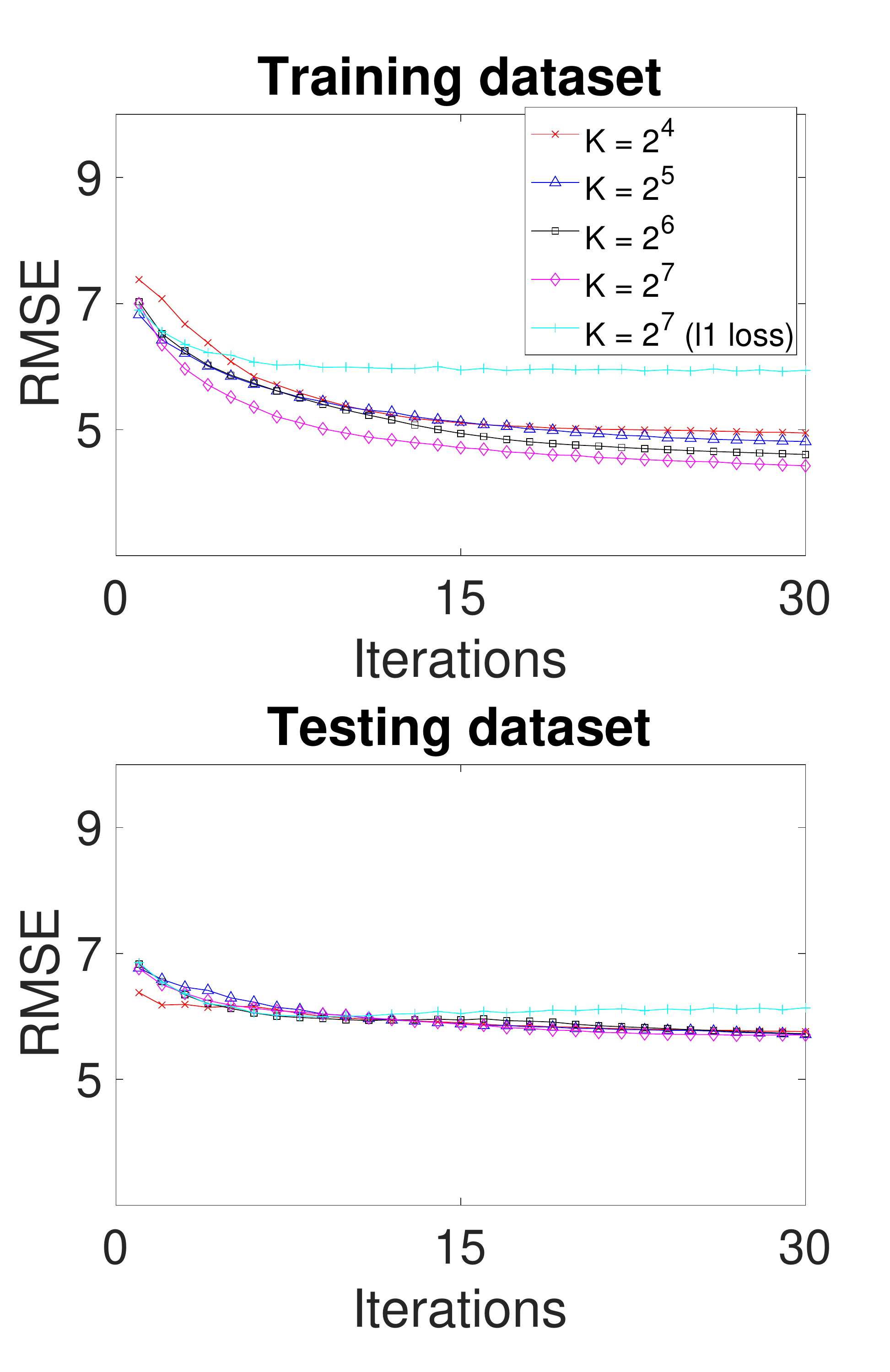} &
\includegraphics[scale=0.24, trim=1em 2em 2em 0em, clip]{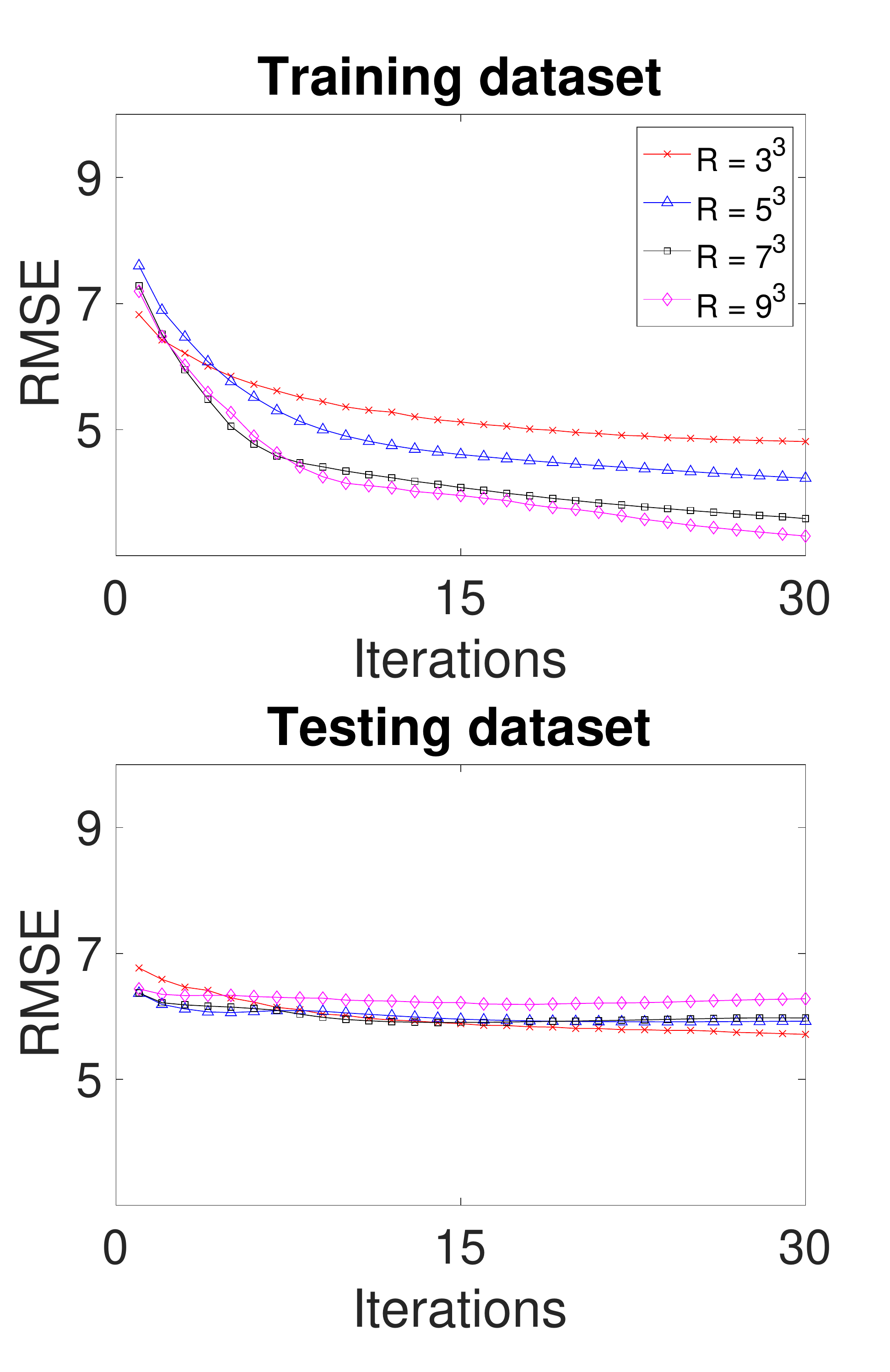} \\
\end{tabular}
\begin{tabular}{ccc}
(c) Trained with $l1$ loss & (d) $K=2^4$ & (e) $R=9^3$ \vspace{-0.2em}\\
\includegraphics[scale=0.185, trim=6.3em 0em 5.2em 0em, clip]{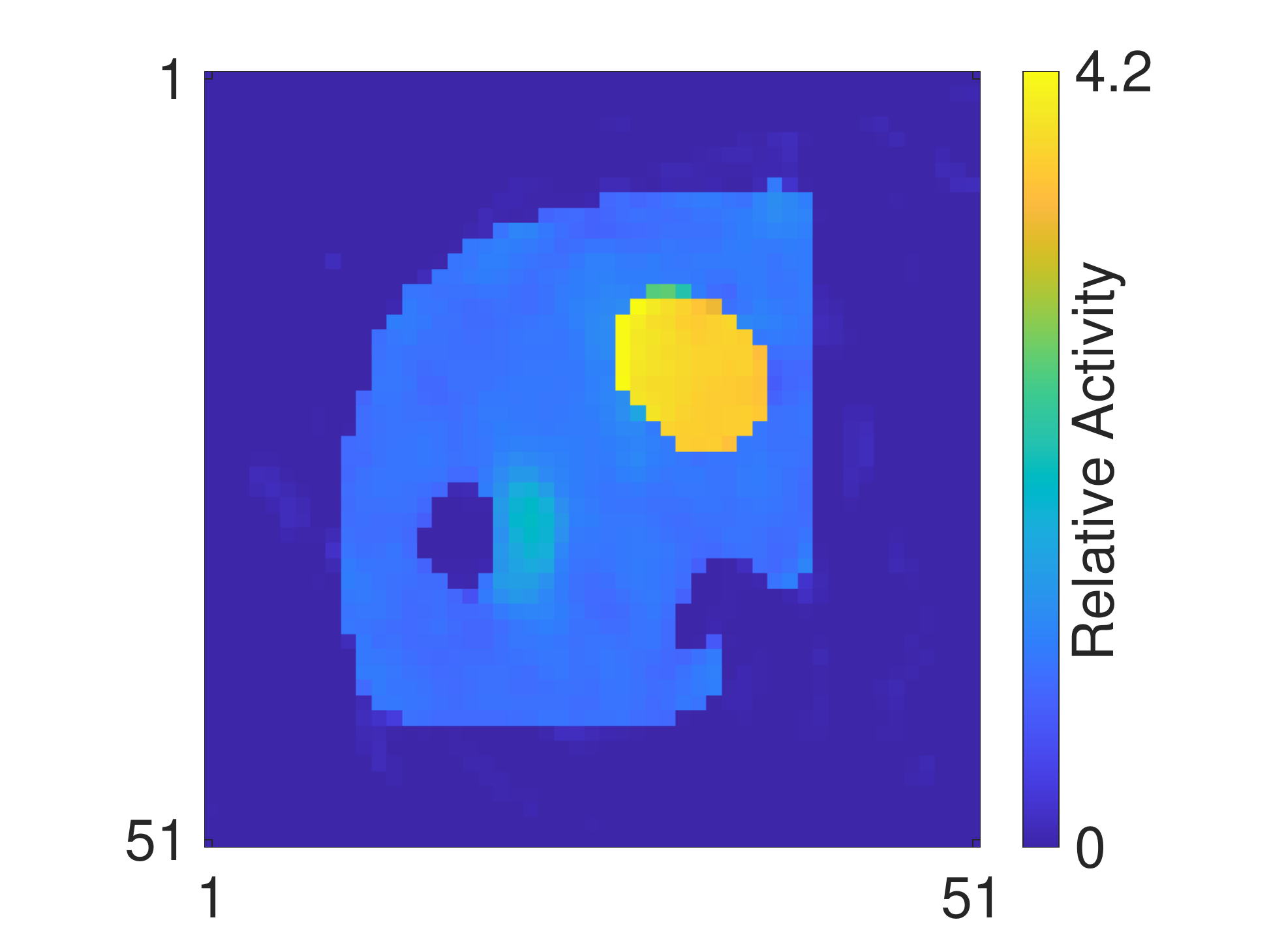} &
\includegraphics[scale=0.185, trim=6.3em 0em 5.2em 0em, clip]{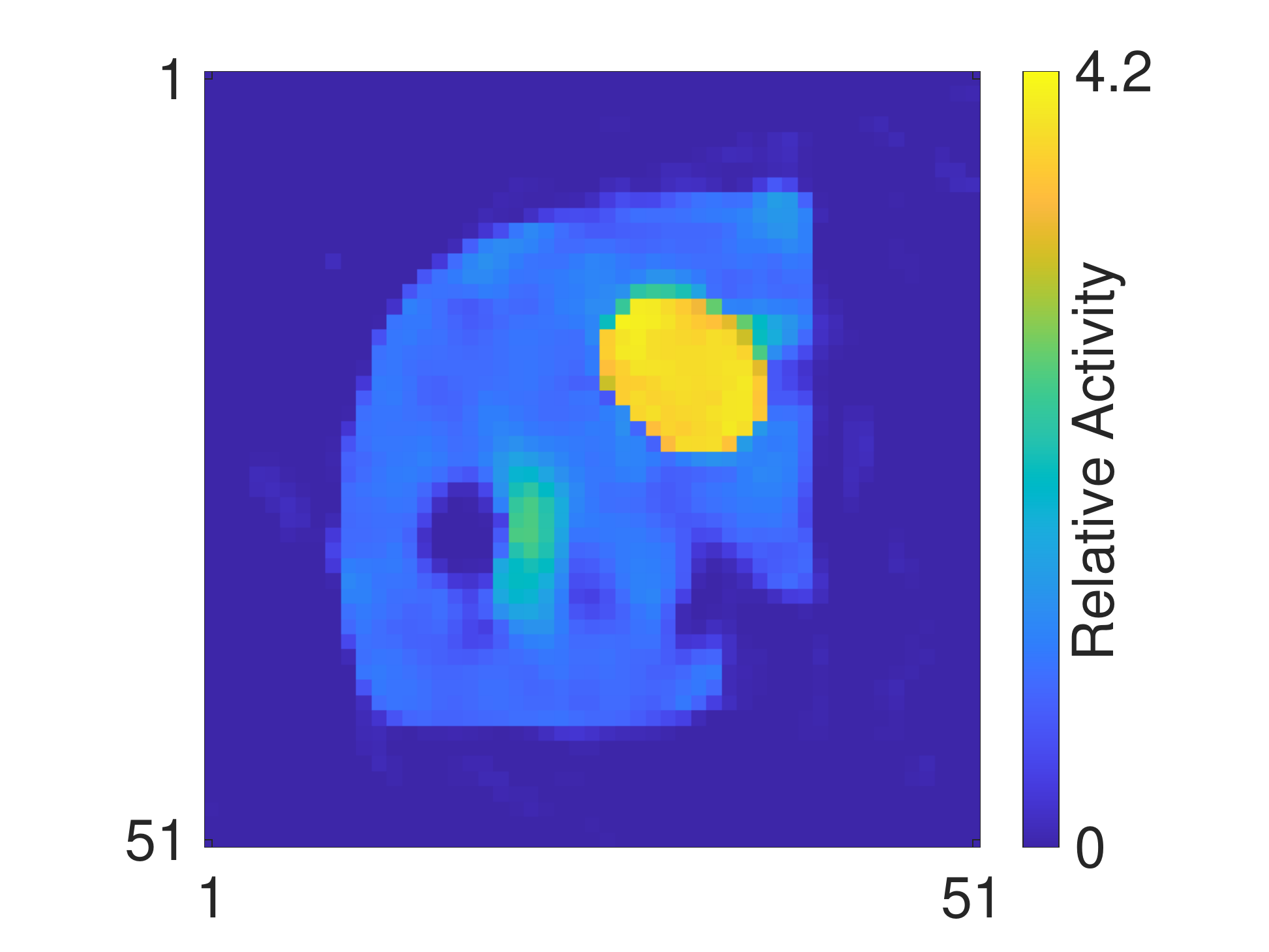} &
\includegraphics[scale=0.185, trim=6.3em 0em 5.2em 0em, clip]{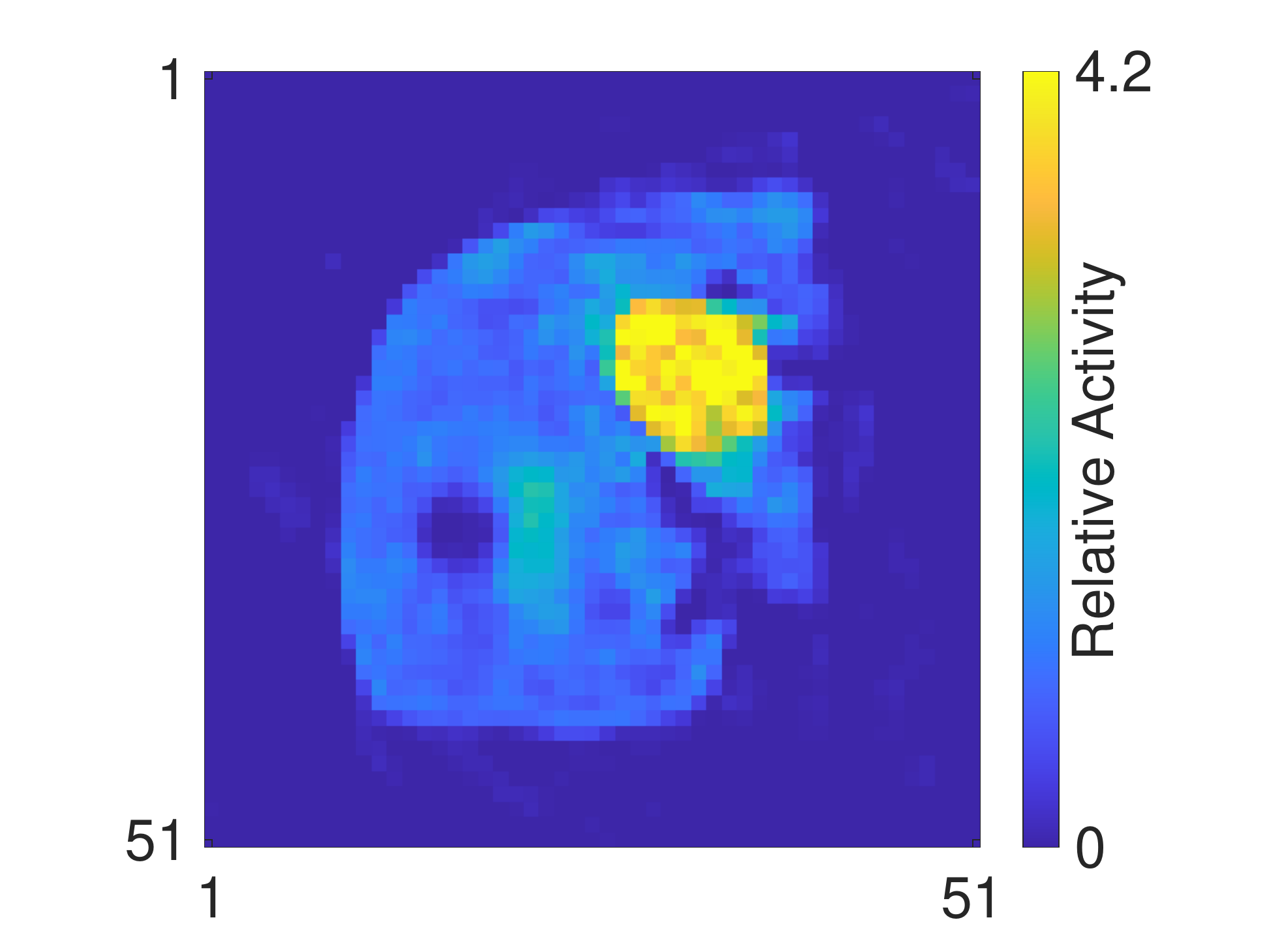} 
\end{tabular}
\caption{(a)-(b) Impact of number/size of filter and training loss on testing dataset RMSE. (c) Reconstructed image from BCD-Net-CID with filters and thresholding values trained with $l1$-loss.  }
\label{fig8}

\end{figure}



\begin{figure}[t!]
\small\addtolength{\tabcolsep}{-6.5pt}
\centering
\begin{tabular}{c}
$\beta$ in training and testing case when $c=0.005$ \\
\includegraphics[scale=0.33, trim=3em 0em 3em 1em, clip]{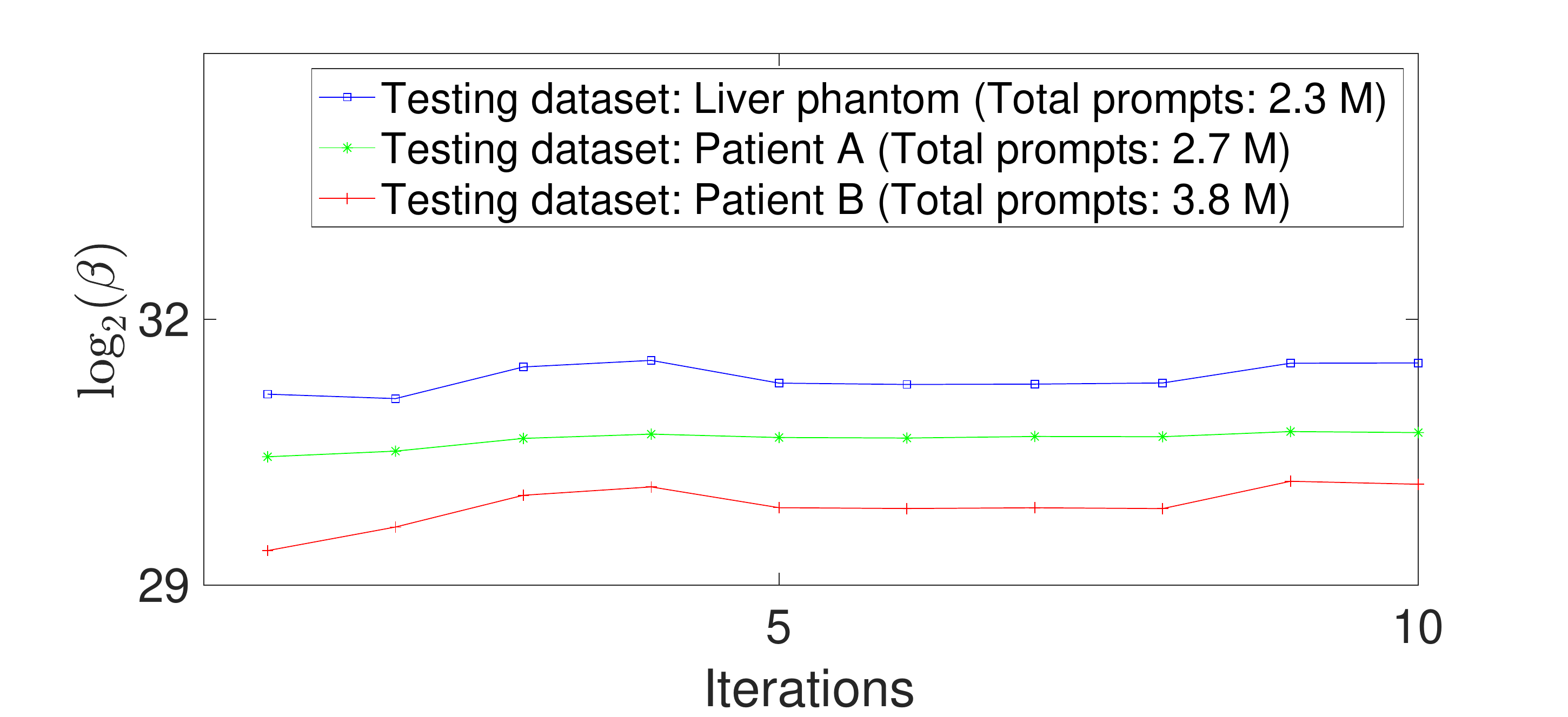} \\
\end{tabular}
\caption{Efficacy of adaptive selection of regularization parameter $\beta$.}
\label{fig10} 
\end{figure}

\begin{table}[t]
    \begin{center}
    \caption{Impact of imaging variable on generalization capability of BCD-Net-CID. }
    \begin{tabular}{|c|c|c|c|c|}
\cline{1-5}
Changed imaging variable & Training & Testing & RMSE & Drop (\%) \\ \cline{1-5} 
Identical & \multicolumn{2}{|c|}{-} & 4.74 & - \\ \cline{1-5} 
Shape and size & \multicolumn{2}{|c|}{See Fig.~\ref{fig2}} & 5.49 & 15.9 \\ \cline{1-5} 
Concentration ratio & 9:1 & 4:1 & 5.55 &  17.1   \\ \cline{1-5} 
Concentration ratio &  1.7:1 & 4:1 & 5.81 &22.5 \\ \cline{1-5} 
Trues Count-level &  $2\times10^5$ & $5\times10^5$ & 5.01  & 5.7 \\ \cline{1-5} 
Trues Count-level & $11\times10^5$ & $5\times10^5$ & 5.71 & 20.5 \\ \cline{1-5} 
\end{tabular}
\label{table4}
    \end{center}
\end{table}

\begin{table}[t]
    \begin{center}
    \caption{Comparison between post-reconstruction processing and iterative NN. }
    \begin{tabular}{|c|c|c|c|}
\cline{1-4}
 \multicolumn{1}{|c|}{Method} & $g_2\left(\bs{u}^{(1)}\right)$ RMSE & $\bs{x}^{(30)}$ RMSE & $\Delta$ (\%) \\ \cline{1-4} 
CID (params: 4~k) & 6.87 & 6.36 & 7.52 \\  \cline{1-4} 
U-Net (params: 4~k) & 7.39 & 6.70 & 9.30 \\  \cline{1-4} 
U-Net (params: 1.4~M) & 7.67 & 7.04 & 8.15 \\  \cline{1-4} 
\end{tabular}
\label{table5}
    \end{center}
\end{table}

\begin{figure}[t!]
\small\addtolength{\tabcolsep}{-6.5pt}
\centering

\begin{tabular}{ccc}
(a) CID   & (b) U-Net & (c) U-Net \\

(params: 4~K) & (params: 4~K) & (params: 1.4~M)\\ 
\includegraphics[scale=0.185, trim=6.3em 0em 5.2em 0em, clip]{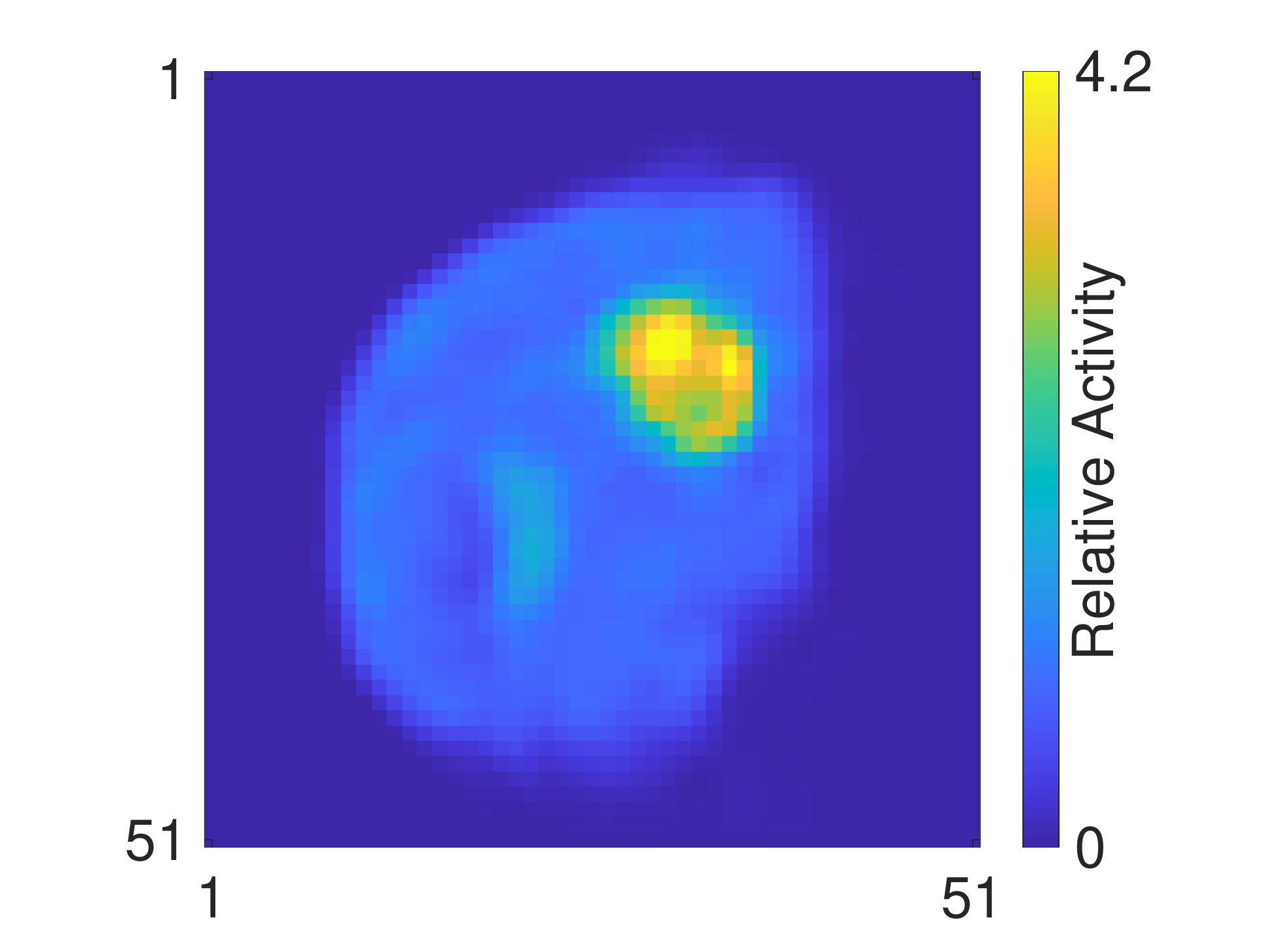} &
\includegraphics[scale=0.185, trim=6.3em 0em 5.2em 0em, clip]{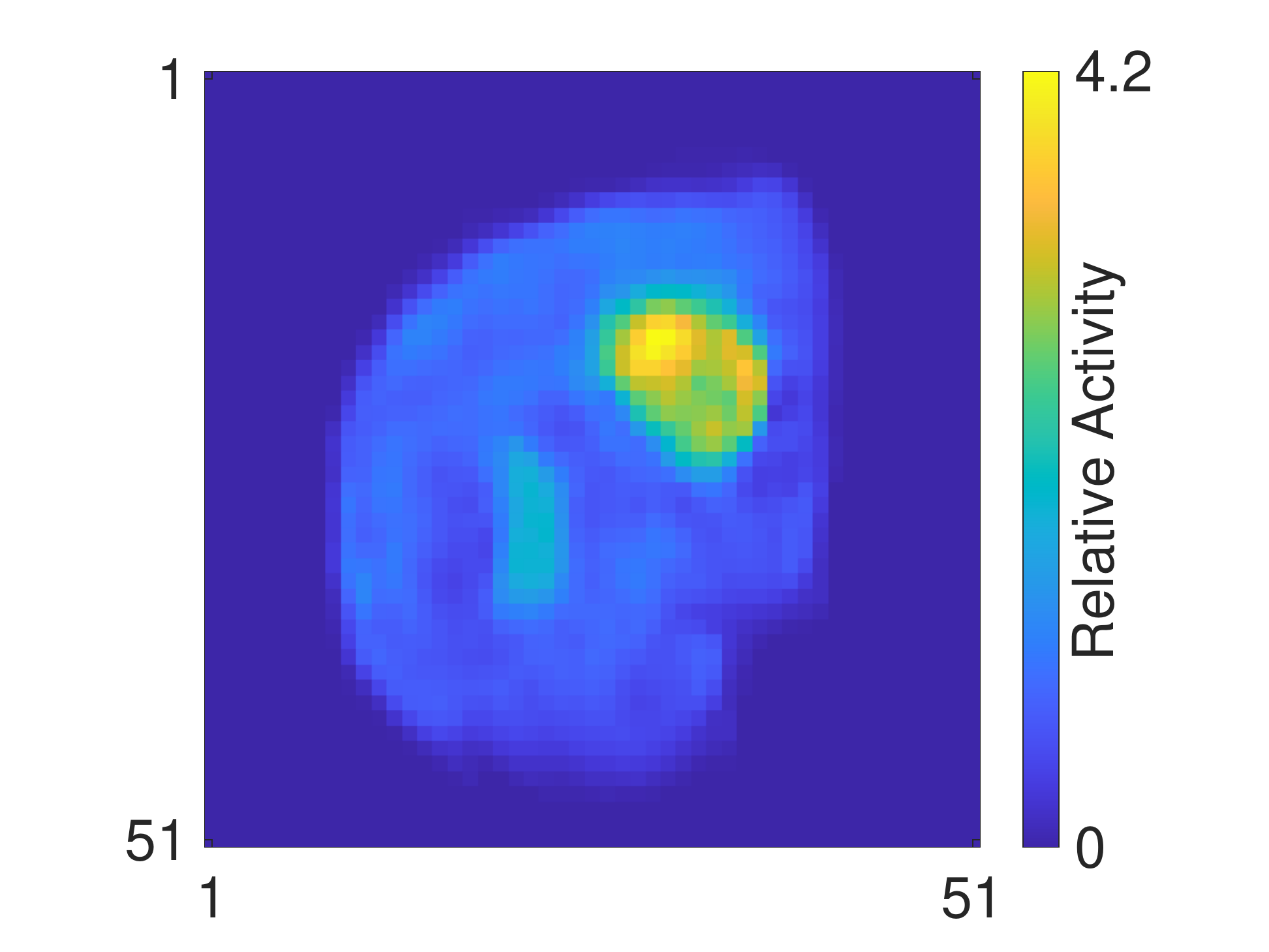} &
\includegraphics[scale=0.185, trim=6.3em 0em 5.2em 0em, clip]{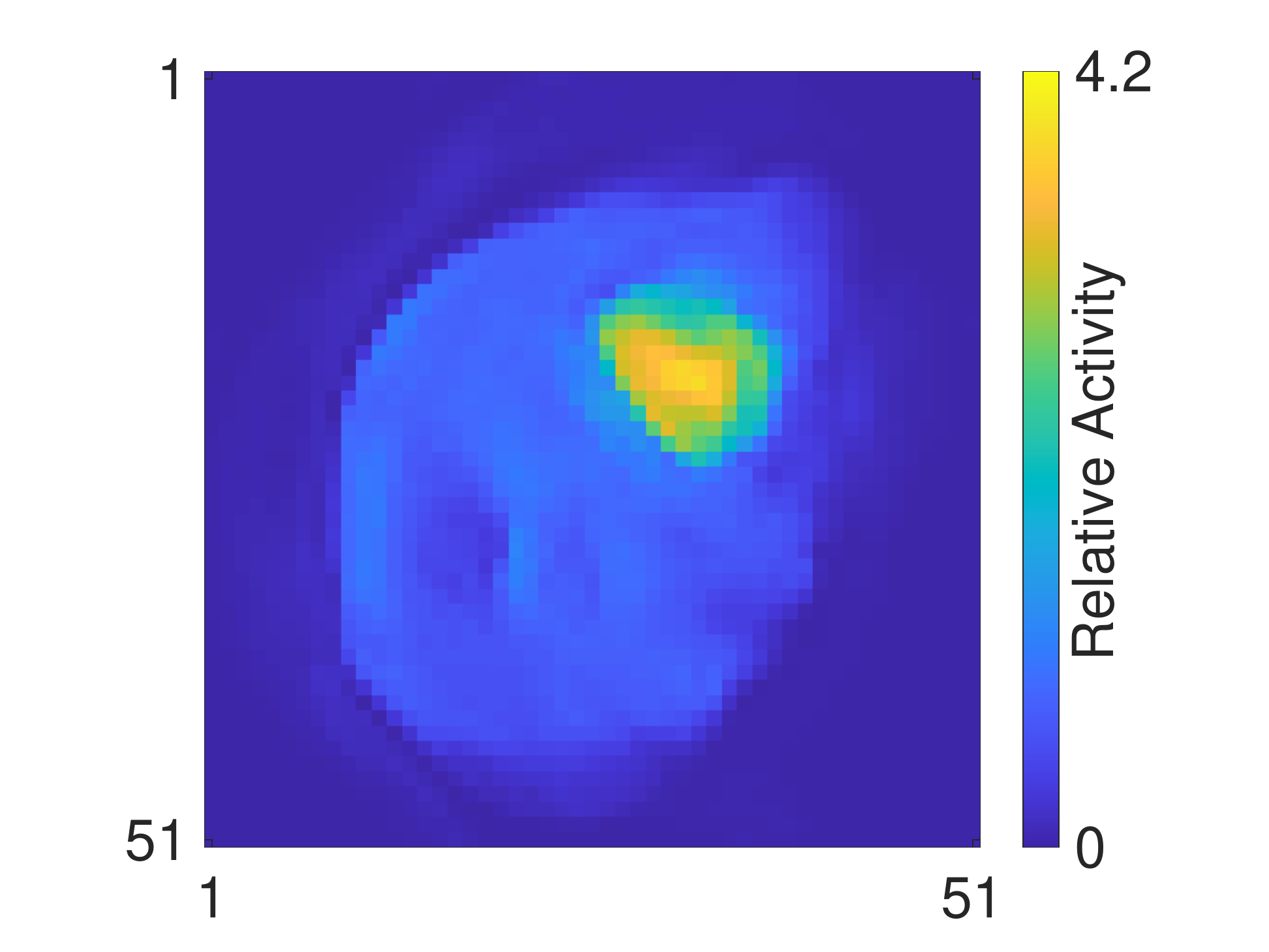} \\
\end{tabular}
\caption{$g_2(\bs{u}^{(1)})$ generated by (a) CID and (b)-(c) U-Net. Using more parameters degraded the visual quality and RMSE value as in the iterative NN.  }
\label{fig11} 

\end{figure}

We tested which imaging variable most affects the generalization performance of the proposed BCD-Net. 
Table~\ref{table4} shows how BCD-Net performs when training and testing data had the same activity distribution and count-level (only difference is Poisson noise) and how the performance of BCD-Net is degraded when each imaging variable is changed between training and testing dataset. We changed one of three factors (shape and size of tumor and liver, concentration ratio, count-level) in training dataset compared to testing dataset. The result shows that generalization performance of the proposed BCD-Net depends largely on all imaging variables. 
However, training with higher contrast and lower count-level dataset (compared to testing dataset) gave less degradation of performance compared to the opposite cases. This result suggests that it is better to have noisier data in training dataset than testing dataset. In other words, training for extra noise reduction than needed is better than less noise reduction than needed. 

We also investigated how each factor in training of denoising module \R{eq:u-update} impacts the generalization capability of BCD-Net. Fig.~\ref{fig8}(a)-(b) show the impact of number and size of filters on performance. Plots show that the proposed BCD-Net achieved lower training RMSE when using larger number and size of filters; however, it did not decrease testing RMSE compared to smaller number and size of filters and BCD-Net with larger size of filter exceedingly blurs image thereby resulting in higher RMSE. See Fig.~\ref{fig8}(e). We also tested $l1$ training loss to see if it improves the performance over the $l2$ loss (MSE) in (\ref{eq:training}). However, it led to unnaturally piece-wise constant images and details in small cold regions were ignored.


Fig.~\ref{fig10} shows how the regularization parameter $\beta$ in (\ref{eq:beta_sel}) changes with iterations in training and testing datasets. The $\beta$ value in each iteration converges to different limits in training and testing cases. The adaptive scheme automatically increases the $\beta$ value when the count-level decreases. This behavior concurs with the general knowledge that more regularization is needed when the noise-level increases. These empirical results underscore the importance of such adaptive regularization parameter selection schemes proposed in Section~\ref{sec:method:meas:reg} in PET imaging. 

Many related works \cite{jin2017deep,kang2017deep,xu2017200x,yang:18:ann} use single image denoising (deep) NN (e.g., U-Net) as a post-reconstruction processing and we investigated how the denoising NN detached from the data-fit term performs compared to iterative NN. Fig.~\ref{fig11} illustrates $\bs{u}^{(1)}$ generated by CID and U-Net. As in the iterative NN, using more trainable parameters degraded the visual quality and RMSE value in U-Net case and CID achieved better result than U-Net. In all cases, iterative NN achieved lower RMSE compared to those post-reconstruction processed images as shown in Table~\ref{table5}.

BCD-Net is trained for a specific number of iterations and its practical use would be akin to how ML-EM is used with a fixed number of iterations in clinical systems.
If one is interested in convergence guarantees with running more iterations,
then one can extend the sequence convergence guarantee of BCD-Net in \cite{chun2019momentum} by setting the $n$th “adaptive” denoiser as $\widetilde{\mathcal{D}}^{(n)} = g_2 ( \mathcal{D}^{(n)} ( g_1 ( \bs{x}^{(n-1)} ) ) )$ with some $n$th denoiser $\mathcal{D}^{(n)}$ (e.g., CID (7) and U-Net), $\forall n$, using sufficient $T'$ (so MAP EM finds a critical point), and additionally assuming that $\beta^{(n)}$ converges. We empirically observed that the $\beta^{(n)}$ tends to converge to some constant in this Y-90 PET as well as another application of Lu-177 SPECT.

To more practically guarantee the convergence, one could use training and testing dataset having similar count-level and a fixed regularization parameter value across iterations using an initial estimated image and a corresponding denoised image as follows:
\begin{align*}
    \beta = \frac{\left\|\nabla_{\bs{x}} f(\bs{x}^{(0)})\right\|_2}{\left\|\nabla_{\bs{x}} \mathsf{R}(\bs{x}^{(0)})\right\|_2} \cdot c \nonumber = \frac{\left\|a_j - e_j(\boldsymbol{x}^{(0)}) \right\|_2}{\left\| \bs{x}^{(0)} - g_2(\bs{u}^{(1)})\right\|_2} \cdot c.
\end{align*}
The convergence properties depend on additional technical assumptions detailed in \cite{chun2019momentum}.


\section{Conclusion}\label{sec:conclusion}
It is important for a ``learned'' regularizer to have generalization capability to help ensure good performance when applying it to an unseen dataset. 
For low-count PET reconstruction, the proposed iterative NN, BCD-Net, showed reliable generalization capability even when the training dataset is small.
The proposed BCD-Net achieved significant qualitative and quantitative improvements over the conventional MBIR methods using ``hand-crafted'' non-trained regularizers: TV and NLM. In particular, these conventional MBIR methods have a trade-off between noise and recovery accuracy, whereas the proposed BCD-Net improves CR for hot regions while not increasing the noise when the regularization parameter is appropriately set. Visual comparisons of the reconstructed images also show that the proposed BCD-Net significantly improves PET image reconstruction performance compared to MBIR methods using non-trained regularizers.

Future work includes investigating performance of BCD-Net trained with end-to-end training principles and adaptive selection of trainable parameter numbers depending on the size of training dataset.

\section{ACKNOWLEDGMENT}
We acknowledge Se Young Chun (UNIST) for providing NLM regularization codes. We also acknowledge Maurizio Conti and Deepak Bharkhada (SIEMENS Healthcare Molecular Imaging) for providing the forward/back projector for TOF measurement data. This work was supported by NIH-NIBIB grant R01EB022075.


{\tiny
\bibliographystyle{IEEEtran}
\bibliography{refs}
}

\end{document}